\documentclass{aa}  
\bibpunct{(}{)}{;}{a}{}{,}

\usepackage{graphicx}
\usepackage{txfonts}
\newcommand{\teff}{\ensuremath{T_{\rm eff}}}

\newcommand{\bz}{\ensuremath{\langle B_z \rangle}}
\newcommand{\bs}{\ensuremath{\langle \vert B \vert \rangle}}

\newcommand{\te}{\ensuremath{T_{\rm eff}}}

\newcommand{\nvar}{\ensuremath{N_{\rm K}}}
\newcommand{\ntot}{\ensuremath{N_{\rm tot}}}
\newcommand{\pphot}{\ensuremath{P_{\rm phot}}}

\begin{document}

   \title{Strong magnetic fields of old white dwarfs \\ are symmetric about the stellar rotation axes}
\titlerunning{Strong fields of old white dwarfs are axisymmetric}

   \author{S. Bagnulo
          \inst{1}
          \and
          J.D. Landstreet
          \inst{1,2}
          }

   \institute{Armagh Observatory and Planetarium, College Hill, Armagh BT61 9DG, Northern Ireland, UK 
         \and
              Dept. of Physics \& Astronomy, University of Western Ontario, London, Ontario N6A 3K7, Canada
             }

   \date{Received July 5, 2024, accepted October 22, 2024}

 
  \abstract{
    Many magnetic white dwarfs exhibit a polarised spectrum that periodically varies as the star rotates because the magnetic field is not symmetric about the rotation axis. In this work, we report the discovery that while weakly magnetic white dwarfs of all ages with $M \le 1\,M_\odot$ show polarimetric variability with a period between hours and several days, the large majority of magnetic white dwarfs in the same mass range with cooling ages older than 2\,Gyr and field strengths $\ge 10$\,MG show little or no polarimetric variability. This could be interpreted as extremely slow rotation, but a lack of known white dwarfs with measured periods longer than two weeks means that we do not see white dwarfs slowing their rotation. We therefore suggest a different interpretation: old strongly magnetic white dwarfs do not vary because their fields are roughly symmetric about the rotation axes. Symmetry may either be a consequence of field evolution or a physical characteristic intrinsic to the way strong fields are generated in older stars. Specifically, a strong magnetic field could distort the shape of a star, forcing the principal axis of maximum inertia away from the spin axis. Eventually, as a result of energy dissipation, the magnetic axis will align with the angular momentum axis. Alternatively, symmetry could be the hallmark of a dynamo that operates after the beginning of core crystallisation.
    We also find that the higher-mass strongly magnetised white dwarfs, which are likely the products of the merging of two white dwarfs, may appear as either polarimetrically variable or constant. This may be the symptom of two different formation channels or the consequence of the fact that a dynamo operating during a merger may produce diverse magnetic configurations. Alternatively, the massive white dwarfs with constant polarisation may be rotating with periods much shorter than the typical exposure times of the observations. 
  }

   \maketitle
%
\section{Introduction}
Most white dwarfs do not show any variability, and more than 97\% of them can be safely considered as suitable flux standard stars \citep{Heretal17}. Because of flux stability, it is generally impossible to measure the rotation period of white dwarfs except when they pulsate or when they have a magnetic field. 

It is well known that magnetic white dwarfs often show periodically variable signals of polarisation and sometimes also periodic photometric variability. In the local 20\,pc volume, more than 20\% of white dwarfs have a magnetic field \citep{BagLan21}, and about 20\% of the magnetic white dwarfs are photometrically variable \citep{Faretal24}. Polarimetric variability is explained in terms of a magnetic field not symmetric about the rotation axis so that the magnetic configuration seen by the observer (as encoded in the polarisation signal) varies as the star rotates. The causes of photometric variability are not well understood \citep[see, e.g.,][]{Bagetal24b}, yet photometric variability of non-pulsating white dwarfs is so clearly correlated with the presence of the magnetic field variations that one may hypothesise that all non-pulsating white dwarfs with light variations are magnetic. If a star belongs to the small class of white dwarfs with a well-detected light curve, then the rotational period may usually be deduced from photometry \citep[see][]{Brietal13,Heretal24,Olietal24}, but, in general, the time series of polarisation measurements may allow one to recover the rotational period of a larger sample of magnetic white dwarfs.

Investigations of magnetic fields and of rotational periods are closely related not only because the presence of a magnetic field enables period measurements but also because one could expect that the occurrence of a magnetic field and its strength is physically related to stellar rotation. For example, should the field be supported by a dynamo, one could predict fields to be stronger in more rapidly rotating stars than in those that are more slowly rotating.  Correlations between the presence of a magnetic field and the rotational period of a star exist even when the field has a fossil origin. Most of the chemically peculiar stars of the main sequence (Ap and Bp stars) have a fossil  magnetic field, and they all rotate more slowly than the non-magnetic non-chemically peculiar stars in the same region of the Hertzsprung-Russel diagram \citep[e.g.][]{DonLan09}. Some of the magnetic Ap and Bp stars have well measured rotational periods as long as several months, years, or even decades \citep[e.g. $\gamma$ Equ,][]{Leretal93}. There is no clear explanation for this well-established correlation between magnetic fields and slow stellar rotation. 

In the case of white dwarfs, various samples of data have gradually become available in the past few years to constrain possible scenarios for the evolution of their magnetic fields. It now appears that among the normal-mass white dwarfs ($M \simeq 0.6 M_\odot$) that largely descend from single-star evolution, magnetic fields are very rare and weak during the first 1--2~Gyr of cooling but then gradually become much more common and often very strong after about 2--3~Gyr of cooling \citep{BagLan21}. This evolution may occur as a result of gradual relaxation to the surface of fields present in stellar cores during earlier evolution and/or as a result of the operation of a dynamo during core crystallisation \citep{Iseetal17,Schetal21,BagLan21,BagLan22,Ginetal22}. Among normal-mass stars, those with a magnetic field seem to rotate faster than non-magnetic white dwarfs \citep{Heretal24}, supporting the idea that {crystallisation-driven} dynamo may play a role in the formation of their magnetic fields. Many highly massive white dwarfs ($M\ge 1.0\,M_\odot$), which are probably the product of mergers of two white dwarfs, appear magnetic very early in their cooling age, and the origin of the magnetic field may be due to a dynamo operating during the merging \citep{Touetal08,Garetal12,Brietal15,BagLan22}. This idea is supported also by the evidence that some white dwarfs have very short rotation periods (e.g. Feige 7: 131\,min, \citeauthor{Lieetal77} \citeyear{Lieetal77}; Cl\,Oct: 12.1\,min,  \citeauthor{Baretal95} \citeyear{Baretal95}; WD\,2209+113: 70\,s, \citeauthor{Kiletal21} \citeyear{Kiletal21}).

At the same time, some polarimetric studies have hinted that a few old magnetic white dwarfs with a field strength of the order of tens of to a hundred megagauss have extremely long rotational periods, showing at most quite small variations even on timescales of decades \citep[e.g.][]{BerPii99,BagLan19a}. Traditionally, this lack of obvious variability is explained by the assumption that such non-varying magnetic white dwarfs have very long rotation periods of the order of centuries \citep{SchNor91}. In the absence of clear observational support or contradiction, this assumption has been widely accepted \citep[e.g.][]{Feretal15}. The conclusion is basically that the known periods, $P$, fall into two very different families: one with $P  \la 2$~weeks, characterising most of the sample, and a few percent with strong fields that have periods of centuries (i.e. no clearly detected rotation). There is no explanation as to why some old white dwarfs would have extremely long rotation periods. Magnetic braking has generally been invoked but without the support of numerical calculations.

In this work, we report new polarimetric observations and combine them with data collected from the literature. We analyse the magnetic variability and thus the rotation of a sample of 74 white dwarfs, and we discuss the constraints that these data place on possible evolution paths for magnetic fields in white dwarfs.

\section{Polarimetric observations of magnetic white dwarfs}\label{Sect_Observations}

\subsection{Observations not previously published}\label{Sect_New_Observations}
We present here 49 unpublished spectropolarimetric observations of 13 magnetic white dwarfs. 
Fourteen spectra of six white dwarfs were recently obtained with the FORS2 instrument \citep{Appetal98} of the ESO VLT in the course of a spectropolarimetric survey of the solar neighbourhood. 
Seven polarised spectra of two other white dwarfs were retrieved from the ESO archive: one of the spectra was obtained with FORS1 and the remaining ones with FORS2 (we note that FORS1 and FORS2 are virtually identical instruments). 
Fourteen new spectra of four white dwarfs were obtained with the ISIS instrument of the \textit{William Herschel} Telescope (WHT). 
For all of these data, the observing strategy and data reduction are identical to the procedures described by \citet{BagLan18} for circular polarisation data and by \citet{BagLan19a} for linear polarisation data.
For two of the four stars recently observed with ISIS, we also present here three previously unpublished spectra obtained in the 1970s
that we used to study the long-term variability of magnetic white dwarfs. These data consist of low-resolution polarised spectra obtained by Angel and Landstreet using the multi-channel spectrophotometer (hereafter MCSP) on the 5-m Palomar telescope. The instrument set-up and data reduction of these observations are described in detail by \citet{AngLan74}.  Finally,  eleven spectra of WD\,1105--340 were obtained with the ESPaDOnS instrument of the Canada-France-Hawaii Telescope (CFHT).
These data were reduced by the automatic CFHT pipeline LibreEsprit. 

The log of these unpublished observations is given in Table \ref{Tab_Log}, and results are described in the sections dedicated to individual stars of Appendix~\ref{Sect_Stars}. Apart from their general purpose of assessing polarimetric variability, our new observations confirm that star WD\,1116$-$470 is magnetic \citep[it was previously considered as suspected to be magnetic by][]{BagLan21} and bring the number of magnetic white dwarfs in the local 20\,pc volume to 34 \cite[see][]{BagLan21}. We have also discovered two new massive magnetic white dwarfs: WD\,1619$+$054 and WD\,1754$-$550 (both rapidly variable).  

\subsection{Literature data}
We searched the literature and collected data for all magnetic white dwarfs that were observed in polarimetric mode at least twice as well as for all magnetic white dwarfs for which multiple spectroscopic intensity observations revealed magnetic variability. The way literature data and new observations have been used is described in the next section.

\section{Determination of the variability of the magnetic white dwarfs}\label{Sect_Results}
We are interested in establishing whether the magnetic field of a white dwarf appears constant or variable with time to the observer. Magnetic variability is ascribed to changes in the apparent field geometry as a star rotates. It can be detected mainly via circular spectro- or broadband polarimetry, which are both sensitive to the longitudinal component of the magnetic field. Linear polarisation, which is sensitive to the transverse component of the magnetic field, is usually much weaker than circular polarisation, and it is rarely detected in white dwarfs. Intensity spectra (if the star is not featureless) are sensitive to the mean field modulus and may also be used to detect magnetic variability, but they are less sensitive to changes of the apparent magnetic configuration than polarimetry. Magnetic white dwarfs that have been repeatedly observed in spectroscopic mode and that do not show sign of variability have not been considered here because having observed a constant intensity spectrum is not a sufficiently strong indication that a star is not magnetically variable. An example is WD\,2047$+$372, as it shows an almost constant Zeeman triplet in multi-epoch intensity spectra and a variable, sign-reversing circular polarisation spectrum \citep{Lanetal17}. 

Photometric variability is observed in magnetic white dwarfs, but in the absence of convincing observational evidence or theoretical arguments, we have chosen not to consider it alone as a proxy for magnetism nor magnetic variability. For example, \citeauthor{Brietal13} (\citeyear{Brietal13}, see in particular their Tables~1 and 3) detected light variability for a number of magnetic stars. For many of them, however, the light amplitude is quite low, and there are no multi-epoch polarimetric data to confirm magnetic variability. Furthermore, recent analysis of TESS data made by \citet{Heretal24} and \citet{Olietal24} failed to detect a periodic light curve for many of these targets. These stars are not included in our sample. In summary:\\
\textit{1)} Two or more intensity spectra, or polarisation spectra, or broadband polarisation measurements that clearly differ from each other are interpreted as being due to magnetic variability.\\
\textit{2)} Repeated observations of circular polarisation (either spectropolarimetry or broadband polarimetry) that are consistent among themselves within uncertainties are interpreted as evidence of a lack of magnetic variability (that is, the observer always sees the same magnetic configuration).\\
\textit{3)} No conclusion will be drawn from a series of intensity spectra obtained in different epochs that appear similar to each other within uncertainties.\\
\textit{4)} Photometric variability that is not associated with observed spectroscopic or polarimetric variability will not be used to decide that a star is magnetically variable.

We gathered reasonable evidence for polarimetric variability, or non-variability, for the sample of 74 magnetic white dwarfs listed in Table~\ref{Tab_Stars}. Admittedly, for some stars, we found that establishing whether or not two or more observations are consistent among themselves was somewhat subjective. In Sect.~\ref{Sect_Field_Symmetric}, we argue that our results are not affected by these uncertain cases.  Appendix~\ref{Sect_Stars} discusses in detail all individual stars. In this section, we summarise the results and flag special cases.

Among the 74 white dwarfs, 44 are magnetically variable (Sect.~\ref{Sect_Var}), while for 27 there is so far no evidence of variability (Sect.~\ref{Sect_NoVar}). The remaining three white dwarfs, which have strong fields and have been observed over several decades, may be further tested for subtle or possibly very slow variability (Sect.~\ref{Sect_Subtle_Var}).

\subsection{Stars that show evidence of magnetic variability}\label{Sect_Var}
For 30 white dwarfs among the 44 stars that show polarimetric variability, a rotational period, or at least a good candidate for it, has been established in the literature. This period is given (in days) in the last column of Table~\ref{Tab_Stars} and is followed by `:' when only tentative or approximated values are known. Rotational periods are generally of the order of hours; about 25\% are of the order of one day or longer, and only two stars have a rotation period longer than one week \citep[the slowest magnetic white dwarf for which the period is known, WD\,2316+123, has $P\simeq 17.4$\,d,][]{SchNor91}. Most of these stars were listed in Table C.1 of \citet{Heretal24}, and they are shown with blue symbols in Fig.~\ref{Fig2} of Sect.~\ref{Sect_Clustering}.

Among the other 14 variable stars, we do not have enough data to estimate the rotation period. For ten of them, the evidence of polarimetric variability has been securely established by two or more observations. These ten variable stars are marked in last column of Table~\ref{Tab_Stars} with `var.' and are also represented with a blue symbol in Fig.~\ref{Fig2}. The remaining four white dwarfs show subtle signs of variability among the observations obtained with the same instrument. They are marked in Table~\ref{Tab_Stars} with `var.:' and are shown with light blue symbols in Fig.~\ref{Fig2}. For WD\,0446--789 and WD\,1009-184, our conclusion is that the field variations are real but small due to a configuration probably nearly symmetric about the rotation axis. WD\,1105--048 was repeatedly observed with FORS1, FORS2, and ESPaDOnS, and a field was detected on only two occasions out of 12 observing epochs. If the star is magnetic, then it is certainly variable with a timescale of the order of days or weeks, but, admittedly, one could suspect that the two detections are in fact spurious. We decided to define the star as a probable variable star. WD\,2049--222 shows a signal of polarisation that is marginally higher than instrumental polarisation, but for the reasons explained in Sect.~\ref{Sect_WD2049}, we decided to consider its subtle variations as real.

\subsection{Stars for which there is no evidence of magnetic variability}\label{Sect_NoVar}
Nineteen stars have been observed three or more times in polarimetric mode, and the observations are always consistent among themselves within uncertainties. We consider these stars as established non-variable white dwarfs. They are marked with `n.v.' in Table~\ref{Tab_Stars} and represented with red symbols in Fig.~\ref{Fig2}. Among these 19 stars, four were observed over a time interval of decades and with different instruments: WD\,0548$-$001 = G\,99-37, WD\,0553$+$053 = G\,99-47, WD\,1036$-$204 = LP\,790$-$29, and WD\,1829$+$547 = G\,227-35. Obviously, it is possible that the other members of this group of `non-variable stars',  for which the observations span a time interval of up to a few months or years, are actually variable on a timescale of decades and that as such they could be stars with (so far undetected) long-term variability. 

Eight stars should be considered simply as candidate non-variable white dwarfs (`n.v.:' in Table~\ref{Tab_Stars} and magenta symbols in Fig.~\ref{Fig2}). For seven of them, the reason for considering them only as candidate non-variable stars is that they have only been observed twice. In particular we mention that for normal-mass WD\,0236--269 and for the massive WD\,0330$-$000, the non-variability is inferred based on comments in the original discovery paper by \citet{Schetal01}, which represents especially weak evidence. Another candidate non-variable star, WD\,1750$-$311, was observed only once (with FORS2). The comparison of spectra obtained within that single observing series shows some change of circular polarisation and certainly a rapidly variable intensity spectrum (timescale of minutes) in the regions of the Balmer lines, particularly around H$\beta$. The star also shows photometric variability with a 35-min period \citep{Macetal17}. Our interpretation is that WD\,1750--311 is probably a magnetic white dwarf with a field symmetric, or nearly symmetric about the rotation axis, showing strong abundance patches of hydrogen.

\subsection{Stars possibly subtly or slowly variable}\label{Sect_Subtle_Var}
\begin{figure*}
\begin{center}
\includegraphics[angle=0,width=8.5cm,trim={0.5cm 8.9cm 1.0cm 3.0cm},clip]{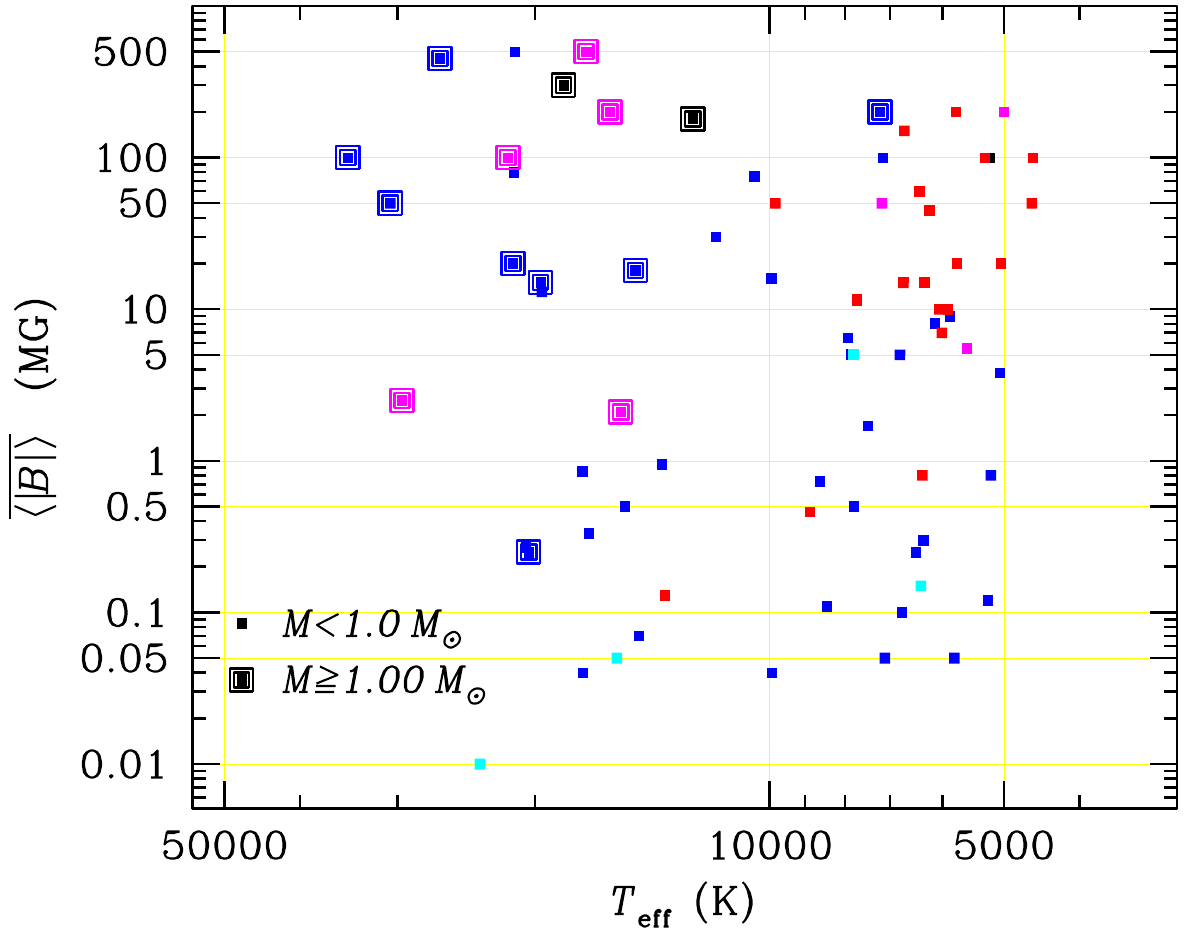}\ \ \ \ \ \ \ \
\includegraphics[angle=0,width=8.5cm,trim={0.5cm 8.9cm 1.0cm 3.0cm},clip]{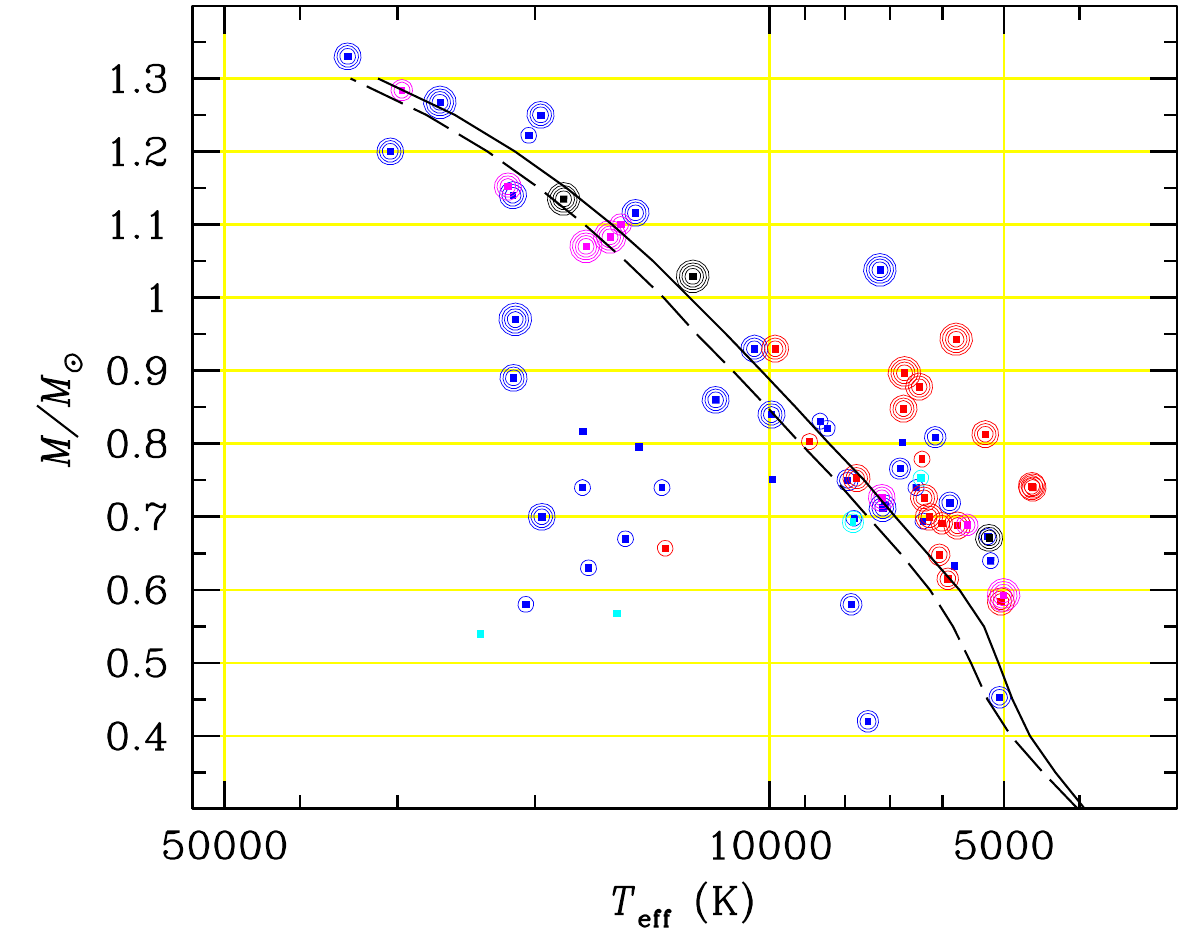}\\
\includegraphics[angle=0,width=8.5cm,trim={0.5cm 8.9cm 1.0cm 3.0cm},clip]{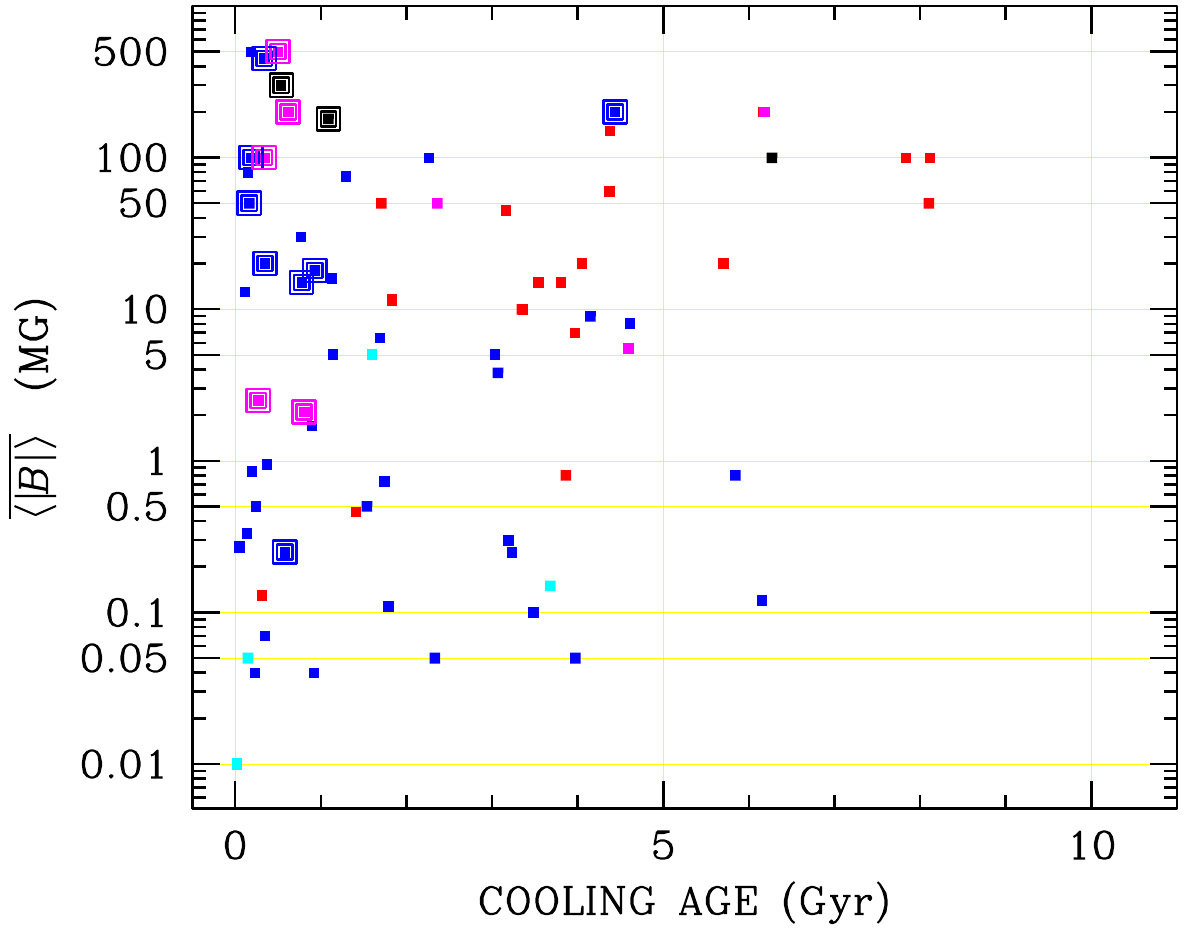}\ \ \ \ \ \ \ \
\includegraphics[angle=0,width=8.5cm,trim={0.5cm 8.9cm 1.0cm 3.0cm},clip]{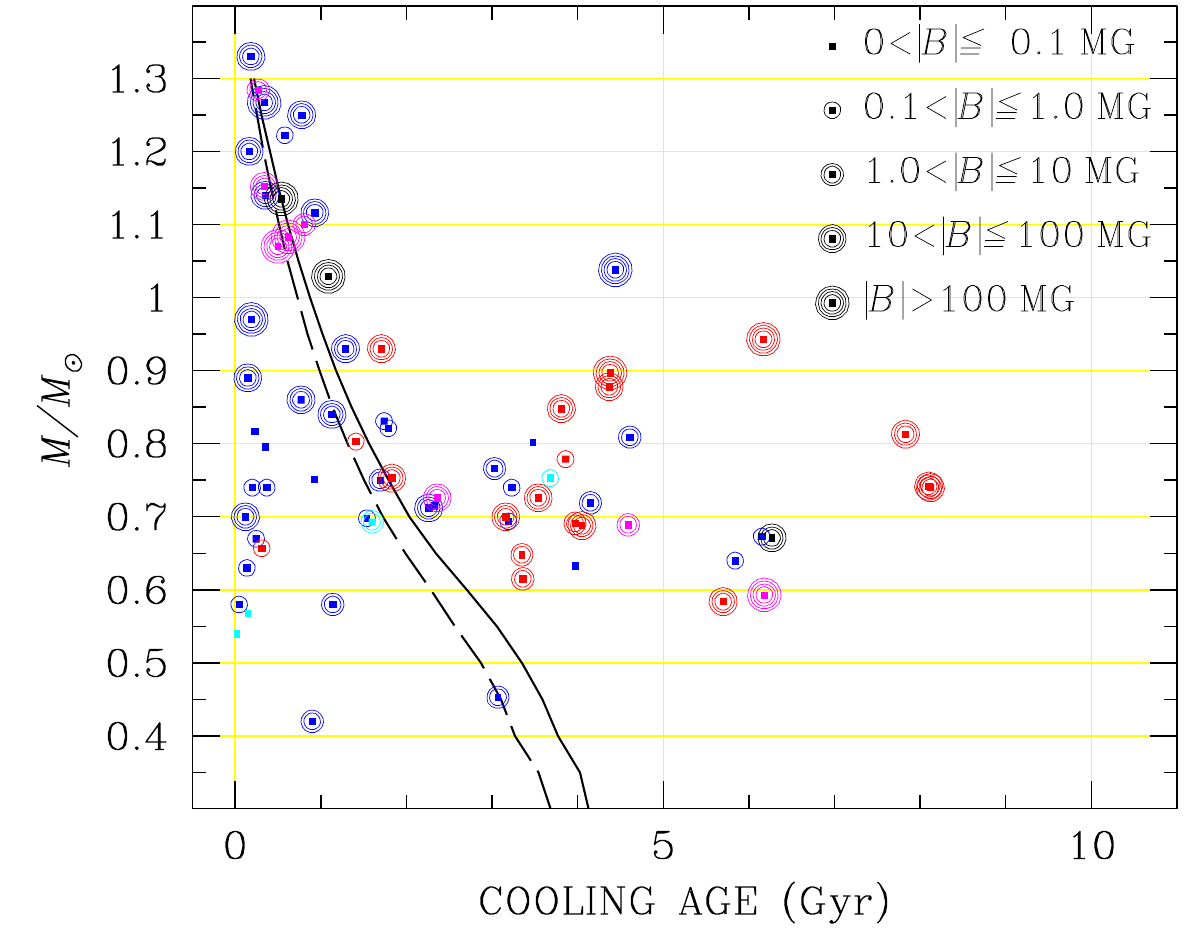}\\
\includegraphics[angle=0,width=8.5cm,trim={0.5cm 8.9cm 1.0cm 3.0cm},clip]{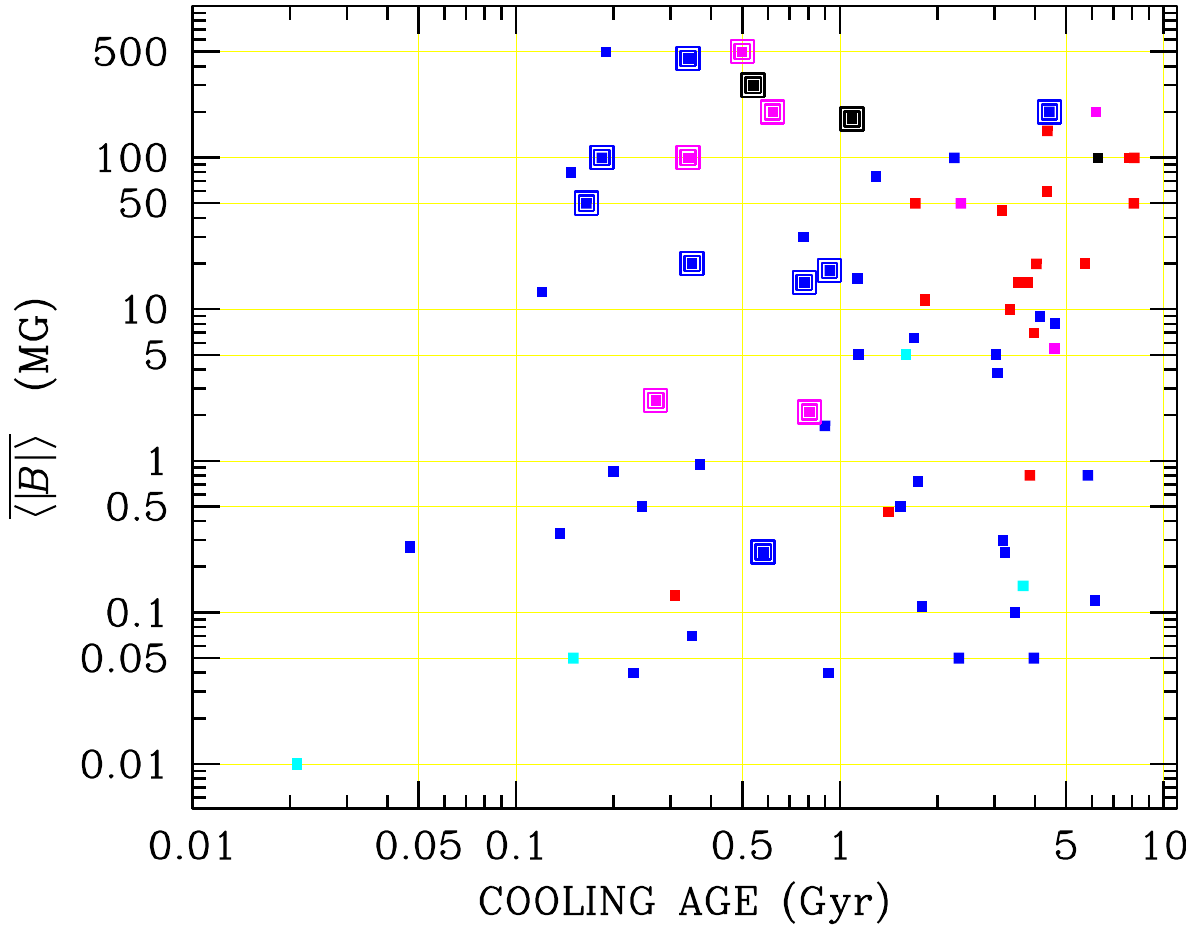}\ \ \ \ \ \ \
\includegraphics[angle=0,width=8.5cm,trim={0.5cm 8.9cm 1.0cm 3.0cm},clip]{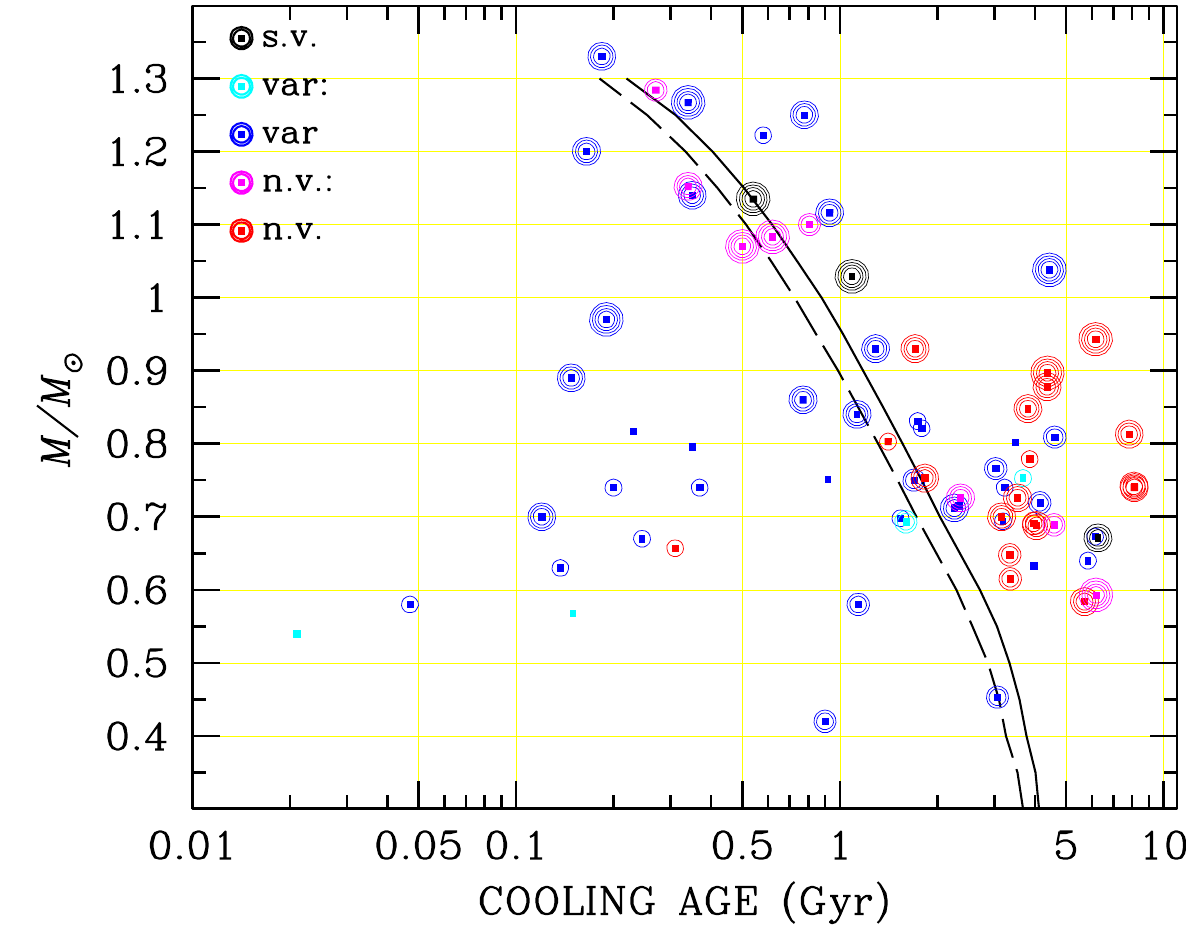}
\vspace{2mm}

\noindent
\end{center}
\caption{\label{Fig2} Correlations between field strength, magnetic variability, and other stellar parameters. {Left panels:} Field strength versus effective temperature and versus cooling age for variable and non-variable stars of Table~\ref{Tab_Stars}. The size of the symbols is related to the mass of the star as shown in the legend. Red symbols refer to stars that were observed at least three times with no evidence for variability; magenta symbols are for stars that were observed only twice and are tentatively assumed non-variable. 
Blue symbols refer to stars that show variability. Light blue symbols indicate stars that show marginal but probably real signs of variability, maybe due to a field nearly aligned with the rotation axis. Black symbols are used for stars that show signs of long-term variability that we were not able to interpret.
{Right panels:} Temperature-mass and cooling age-mass diagrams for the same sample of stars.
The size of the symbols is related to the field strength as shown in the legend. The meaning of the colour is the same as for the left panels. 
Black lines represent the onset of crystallisation (solid line for H-thick envelop and dashed line of H-thin envelop) obtained via interpolation of the cooling tables by \citet{Bedetal20}. In all panels, the yellow vertical and horizontal lines highlight the position of the box tick marks. We recall that these diagrams do not refer to a volume-limited sample and that both young normal-mass magnetic white dwarfs and highly massive magnetic white dwarfs, which are rare objects in space, are over-represented.
} 
\end{figure*}

There are three stars that have been monitored for decades, 
WD\,1748$+$708 = G\,240$+$708, 
WD\,1900$+$310 = Grw+70$^\circ$\,8247, and 
WD\,2010$+$310 = GD\,229, for which our conclusions are uncertain. These stars, marked with `s.v.:' in Table~\ref{Tab_Stars} and shown with black symbols in Fig.~\ref{Fig2}, may indeed show some low amplitude and possible long-term variability, but there might be room for a different interpretation.
The recent detailed comparison of old and new polarisation spectra of Grw+70\degr\,8247 by \citet{BagLan19a} identifies small changes between spectra taken decades apart (see in particular their Fig.~5), but for the two other stars, it is not entirely clear which changes are real and which are due to differences between observational equipment and techniques. \citet{BagLan19a} assumed that timescales of the observed small changes were those corresponding to the large time gaps between isolated data sets. However, because the changes are generally small, one could argue that the subtle changes have timescales as short as weeks and have not have been noticed because the stars were not adequately monitored. For example, as previously discussed, WD\,1105$-$048 appears constant (and non-magnetic) in ten out of 12 observations, and a magnetic field was detected in only two epochs. In case of  WD\,1748$+$708, photometry leads to ambiguous results. The star was found photometrically variable with a period between 5\, and 48\,h by \citet{Antetal16}. \citet{Brietal13} did not confirm its short-term photometric variability \citep[nor did][]{Heretal24} but instead claimed detection of light changes over a period of ten months.

\section{Clustering in parameter space}\label{Sect_Clustering}
In this section, we correlate the variability or non-variability of the stars with their mass, field strength, and stellar temperature. The left panels of Fig.~\ref{Fig2} show the field strength versus effective temperature and cooling age for the stars of our sample, with the cooling age on both a linear scale and a logarithmic scale. In these plots, stars are represented with symbols that increase in size with a star's mass, while different colours are used to mark the characteristics in terms of variability. The right panels of Fig.~\ref{Fig2} show the cooling age-mass diagrams for the same sample, with cooling age reported again both with linear and logarithmic scales. The size of the symbols is proportional to the field strength. With the help of these figures and Table~\ref{Tab_Stars}, we studied the relationships between field strength, mass, age, and variability of the magnetic white dwarfs. We recall that our data do not come from a volume-limited sample of stars and do not reflect the relative density of magnetic white dwarfs in different regions of the diagrams of Fig.~\ref{Fig2}. In fact, both young magnetic normal-mass white dwarfs and ultra-massive white dwarfs are quite rare objects in space, and they are over-represented in our sample because of biases of the surveys \citep[see][]{BagLan22}. The following analysis is about the relative frequency of variable and non-variable stars in different regions of the diagrams where we noticed the existence of statistically remarkable differences.

\subsection{Normal-mass and weakly magnetic white dwarfs}\label{Sect_Younger}
There are 25 white dwarfs with a field strength of $\le 1$\,MG, and all of them but one (WD\,2051--208) have $M\le 1.0\,M_\odot$. These `normal-mass' `weak-field' white dwarfs span an age range up to $\tau \approx 6$\,Gyr. Twenty-one of these 24 stars are variable. Among the three that are non-variable, one is younger than 1\,Gyr: WD\,1105$-$340 ($\tau = 0.34$\,Gyr).

\subsection{Normal-mass and strongly magnetic white dwarfs}\label{Sect_Older}
There are 25 magnetic white dwarfs with $M \le 1.0\,M_\odot$ and field strength of $\ge 10$\,MG. 
Among those that are younger than 2\,Gyr, six out of eight white dwarfs are variable; however, we 
recall that young strongly magnetic white dwarfs are rare in
space \citep{BagLan22}. Among the magnetic white dwarfs older than 2\,Gyr, only one out of the 17 shows
polarimetric variability.

\subsection{Massive magnetic white dwarfs}\label{Sect_Massive}
Fifteen stars of our sample have $M > 1.0\,M_\odot$, and all but one are younger than about 1\,Gyr.  Eight of them are magnetically variable, including the only old ($\tau \simeq 4$\,Gyr) star WD\,0756$+$437. Nearly all of these massive magnetically variable white dwarfs have a strong field (tens to hundreds megagauss), except for WD\,2051--208, which is the only ultra-massive star ($M = 1.2\,M_\odot$) with a sub-MG field strength (0.25\,MG) in our sample.

In contrast, seven massive magnetic white dwarfs show no obvious signs of variability, although none of them may be safely considered as non-variable, either because they were observed only twice (or even just once, i.e. WD\,1750--311; see Appendix~\ref{Sect_WD1750}) or because some subtle sign of variability may have been detected (in WD\,1900$+$705 and WD\,2010$+$310). 
WD\,1658$+$440 and WD\,1750--311 are the only massive non-variable white dwarfs in our sample with a field as low as $\approx 2$\,MG. The remaining five non-variable stars have magnetic fields with strengths of the order of hundreds of megagauss.

\subsection{Statistical analysis of normal-mass white dwarfs}
\begin{figure}
\begin{center}
\includegraphics[angle=0,width=8.5cm,trim={1.5cm 5.8cm 1.0cm 3.1cm},clip]{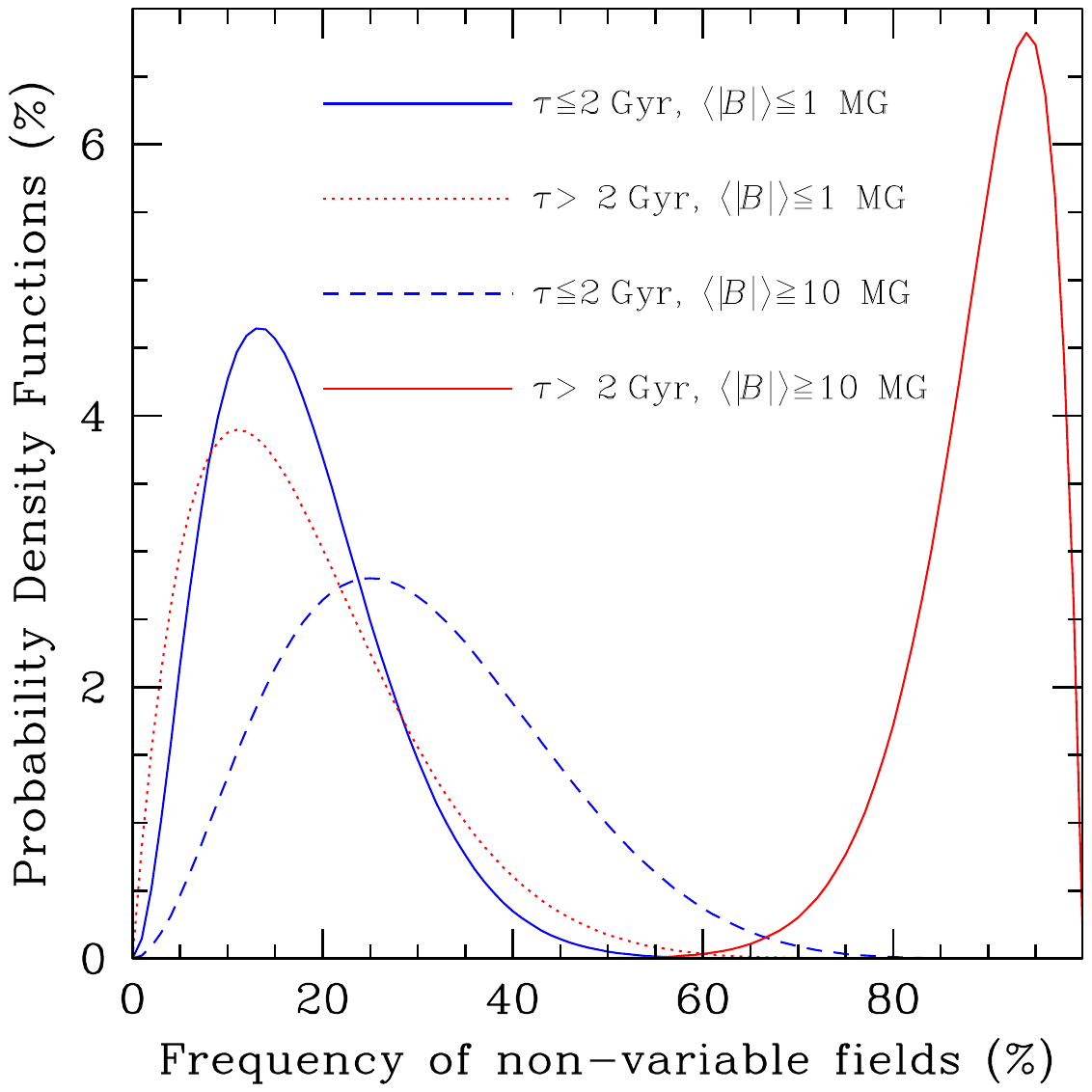}\ \ \ \ \ \ \ \
\vspace{2mm}

\noindent
\end{center}
\caption{\label{Fig_Distributions} Probability density functions of the non-variable fields in normal-mass white dwarfs with the age and field strength
specified in the legend.
} 
\end{figure}
\begin{figure}
\begin{center}
\includegraphics[angle=0,width=8.5cm,trim={1.2cm 6.2cm 0.7cm 3.0cm},clip]{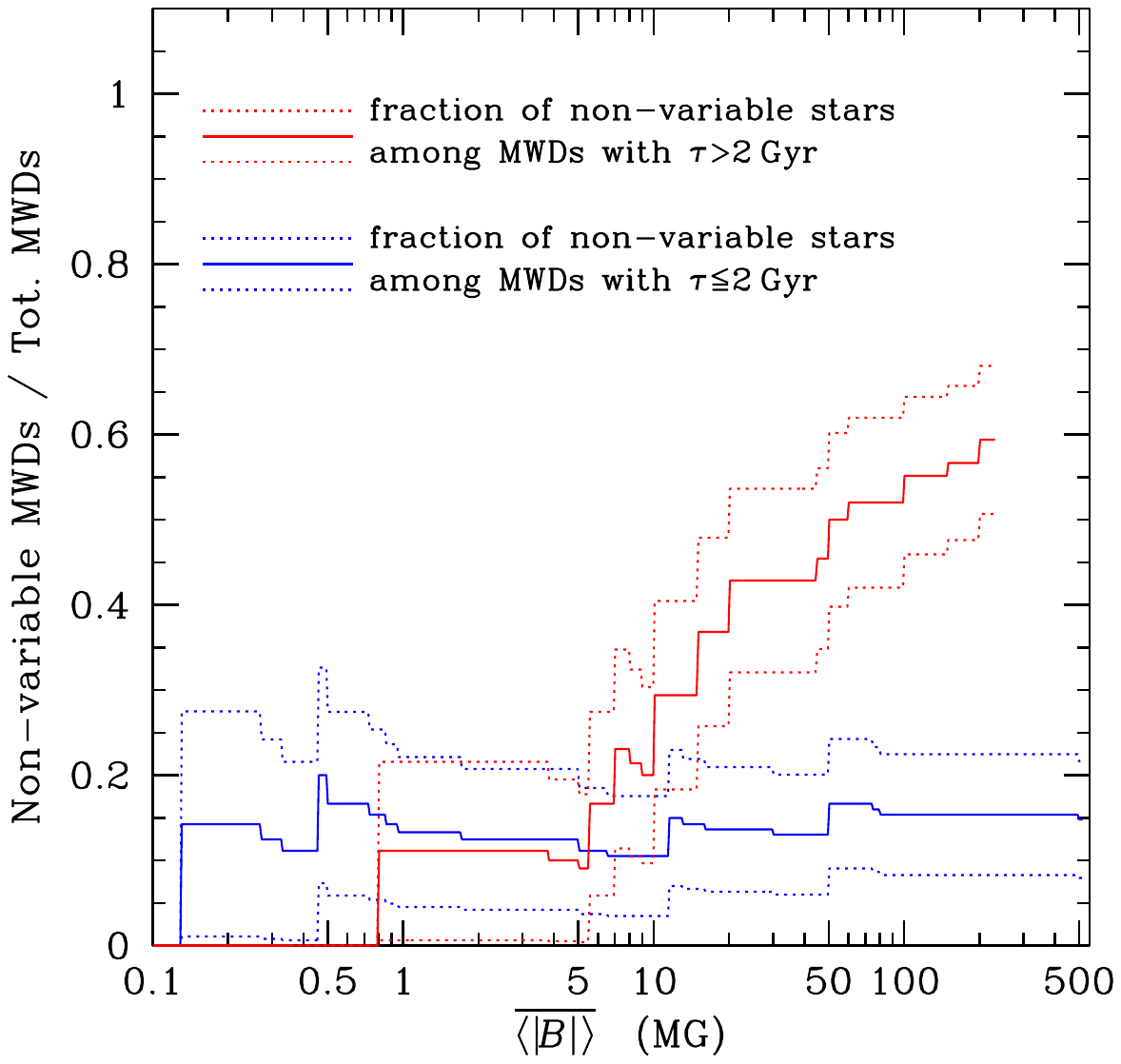}\ \ \ \ \ \ \ \
\vspace{2mm}

\noindent
\end{center}
\caption{\label{Fig_Cumul} 
Cumulative fractions of variable and non-variable magnetic white dwarfs with $M \le 1\,M_\odot$. The blue solid line represents the number of non-variable stars  younger than\,2 Gyr divided by the total number of stars younger than 2\,Gyr with an average field strength lower than the value in the $x$-axis. The red solid line refers to the same quantity for magnetic white dwarfs older than 2\,Gyr. Dotted lines show the estimate of the uncertainties.
} 
\end{figure}

\begin{table}
\caption{\label{Tab_Summary} Frequency of polarimetrically non-variable stars among normal-mass white dwarfs ($M \le 1\,M_\odot$).  }
\begin{center}
\tabcolsep=0.08cm
\begin{tabular}{l|cc}
\hline
                                   & $\tau \le 2$\,Gyr       & $\tau > 2$\,Gyr     \\
                                   & ('young stars')         & ('old stars')       \\
\hline
$\bs \le  1$\,MG ('weak fields')   &  $ 13 \pm 9$\%  (2/15)  & $11 \pm 10$\%   (1/9) \\ 
1\,MG $\le \bs \le 10$\,MG         &       0\%  (0/4)        & $50 \pm 18$\%   (4/8) \\ 
$\bs \ge 10$\,MG ('strong fields') &  $ 25 \pm 15$\%  (2/8)  & $94 \pm  6$\%  (16/17) \\ 
\hline
\end{tabular}
\end{center}
\tablefoot{The full list includes 59 stars, 23 of which are non-variable. Between brackets, the table also gives the number of non-variable stars \nvar\ divided by the number of stars \ntot\ in the subset as defined by the age and field values.}
\end{table}

Because we have only one example of an old massive magnetic white dwarf, we do not know how variability in massive magnetic white dwarfs evolves with time. Vice versa, an evolutionary path is clearly seen in normal-mass ($M \la 1\,M_\odot$) white dwarfs. In our sample, field variability is nearly ubiquitous among weakly magnetic white dwarfs of all ages and in young white dwarfs regardless of the field strength, it is but very rare in 
old ($\tau \ge 2$\,Gyr) strongly magnetic ($B \ge 10$\,MG)  white dwarfs. Next, we assess the statistical significance of this pattern, specifically whether it can be attributed to small number statistics or if it reflects a genuine correlation between cooling age, field strength, and field variability.

We first split the sample of stars with $M \le 1\,M_\odot$ into two groups: those younger than 2\,Gyr and those older than 2\,Gyr.
Each of these two groups was divided into two subsets: stars with a field strength $\bs \le 1$\,MG and stars with $\bs > 10$\,MG. For each of these four subsets, we considered the ratio between the number of non-variable magnetic stars, \nvar, and the total number of magnetic stars, \ntot. From these numbers, we estimated the probability, $P_r$, that the sample frequency of non-variable stars is between $r$ and $r + {\rm d}r$, normalised to one and obtained assuming that all numbers between zero and one are a priori equally probable, using
\begin{equation}
  P_r = \frac{(\ntot+1)!}{\nvar!\ (\ntot-\nvar)!}\ r^{\,\nvar}\ (1-r)^{(\ntot-\nvar)} \;.
\label{Eq_Bayes}
\end{equation}
Figure~\ref{Fig_Distributions} shows the probability density functions for the stars belonging to these sets. It clearly appears that there is little to no overlap between the density functions of the symmetric field in old strongly magnetic white dwarfs and of young white dwarfs. Table~\ref{Tab_Summary} reports the points of maximum for these distributions, $f = \nvar/\ntot$, and their uncertainties
\begin{equation}
\sqrt{\frac{f(1-f)}{\ntot}} \; .
\label{Eq_Uncertainty}
\end{equation}
Table~\ref{Tab_Summary} also includes the results for the smaller sets of stars with intermediate strength, $1 \le \bs \le 10$\,MG (with endpoints overlapping with the other two sets), which could possibly be considered a `transitioning' field strength range.

Figure~\ref{Fig_Cumul} shows the ratio between the number of non-variable magnetic white dwarfs with a field strength lower than a given value $\bar{\bs}$ and the total number of magnetic white dwarfs with a field strength lower than that value as a function of $\bar{\bs}$  for stars with cooling ages $\tau \le 2$\,Gyr (blue lines) and stars with $\tau > 2$\,Gyr (red lines). This plot supports our claim that non-variable fields are much more common in old, strongly magnetic white dwarfs than in young strongly magnetic white dwarfs, and than in weakly magnetic white dwarfs of all ages. It suggests also that the minimum field strength required for a field to become non-variable is in the range of 5 to 10\,MG.

\section{Explanation for the lack of observed variability}\label{Sect_Aligned}
There are three possible reasons for the observed non-variability of so many magnetic white dwarfs. We examine them in the following sections.

\subsection{Extremely slow rotation}
We first consider the possibility that the rotation of magnetic white dwarfs slows with age and/or magnetic field strength. Perhaps, as the surface field becomes stronger, the white dwarf loses angular momentum to its environment by electromagnetic dipole radiation \citep{Garetal12}, by coupling with gas clouds in the ISM, or by a very weak magnetically coupled wind. Each of these possibilities would shed stellar angular momentum faster from a magnetic white dwarf with a strong field than from one with a weak field, preferentially slowing the magnetic white dwarfs  with strong fields.  However, all of these mechanisms are expected to exert at most a very weak influence on the rotation of a white dwarf, and it seems quite unlikely that any of them could slow the rotation of a magnetic white dwarf so much that repeated observations even a year apart would not show any rotation. Nevertheless, the idea of extremely slowly rotating white dwarfs has been generally accepted. This hypothesis likely originates from Sect.~3 of \citet{SchNor91}, which says: `We therefore assign long periods to these stars, with the recognition that other explanations for their polarimetric constancy are possible.' This statement was accompanied by the first plots of field strength versus rotational period, a plot that in updated form has reappeared in numerous reviews \citep[e.g.][]{Feretal15,Kawka20,Feretal20} but has never been critically revisited.

We note that there is a complete lack of white dwarfs that are known to vary on a timescale between weeks and decades. In this respect, there is a profound difference between candidate long-term variable magnetic white dwarfs and long period magnetic Ap and Bp stars. While we do not know of any white dwarf with a firmly established polarimetric variability with a period longer than approximately two weeks, we are sure that a number of Ap and Bp stars are very slowly rotating stars because they definitely show clearly periodic field variation, even \bz\ sign reversals, on a very long but measured timescale that may be months, years, or decades. Furthermore, the distribution of periods is roughly continuous between periods of less than one day and periods of several years \citep{Mat08}. If we wanted to interpret the lack of polarimetric variability in white dwarfs as the effect of extremely long rotation periods, we would need to accept that the rotation periods of white dwarfs show an extremely bimodal distribution that peaks around hours and days and around centuries with nothing in between. This appears to be a very unlikely scenario.

\subsection{Extremely fast rotation}
A second possibility is that some or all of the non-variable white dwarfs actually are very rapidly rotating stars. If the star has a rotation period much shorter than the individual frame exposure time, then the polarimetric variability would be smeared out, and the star would appear constant. 

In fact, it has been possible to probe variability on the timescale of the exposure time of each individual frame (typically 10\,min or less). In very general terms, our FORS2 and ISIS polarimetric observations allowed us to identify a star as variable provided that its rotation period is longer than $\approx 10$\,min and its magnetic configuration is clearly not symmetric about the rotation axis (for example, WD\,1712--590 in Sects.~\ref{Sect_WD1712}). Isolated white dwarfs that are the product of single-star evolution can hardly have rotation periods shorter than that \citep{Kawetal15}. This is confirmed by the results of \citet{Heretal24}, who have shown that rotation periods of young magnetic white dwarfs in the normal mass range are only marginally shorter than the rotation periods of non-magnetic white dwarfs. Merger products, in contrast, can have rotation periods of the order of 1\,min \citep{Schwab21,Kiletal21} if most of the binary angular momentum is retained by the merger product. It is therefore possible that the non-variable massive white dwarfs have non-axisymmetric fields but are rotating very rapidly. Periods down to 1\,min or less can be probed via rapid cadence photometry. None of the massive, magnetically non-variable white dwarfs show TESS light variability (Hernandez, priv.\ comm.; Ramsay, priv.\ comm.), although of course one cannot rule out that more accurate photometry could reveal some extremely rapid rotators. Provisionally, we rule out very rapid rotation for the massive magnetic white dwarfs that show a constant polarisation.

\subsection{Magnetic fields are symmetric about the stellar rotation axes}\label{Sect_Field_Symmetric}
After ruling out both extremely slow and extremely fast rotation, we are left with the hypothesis that a non-variable field is approximately axisymmetric about the rotation axis. This means that as the star rotates with a normal period between a fraction of an hour and several days, the observer does not see any significant change in the signal of circular polarisation. 

This interpretation naturally accounts for some outliers, such as the non-variable weakly magnetic star WD\,1105--340. It is reasonable to hypothesise that this one star does not appear variable simply because its rotation axis is tilted at a small angle with respect to the line of sight. 

In Sect.~\ref{Sect_Results}, we highlighted that it may be difficult to firmly assess whether a star is magnetically variable because changes could be too subtle to be detected. However, the risk of classifying as 'non-variable' a star that in fact exhibits subtle but real changes does not weaken our analysis because small changes of the circular polarisation spectrum are still symptoms of a field nearly symmetric about the rotation axis.

\section{Discussion}
Here, we first consider the case of normal-mass white dwarfs (Sect.~\ref{Sect_Normal}) and then that of massive stars (Sect.~\ref{Sect_Massive_WDs}). We note that our analysis applies only to isolated white dwarfs; symmetric fields seem uncommon among magnetic white dwarfs accreting material from a companion \citep{Cropper88,Reietal99,Schetal95}.  

\subsection{Normal-mass white dwarfs}\label{Sect_Normal}
It is clear that most of the weak fields of normal-mass stars of all ages are not symmetric about the stellar rotation axis. Most of the strong fields of normal-mass, young white dwarfs are also not axisymmetric. Most of the strong field of stars older than 2\,Gyr are symmetric about the star's rotation axis. We now investigate what are the relationships between these weak and strong fields.

Our hypothesis is that the weak and non-symmetric magnetic fields of the young normal-mass white dwarfs are due to a relaxation process of a field that was produced in a previous evolutionary stage of the star, a field that was buried below the surface of the newly formed white dwarf, and that it starts to reveal itself with time as the star cools down. Such seed fields could be those found in red-giant stars, in the vicinity of the hydrogen-burning shell \citep[e.g.][]{Ligetal22,Ligetal23}. The question is whether the older, generally stronger fields share essentially the same origin as the weak, younger ones, having just evolved for longer time, or if the strong fields in older stars formed by a totally different mechanism than the weaker, younger ones, for instance by a crystallisation dynamo.  Whatever its origin is, we consider first the possibility that
the small distortion of the shape of a magnetic white dwarf produced by magnetic forces in outer layers might actually drive the evolution of the global field towards axisymmetry. In Sect.~\ref{Sect_Alignment}, we discuss one example of a physical effect that might drive such evolution.

\subsubsection{The magnetic field structure of white dwarfs may evolve towards rotational axisymmetry during cooling}\label{Sect_Alignment}

In the absence of a magnetic field, the rotation of the white dwarf will lead to an equatorial bulge, increasing the component of the principal moment of inertia that is parallel to the rotation axis. In this case, the rotation about the spin axis, which coincides with the largest moment of inertia, is stable. Next, we introduce a dipolar magnetic field inclined to the axis of rotation by, say, $45^\circ$. We suppose that the visible surface field is maintained by a fossil current system deep in the star, which is gradually decaying due to Ohmic losses. As the interior field strength decreases, an azimuthal electric field will be produced by Faraday induction outside the white dwarf's initial current loop. This field will generate a current around the magnetic axis that will in turn interact with the local meridional magnetic field \citep{Lan87} to produce an outward-directed force, distorting the stellar outer layers into a slightly oblate shape, with the largest radius around the magnetic equator. Because of this effect, the principal moment of inertia of the star will be rotated towards the symmetry axis of the field, which is not aligned with the rotation axis or the angular momentum axis of the star. The white dwarf effectively becomes an unstable free asymmetric rotator (an asymmetric top). If such objects can dissipate some of the energy supporting the asymmetric shape, the field axis will gradually shift to bring the principal moment of inertia closer to the angular momentum axis, which is a stable minimum energy state of an oblate system. Energy dissipation will gradually lead to the alignment of the magnetic axis with the rotation axis, which is the effect we observe in old normal-mass strongly magnetic white dwarfs. 

This basic physical effect was first identified in the 1970s as possibly operating in main-sequence magnetic Ap stars, which, similar to magnetic white dwarfs, possess global fossil magnetic fields that are usually roughly dipolar and are generally oblique to the rotation axis \citep{Stibbs50,Pre71}. A number of efforts to estimate the timescale of possible evolution of an oblique magnetic field to a state of small or vanishing obliquity have been published, for example, by \citet{MesTak72}. Much of this work is summarised in Chapter 9 of \citet{Mes99}. However, this line of investigation did not lead to a convergence of clear results about possible timescales or about dependence on basic input physics or parameters, such as interior field strength, or details of induced forces in outer layers and their consequences. As the increasing sample of magnetic Ap star models failed to reveal an obliquity distribution suggestive of this effect, possibly because of the relatively short (of order $10^8$\,yr) main-sequence lifetimes of magnetic Ap stars, theoretical studies of stellar fossil fields moved on to other effects. However, a similar effect may be at work aligning the magnetic fields to the rotation axis in white dwarfs, which have a very different structure and much longer evolutionary times than Ap stars. This hypothesis could be explored with the aid of numerical modelling of the global (interior and surface) structure and evolution of a rotating white dwarf with an oblique magnetic field. Modelling of comparably complex white dwarf states (e.g. polars, mergers) that include magnetic fields \citep[e.g.][]{FraSch15,Bisetal21,Zhoetal24} suggest that numerical methods for exploring this problem are already available.

\subsubsection{Crystallisation-driven dynamo and axisymmetric fields}\label{Sect_Crystalisation}
The line that marks the beginning of core crystallisation in the age-mass diagram separates variable from non-variable magnetic white dwarfs perhaps in a cleaner way compared to a mass-independent age threshold (see Fig.~\ref{Fig2}). Before core crystallisation begins, magnetic fields are almost always non-symmetric about the rotation axis. From volume-limited surveys, we know that in normal-mass white dwarfs, before crystallisation, strong fields are rare \citep{BagLan21,BagLan22}, but those that have been discovered and monitored show polarimetric variability. After the beginning of core crystallisation, many strong fields appear, and most of them are symmetric about the rotation axis. White dwarfs with weak and non-symmetric fields continue to appear also after the beginning of core-crystallisation. 

The strong magnetic fields of old normal-mass white dwarfs could be generated by the crystallisation convective dynamo mechanism \citep{Iseetal17,Schetal21}, as almost all have cooling ages longer than the cooling age required for the onset of this dynamo. In this case, the conclusion would be that the crystallisation dynamo is responsible for approximate symmetry about the rotation axis. This situation could have similarities with what has been observed in fully convective late type stars with no differential rotation, which are known to be able to generate strong and simple large-scale, mostly axisymmetric, poloidal fields \citep[][Fig.~14]{Donetal06,Moretal08a,Moretal08b,Koc21}. However, how the crystallisation dynamo alone can produce fields stronger than $\sim 1$\,MG is still unclear \citep{Iseetal17,Casetal24}. In fact, \citet{MonDun24} have argued that fluid mixing by phase separation is not a viable mechanism to produce the strong fields observed in old white dwarfs. Perhaps the crystallisation dynamo could be more powerful if it were to amplify a pre-existing internal field, such as the same seed field that is seen in some of the white dwarfs prior to the beginning of crystallisation. 

\citet{MonDun24} have proposed that core-crystallisation could trigger temporary differential rotation. Therefore, one could suspect that rapid rotation in normal-mass white dwarfs might generate a magnetic field in the stellar core using rotational shear on a seed field left from an earlier point in the star’s evolutionary history. In this situation, we might also expect that the field would be roughly axisymmetric, although it is not clear why such fields would always be strong or how they might be related to the weaker oblique fields of stars of similar mass. Furthermore, \citet{Spruit99} has shown that differential rotation would suppress the non-axisymmetric field component of weak fields, but not in the stronger fields. This is the opposite of what we observed.

The origin of strong non-axisymmetric fields of normal-mass white dwarfs that appear before crystallisation could possibly be similar to that of higher-mass white dwarfs (see Sect.~\ref{Sect_Massive_WDs} below).

\subsection{The variability of massive magnetic white dwarfs}\label{Sect_Massive_WDs}
Compared to normal-mass white dwarfs, the magnetic fields of massive white dwarfs present a different behaviour. Magnetic fields appear when the stars are still very young, and fields may be either symmetric or non-symmetric about the rotation axis. In massive white dwarfs, strong fields (tens or hundreds of megagauss) seem much more common than weaker sub-megagauss fields. It is widely thought that many of these high-mass objects are the result of WD-WD mergers that cause rapid generation of a strong magnetic field \citep{Garetal12,BagLan22}. However, \citet{Cametal22} and \citet{BlaGin24} have shown that at least some of them, depending on their core composition, may have started the process of core crystallisation. Hence, the origin of their field could be linked to the crystallisation dynamo, at least in some cases. 

The bimodal distribution of the morphologies of the fields of the massive magnetic white dwarfs could indeed reflect two different channels of formation, one from WD-WD merger (or by a different binary evolution path), and one by massive single-star evolution, such as that of normal-mass white dwarfs. Alternatively, the dynamo stimulated by WD-WD merging may be capable of generating both axisymmetric and non-axisymmetric global fields, perhaps depending on the initial angular momentum vectors of the individual merging white dwarfs relative to the orbital angular momentum vector or the mass ratio of the two merging white dwarfs. 

A viable alternative explanation is that the massive white dwarfs that do not show variability are actually rotating with a period much shorter than the typical exposure time of individual polarimetric measurements. In any case, the large mass of stars in this sample and the occurrence of some rotation periods as short as minutes make it very reasonable to suppose that the evolution of the magnetic fields and rotation periods follows a different course from the evolution of fields in normal-mass magnetic white dwarfs. 

Because almost all of the massive magnetic white dwarfs of our sample, except one, are very young, we cannot test whether the morphology of older stars is generally axisymmetric. Remarkably, however, the only example we know of a massive star older than 2\,Gyr shows a non-axisymmetric field. With an age of more than 4\,Gyr, high mass, and a very strong and variable field, the very unusual WD\,0756$+$437 may be considered further evidence that the origin of the fields in massive stars is different than in normal-mass stars.

\section{Conclusions}
Since the earliest discoveries of magnetic white dwarfs \citep{Kemetal70,AngLan70Further,AngLan71-Periodic}, two quite different categories of them have been known to exist. Some magnetic white dwarfs show periodic variations of circular polarisation (which probes the longitudinal magnetic field) with timescales ranging from minutes to days.  Classically, a signal of circular polarisation constant with time has been interpreted as indicating either a very long rotation period or a lack of rotation.

Using both literature data and new observations, we have studied the polarimetric variability of a sample of 74 magnetic white dwarfs. We find  that among white dwarfs with $M \le 1.0\,M_\odot$ (`normal-mass white dwarfs'), nearly all stars with fields weaker than about 1\,MG show circular polarisation varying with time.  
Furthermore, the rare normal-mass white dwarfs younger than $\approx 2$\,Gyr with strong magnetic fields also show polarimetric variability. In striking contrast, 16 out of the 17 normal-mass stars older than 2\,Gyr with fields stronger than about 10\,MG in our sample show constant polarisation. Magnetic white dwarfs with $M \ge 1.0\,M_\odot$ (`massive white dwarfs'), many of which are the product of WD-WD merging, show a mixed behaviour. Many of them have a strong magnetic and variable field, but some have a strong and constant magnetic field. In our sample, nearly all the massive magnetic white dwarfs are younger than $\approx 1$\,Gyr, with the exception of WD\,0756+437, an old strongly magnetic variable star. 

The lack of evidence for major variations of circular polarisation on any timescale longer than about two weeks suggests that the interpretation of non-variability arising from extremely long rotational periods is incorrect. We have argued that the non-variability is due to magnetic field structures roughly symmetric around the star's rotation axis.

A possible explanation for the observed symmetry could be that most or all of the magnetic fields evolved from pre-existing fields that formed during pre-white dwarf evolution stages \citep{Ligetal22,Ligetal23} and gradually relaxed to the stellar surface over a relation time of 1 or 2\,Gyr; in fact, very recently, \citet{Cametal24} have shown that the magnetic fields of old white dwarfs with $M \ge 0.65\,M_\odot$ may have been generated by a core-convection dynamo when the star was in the main sequence, and emerged at the stellar surface during the white dwarf cooling phase. As the surface field increases in strength, at some point the Lorentz forces in the outer layers, together with the Coriolis forces acting on rotating convective flows in what are asymmetric rotators, could lead to a gradual relaxation, through energy dissipation, of the global field structure to a form that is symmetric about the rotation axis. The timescale would depend on the field strength, and for a field weaker than several megagauss, it could require a timescale too long to be observed in white dwarfs that still show spectral lines.

Alternatively, because these normal-mass large-field magnetic white dwarfs mostly occur after the start of crystallisation \citep{BagLan21,BagLan22}, one could speculate that the crystallisation dynamo \citep{Iseetal17,Schetal21,Ginetal22} may generate some fraction of the observed fields and further that this dynamo produces essentially axisymmetric surface magnetic field structures with extremely strong fields. This kind of origin would have similarities with the axisymmetric fields that are commonly found in fully convective strongly magnetic M-dwarfs and, much weaker, in the planets Earth and Jupiter. However, it has been argued that the instability of the mantle surrounding the core that starts to crystallise cannot produce fields much stronger than 1\,MG \citep{Iseetal17,MonDun24,Casetal24}. Furthermore, \citet{Cametal24} find that, even if the crystallisation-driven dynamo could generate a strong magnetic field, this field would take too long to emerge at the stellar surface. On the other hand, \citet{MonDun24} have suggested that crystallisation may still play a role by triggering a temporary phenomenon of differential rotation, which in turn would generate a magnetic field -- the diffusion timescale of which has not been discussed.

The situation for ultra-massive white dwarfs (with $M\ga 1.0\,M_\odot$) is different, as both strongly magnetic variable and non-variable white dwarfs are found among them. Some of these massive white dwarfs are the product of WD-WD merging, during which a dynamo could create a strong magnetic field \citep{Garetal12}, though it would not necessarily be symmetric about the rotation axis. In addition, some massive magnetic white dwarfs could be the result of single-star evolution and have acquired an axisymmetric field following the same mechanism acting in normal-mass white dwarfs.

Further investigation into the variability of magnetic white dwarfs is necessary. Nevertheless, current observations have already imposed significant constraints that any theory explaining the origin and evolution of magnetic fields in degenerate stars must take into account.

\onecolumn
\begin{center}
\begin{small}
\tabcolsep=0.14cm
\begin{longtable}{llllrrrrrccl}
\caption{\label{Tab_Stars} Stars used in this work, their main physical parameters, and a note on their magnetic variability.} 
\\
\hline\hline
\multicolumn{2}{c}{STAR}            & Spect. &Atm. & $G$ & $d$ &\teff &$M$        & Age & \bs\ &$P$\\
               &                    & Type   &comp.&     & (pc)& (K)  &($M_\odot$) &(Gyr)&(MG)  &(d)\\   
\hline
\endfirsthead \caption{Continued.}\\
  \hline\hline
\multicolumn{2}{c}{STAR}            & Spect. &Atm. & $G$ & $d$ &\teff &$M$        & Age & \bs\ &$P$\\
               &                    & Type   &comp.&     &(pc) & (K)  &($M_\odot$) &(Gyr)&(MG)  &(d)\\
\hline
\endhead
\hline
\endfoot
 WD\,0004$+$122  &LP 464-57                    &DCP    &H   &16.25& 17.5& 5285& 0.813 & 7.830&100   &   n.v. \\
 WD\,0009$+$501  &EGGR 381                     &DA     &H   &14.23& 10.9& 6483& 0.740 & 3.230&  0.25&   0.334\\
 WD\,0011$-$134  &G 158-45                     &DAH    &H   &15.75& 18.6& 5869& 0.719 & 4.150&  9   &   0.031\\
 WD\,0041$-$102  &Feige 7                      &DBAH   &He  &14.53& 31.1&21334& 1.140 & 0.350& 20   &   0.091\\
 WD\,0051$+$117  &PHL 886                      &DAH    &H   &15.96&115.7&20534& 0.58  & 0.047&  0.27&   4.703\\
 WD\,0058$-$044  &GD 9                         &DAH    &H   &15.43& 72.1&17062& 0.63  & 0.137&  0.33&   1.96 \\
 WD\,0232$+$525  &EGGR 314                     &DAH    &H   &13.87& 28.8&17351& 0.817 & 0.230&  0.04&   var. \\
 WD\,0233$-$242A &LP 830-14                    &DAH    &H   &15.67& 18.5& 5066& 0.453 & 3.070&  3.8 &   0.066\\
 WD\,0236$-$269  &PHL 4227                     &DXP    &He  &16.48& 35.0& 7175& 0.726 & 2.360& 50   &   n.v.:\\
 WD\,0253$+$508  &KPD 0253+5052                &DAH    &H   &15.26& 68.7&19590& 0.70  & 0.120& 13   &   0.170\\
 WD\,0301$+$059  &SDSS J030350.63+060748.9     &DXP    &H   &14.96& 31.3&16018& 1.083 & 0.620&200   &   n.v.:\\
 WD\,0313$-$084  &GALEX J031613.8-081637       &DA     &H   &16.69& 31.4& 6369& 0.779 & 3.860&  0.8 &   n.v. \\
 WD\,0316$-$849  &V* CL Oct                    &DAH    &H   &14.75& 29.4&26465& 1.267 & 0.340&450   &   0.008\\
 WD\,0322$-$019  &EGGR 566                     &DAZH   &H   &15.90& 16.9& 5248& 0.673 & 6.150&  0.12&   var. \\
 WD\,0410$-$114  &G 160-51                     &DAH    &H   &15.38& 35.1& 7474& 0.420 & 0.900&  1.7 &   0.016\\
 WD\,0446$-$789  &WG 47                        &DA     &H   &13.44& 44.0&23514& 0.54  & 0.021&  0.01&   var.:\\
 WD\,0548$-$001  &EGGR 248                     &DQP    &He  &14.42& 11.2& 6053& 0.648 & 3.350& 10   &   n.v. \\
 WD\,0553$+$053  &EGGR 290                     &DAH    &H   &13.96&  8.1& 5741& 0.688 & 4.050& 20   &   n.v. \\
 WD\,0637$+$478  &GD 77                        &DAH    &H   &14.82& 40.2&13737& 0.74  & 0.369&  0.95&   1.362\\
 WD\,0654$+$059  &2MASS J06572938+0550479      &DC     &He  &17.26& 39.0& 6010& 0.691 & 3.970&  7.0 &   n.v. \\
 WD\,0708$-$670  &SCR J0708-6706               &DCP    &H   &15.95& 17.0& 5007& 0.593 & 6.180&200   &   n.v.:\\
 WD\,0745$+$115  &GALEX J074842.4+112502       &DC     &He  &16.50& 46.7& 9831& 0.93  & 1.708& 50   &   n.v. \\
 WD\,0756$+$437  &EGGR 428                     &DCP    &H   &16.14& 22.0& 7215& 1.038 & 4.440&200   &   0.278\\
 WD\,0810$-$353  &UPM J0812-3529               &DAH    &H   &14.34& 11.2& 6239& 0.700 & 3.160& 45   &   n.v. \\
 WD\,0816$-$310  &SCR J0818-3110               &DZH    &He  &15.38& 19.4& 6754& 0.802 & 3.480&  0.1 &   10:  \\
 WD\,0850$+$192  &EGGR 904 = LB 8915 (not 8827)&DBAH   &He  &15.60& 72.1&17368& 0.74  & 0.20 &  0.85&   0.1: \\
 WD\,0907$+$213  &GALEX J091016.5+210555       &DBAH   &He  &16.36& 96.3&15302& 0.67  & 0.245&  0.5 &   0.3: \\
 WD\,0912$+$536  &EGGR 250                     &DCP    &He  &13.78& 10.3& 7155& 0.712 & 2.260&100   &   1.331\\
 WD\,1008$-$242  &UCAC4 328-061594             &DAP    &H   &15.19& 39.7&21653& 1.152 & 0.340&100   &   n.v.:\\
 WD\,1008$+$290  &LP 315-42                    &DQpecP &H   &16.54& 14.7& 4595& 0.74  & 8.12 &100   &   n.v. \\
 WD\,1009$-$184  &WT 1759                      &DZH    &He  &15.33& 18.1& 6394& 0.753 & 3.680&  0.15&   var.:\\
 WD\,1015$+$014  &PG 1015+014                  &DA     &H   &16.33& 49.4&10456& 0.93  & 1.29 & 75   &   0.069\\
 WD\,1031$+$234  &Ton 527                      &DAH    &H   &15.65& 79.8&21185& 0.97  & 0.190&500   &   0.140\\
 WD\,1036$-$204  &EGGR 535                     &DQpecP &H   &15.83& 14.1& 5761& 0.943 & 6.170&200   &   n.v. \\
 WD\,1043$-$050  &HE 1043-0502                 &DBH    &He  &16.82& 82.6&17190& 1.07  & 0.50 & 500  &   n.v.:\\
 WD\,1045$-$091  &HE 1045-0908                 &DAH    &H   &16.45& 54.1& 9938& 0.84  & 1.13 & 16   &   0.113\\
 WD\,1105$-$340  &SCR J1107-3420A              &DAH    &H   &13.69& 26.2&13603& 0.657 & 0.310&  0.13&   n.v. \\
 WD\,1105$-$048  &EGGR 76                      &DAH    &H   &13.09& 24.8&15703& 0.568 & 0.150&  0.05&   var.:\\
 WD\,1116$-$470  &SCR J1118-4721               &DC     &He  &15.36& 17.0& 5903& 0.615 & 3.360& 10   &   n.v. \\
 WD\,1211$-$171  &HE 1211-1707                 &DBH    &He  &16.70& 90.7&30650& 1.20  & 0.165& 50   &   0.083\\
 WD\,1249$-$022  &GALEX J125230.9-023417       &DAHe   &H   &17.46& 77.7& 7856& 0.58  & 1.14 &  5   &   0.0037\\
 WD\,1312$+$098  &PG 1312+099                  &DAH    &H   &16.42&101.5&21293& 0.89  & 0.148& 80   &   0.226\\
 WD\,1315$-$781  &LAWD 45                      &DAH    &H   &15.96& 19.3& 5575& 0.689 & 4.590&  5.5 &   n.v.:\\
 WD\,1315$+$222  &LP 378-956                   &DCP    &He  &16.69& 31.8& 6334& 0.726 & 3.540& 15   &   n.v. \\
 WD\,1328$+$307  &G 165-7                      &DZH    &He  &15.97& 25.5& 6351& 0.694 & 3.190&  0.3 &   var. \\
 WD\,1346$+$121  &LP 498-66                    &DCP    &He  &17.81& 28.3& 4607& 0.742 & 8.100& 50   &   n.v. \\
 WD\,1350$-$090  &PG 1350-090                  &DAH    &H   &14.54& 19.7& 8881& 0.803 & 1.410&  0.46&   n.v. \\
 WD\,1532$+$129  &G 137-24                     &DZH    &He  &15.59& 19.3& 5796& 0.633 & 3.970&  0.05&   var. \\
 WD\,1556$+$044  &PM J15589+0417               &DCP    &He  &15.99& 22.5& 6730& 0.848 & 3.810& 15   &   n.v. \\
 WD\,1615$+$542  &GALEX J161634.4+541011       &DAHe   &H   &18.24& 93.7& 7938& 0.75  & 1.694&  6.5 &   0.067\\
 WD\,1619$+$046  &GALEX J162157.7+043219       &DUH    &X   &16.86& 61.6&19655& 1.25  & 0.777& 15   &   0.028:\\
 WD\,1639$+$537  &GD 356                       &DAHe   &H   &14.96& 20.1& 7729& 0.753 & 1.830& 11.5 &   n.v. \\
 WD\,1658$+$440  &PG 1658+441                  &DAH    &H   &14.78& 31.7&29613& 1.284 & 0.270&  2.5 &   n.v.:\\
 WD\,1703$-$266  &UCAC4 317-104829             &DAH    &H   &15.00& 13.0& 6128& 0.809 & 4.610&  8   &   var. \\
 WD\,1712$-$590  &Gaia DR3 5915797694789556096 &DAH    &H   &15.62& 29.8& 8608& 0.831 & 1.740&  0.73&   var. \\
 WD\,1743$-$521  &L 270-31                     &DAH    &H   &15.67& 38.9&14852& 1.116 & 0.930& 18   &   2.833\\
 WD\,1748$+$708  &EGGR 372                     &DXP    &He  &13.77&  6.2& 5224& 0.671 & 6.270&100   &   s.v.:\\
 WD\,1750$-$311  &UCAC4 295-140552             &DQ     &He  &15.85& 44.6&15518& 1.10  & 0.806& 2.1  &   n.v.: \\
 WD\,1754$-$550  &GALEX J175845.9-550117       &DU     &?   &15.59& 51.5&34753& 1.33  & 0.184&100   &   var. \\
 WD\,1814$+$248  &G 183-35                     &DAP    &H   &16.82& 37.6& 6800& 0.766 & 3.030&  5   &   var. \\
 WD\,1829$+$547  &EGGR 374                     &DXP    &H   &15.45& 17.0& 6711& 0.897 & 4.380&150   &   n.v. \\
 WD\,1900$+$705  &LAWD 73                      &DAH    &H   &13.24& 12.9&12540& 1.029 & 1.090&180   &   s.v.:\\
 WD\,1953$-$011  &LAWD 79                      &DAH    &H   &13.59& 11.6& 7790& 0.698 & 1.540&  0.5 &   1.448\\
 WD\,2010$+$310  &GD 229                       &DBP    &He  &14.77& 30.8&18365& 1.135 & 0.540&300   &   s.v.:\\
 WD\,2047$+$372  &EGGR 261                     &DAP    &H   &13.03& 17.6&14710& 0.796 & 0.350&  0.07&   0.243\\
 WD\,2049$-$222  &LP 872-48                    &DCP    &He  &14.93& 20.3& 7811& 0.693 & 1.600&  5   &   var.:\\
 WD\,2049$-$253  &UCAC4 325-215293             &DCP    &H   &16.03& 18.0& 5048& 0.584 & 5.700& 20   &   n.v. \\
 WD\,2051$-$208  &BPS CS 22880-0134            &DAH    &H   &15.11& 31.3&20362& 1.222 & 0.580&  0.25&   0.059\\
 WD\,2105$-$820  &LAWD 83                      &DAZH   &H   &13.60& 16.2& 9914& 0.751 & 0.920&  0.04&   var. \\
 WD\,2138$-$332  &L 570-26                     &DZAH   &He  &14.44& 16.1& 7113& 0.715 & 2.330&  0.05&   0.258\\
 WD\,2150$+$591  &UCAC4 747-070768             &DAH    &H   &14.37&  8.5& 5202& 0.640 & 5.840&  0.80&   var. \\
 WD\,2211$+$372  &LP 287-35                    &DCP    &He  &16.82& 29.2& 6424& 0.878 & 4.370& 60   &   n.v. \\
 WD\,2316$+$123  &KUV 23162+1220               &DAH    &H   &15.42& 40.3&11721& 0.86  & 0.77 & 30   &   17.86\\
 WD\,2359$-$434  &LAWD 96                      &DA     &H   &12.89&  8.3& 8428& 0.821 & 1.790&  0.11&   0.112\\
 \hline
\end{longtable}
\end{small}
\tablefoot{A number in the last column represents the established rotation period (in d) of a star that shows polarimetric variability;
tentative periods are followed by the symbol `:'.  Other symbols have the following meaning:
`var.'  means that the star is certainly polarimetric variable but the period is still unknown; 
`var.:' means that hints of subtle variability have been detected over a short timescale; 
`n.v.'  means that the observed polarisation was constant (within uncertainties) and the star was observed at least three times; 
`n.v.:' means that observed polarisation was constant (within uncertainties) but the star was observed only twice; 
`s.v.'  means that polarisation shows some sign of subtle variability over a timescale of a decade or longer.
For stars within the local 40\,pc volume, the parameters of stellar magnitude, distance, spectral type, atmospheric composition, temperature, mass, and
age are from \citet{OBretal24}; for the remaining stars, we used the catalogue from \citet{Genetal21} with ages interpolated from
the tables by \citet{Bedetal20}.}
\end{center}
\twocolumn

\begin{acknowledgements}
The new observations presented in this work were made with the FORS2 instrument at the ESO Telescopes at the La Silla Paranal Observatory under program ID 110.243J.001, 110.23XV.001, 110.23XV.002, and 112.25C9.001, and with the ISIS instrument at the \textit{William Herschel} Telescope (operated on the island of La Palma by the Isaac Newton Group), under programmes P10 in 19A and P8 in 19B. This research has made use also of additional FORS1 and FORS2 data obtained from the ESO Science Archive Facility: data for WD\,1036--204 were obtained under programmes IDs 70.D-0259(A), 087.D-0714(A), 089.D-0612(A), 090.D-0269(A). Data for WD\,1105--340 was acquired with ESPaDOnS on the Canada-France-Hawaii Telescope (CFHT) (operated by the National Research Council (NRC) of Canada, the Institut National des Sciences de l’Univers of the Centre National de la Recherche Scientifique (CNRS) of France, and the University of Hawaii), under programmes 17AC01, 19AC04, 19BC02, and 21BC02. We thank the anonymous referee for their very constructive criticism. We thank Matthias Schreiber and Antonino Lanza for very useful comments. JDL acknowledges the financial support of the Natural Sciences and Engineering Research Council of Canada (NSERC), funding reference number 6377-2016. 
\end{acknowledgements}
\bibliography{sbabib}
\appendix

\onecolumn

\section{New observations}

\begin{center}
\tabcolsep=0.08cm
\begin{longtable}{llccrlcccc}
\caption{\label{Tab_Log} Newly presented spectropolarimetric observations}\\
\hline\hline
\multicolumn{2}{c}{STAR}   &  DATE       &  UT       & EXP  & Instrument &grism/  & spectral  & $R$  & Stokes   \\
            &              &             &           & (s)  &            & grating& coverage  &      & parameters\\
\hline
\endfirsthead \caption{Continued.}\\
\hline\hline
\multicolumn{2}{c}{STAR}   &  DATE       &  UT       & EXP  & Instrument &grism/  & spectral  & $R$  & Stokes   \\
            &              &             &           & (s)  &            & grating& coverage  &      & parameters\\
\hline
\endhead
\hline
\endfoot
WD\,0313$-$084 & GALEX J031613.8-081637& 2023-01-19 & 02:42 & 3600 & FORS2     & 1200R & 5700-7100 & 2100 &$IV$ \\
               &                       & 2023-01-20 & 02:13 & 3600 & FORS2     & 1200R & 5700-7100 & 2100 &$IV$ \\
               &                       & 2023-01-23 & 02:07 & 3600 & FORS2     & 1200R & 5700-7100 & 2100 &$IV$ \\
               &                       & 2023-09-18 & 05:53 &  900 & FORS2     & 1200R & 5700-7100 & 2100 &$IV$ \\
               &                       & 2023-09-19 & 04:40 & 3600 & FORS2     & 1200R & 5700-7100 & 2100 &$IV$ \\ [2mm] 
WD\,0548$-$001  & G\,99--37            & 2012-11-17 & 06:16 & 720  & FORS2     & 1200B  & 3700-5100 & 1400 & $IV$ \\
                &                      & 2012-11-18 & 04:17 & 960  & FORS2     & 1200B  & 3700-5100 & 1400 & $IV$ \\
                &                      & 2012-11-19 & 08:42 & 480  & FORS2     & 1200B  & 3700-5100 & 1400 & $I$ \\ [2mm] 
WD\,1036$-$204 &EGGR 535               & 2003-02-03 & 05:11 & 2900 & FORS1     & 600B  & 3500-5900 &  780   & $IV$ \\ 
               &                       & 2011-04-09 & 00:39 & 1000 & FORS2     & 600B  & 3500-6200 &  780   & $IV$ \\
               &                       & 2012-04-11 & 02:46 &  360 & FORS2     & 600B  & 3500-6100 &  780   & $IV$ \\ 
               &                       & 2013-02-21 & 08:55 & 1000 & FORS2     & 600B  & 3500-6100 &  780   & $IV$ \\ [2mm] 
WD\,1105$-$340 &SCR J1107-3420A        & 2019-03-15 & 10:05 & 4188 &  ESP      &       & 4000--9000& 65000  & $IV$ \\
               &                       & 2019-03-18 & 10:26 & 4188 &  ESP      &       & 4000--9000& 65000  & $IV$ \\  
               &                       & 2019-03-19 & 07:38 & 4188 &  ESP      &       & 4000--9000& 65000  & $IV$ \\  
               &                       & 2019-03-19 & 11:50 & 4188 &  ESP      &       & 4000--9000& 65000  & $IV$ \\  
               &                       & 2019-03-20 & 08:24 & 4188 &  ESP      &       & 4000--9000& 65000  & $IV$ \\  
               &                       & 2019-03-20 & 11:27 & 4188 &  ESP      &       & 4000--9000& 65000  & $IV$ \\  
               &                       & 2019-03-21 & 07:47 & 4188 &  ESP      &       & 4000--9000& 65000  & $IV$ \\    
               &                       & 2019-03-21 & 10:40 & 4188 &  ESP      &       & 4000--9000& 65000  & $IV$ \\  
               &                       & 2019-03-22 & 10:23 & 4188 &  ESP      &       & 4000--9000& 65000  & $IV$ \\  
               &                       & 2019-05-31 & 06:20 & 4188 &  ESP      &       & 4000--9000& 65000  & $IV$ \\ 
               &                       & 2022-02-22 & 00:07 & 4188 &  ESP      &       & 4000--9000& 65000  & $IV$ \\[2mm] 
WD\,1116$-$470 &SCR J1118$-$4721.      & 2023-01-19 & 05:44 & 1800 & FORS2     &  300V & 3800--9200&  365   & $IV$ \\[2mm] 
WD\,1619$+$046 &GALEX J162157.7+043219 & 2023-06-16 & 04:07 & 2760 & FORS2     & 1200R & 5700-7100  & 2140  &$IV$  \\
               &                       & 2023-06-16 & 04:43 & 1000 & FORS2     & 1200R & 5700-7100  & 2140  &$IV$  \\ [2mm]
WD\,1658$+$440 & PG~1658+441           & 2019-04-21 & 02:14 & 3600 & ISIS      & R600B & 3700-5200 & 2600  & $IV$  \\
               &                       &            &       & 3600 &           & R1200R& 5900-6670 & 8600  & $IV$  \\ [2mm] 
WD\,1712$-$590&Gaia DR3 5915797694789556096
                                       & 2022-05-16 & 03:34 & 6000 & FORS2     & 1200R & 5700-7100 & 2140 & $IV$  \\
              &                        & 2022-05-25 & 07:41 & 3000 & FORS2     & 1200R & 5700-7100 & 2140 & $IV$  \\ [2mm] 
WD\,1748$+$708&  EGGR~372 =  G240-72   & 1974-09-06 & 06:30 &18000:& MCSP      &       & 3600-9800 & 31-25& $IV$  \\
              &                        & 1974-09-07 & 06:30 &18000:& MCSP      &       & 3700-7600 & 31-25& $IQU$ \\
              &                        & 2019-04-20 & 02:07 & 1200 & ISIS      & R600B & 3650-5250 & 2600 & $IV$  \\
              &                        &            &       & 1200 &           & R1200R& 5900-6650 & 8600 & $IV$  \\
              &                        & 2019-04-20 & 02:15 & 2400 & ISIS      & R600B & 3650-5250 & 2600 & $IQU$ \\
              &                        &            &       & 2400 &           & R1200R& 5900-6650 & 8600 & $IQU$ \\
              &                        & 2019-10-03 & 22:44 & 2400 & ISIS      & R300B & 3650-5350 & 1100 & $IV$  \\ 
              &                        &            &       & 2400 &           & R158R & 5300-9000 & 1000 & $IV$  \\ 
              &                        & 2019-10-03 & 02:36 & 3200 & ISIS      & R300B & 3650-5350 & 1100 & $IQU$ \\ 
              &                        &            &       & 3200 &           & R158R & 5300-9000 & 1000 & $IQU$ \\ [2mm] 
WD\,1754$-$550&GALEX J175845.9-550117  & 2023-06-12 & 07:40 & 2760 &FORS2      & 1200B & 3650-5130 & 1420 & $IV$  \\ [2mm]
WD\,1829$+$547& EGGR~374 = G227-35     &  2019-04-20 &  03:05 & 2000 & ISIS       & R600B  & 3650-5300 & 2600 & $IV$  \\             
              &                        &             &        & 2000 &            & R1200R & 5900-6650 & 8600 & $IV$  \\
              &                        &  2019-04-20 &  03:59 & 4000 & ISIS       & R600B  & 3650-5300 & 2600 & $IQU$ \\
              &                        &             &        & 4000 &            & R1200R & 5900-6650 & 8600 & $IQU$ \\
              &                        &  2019-10-03 &  00:24 & 1800 & ISIS       & R300B  & 3650-5300 & 2600 & $IV$  \\
              &                        &             &        & 1800 &            & R158R  & 5900-6650 & 8600 & $IV$  \\ [2mm]
WD\,2010$+$310&  GD\,229   &  1974-09-08 &  06:07 & 17460&  MCSP      &       & 3200-9400 &31-22 & $IQU$ \\
              &            &  2018-09-23 &  23:52 & 1800 &  ISIS      & R600B & 3650-5200 & 2600 & $IV$  \\
              &            &             &        & 1800 &            & R1200R& 6100-6900 & 8600 & $IV$  \\
              &            &  2018-09-23 &  22:57 & 3200 &  ISIS      & R600B & 3650-5200 & 2600 & $IQU$ \\
              &            &             &        & 3200 &            & R1200R& 6100-6900 & 8600 & $IQU$ \\              
              &            &  2019-04-20 &  04:50 & 1200 &  ISIS      & R600B & 3650-5200 & 2600 & $IV$  \\
              &            &             &        & 1200 &            & R1200R& 5850-6650 & 8600 & $IV$  \\
              &            &  2019-04-20 &  05:33 & 2400 &  ISIS      & R600B & 3650-5200 & 2600 & $IQU$ \\
              &            &             &        & 2400 &            & R1200R& 5850-6650 & 8600 & $IQU$ \\              
              &            &  2019-04-21 &  04:12 & 1800 &  ISIS      & R600B & 3650-5200 & 2600 & $IV$  \\
              &            &             &        & 1800 &            & R1200R& 5850-6650 & 8600 & $IV$  \\
              &            &  2019-04-21 &  05:03 & 3600 &  ISIS      & R600B & 3650-5200 & 2600 & $IQU$ \\
              &            &             &        & 3600 &            & R1200R& 5850-6650 & 8600 & $IQU$ \\ [2mm] 
WD\,2049$-$222& LP~872-48  & 2022-05-16  &  07:30 & 1600 &  FORS2     &  300V & 3800-9200 &  365 & $IV$  \\
              &            & 2022-05-17  &  06:53 &  800 &  FORS2     &  300V & 3800-9200 &  365 & $IV$  \\
              &            & 2022-05-25  &  08:31 & 1600 &  FORS2     &  300V & 3800-9200 &  365 & $IV$  \\ 
\hline
\end{longtable}
\end{center}

\twocolumn

\section{Variability of magnetic white dwarfs: Comments on individual stars}\label{Sect_Stars}
We preliminary note that to compare observations taken by different authors, we need to know how circular polarisation was defined in different works. This problem has been highlighted in detail, for example in Sect.~2 of \citet{BagLan20}. Here, we have adopted the definition of positive handedness of circular polarisation given by \citet{LanLan04}, and converted some of the literature data to it by changing the sign of circular polarisation with respect to the originally published data.

Numerous examples in the literature \citep[e.g.][]{BagLan20} and in this section  show that many magnetic white dwarfs with very strong fields have a circular polarisation spectrum that varies rapidly with wavelength. The consequence is that the same source observed with even slightly different broadband filters may result in differences of the measured broadband polarisation signal that are larger than the formal uncertainties. This specific issue is discussed in Sect.~7 of \citet{BagLan19a}, where some numerical examples are presented. As a consequence, small differences between broadband polarimetric measurements obtained with different instruments cannot reliably provide strong evidence that a star is magnetically variable. 

Spectropolarimetry should allow comparisons that are less instrument dependent, provided the observations overlap in wavelength, and that the resolving power of different instrument is similar. There is no doubt, however, that the safest way to establish variability or constancy of polarisation is to repeatedly observe the star with the same instrument and instrument setting, and to always perform the data reduction using exactly the same technique.

In the following, we discuss all stars of our sample (including two stars that did not make it into the final list of Table~\ref{Tab_Stars}). The titles of the individual subsections contain the star's name in the Villanova system \citep{McCSio77,McCSio99},  the main SIMBAD identifier, and our conclusions about the magnetic variability (using the same designation as in Table~\ref{Tab_Stars}, followed by the rotation period $P$, if this is known). When the rotation period is obtained only from photometric data, we use the designation \pphot.

\begin{figure*}
\begin{center}
\includegraphics[angle=0,width=\textwidth,trim={0.5cm 1.4cm 1.0cm 2.0cm},clip]{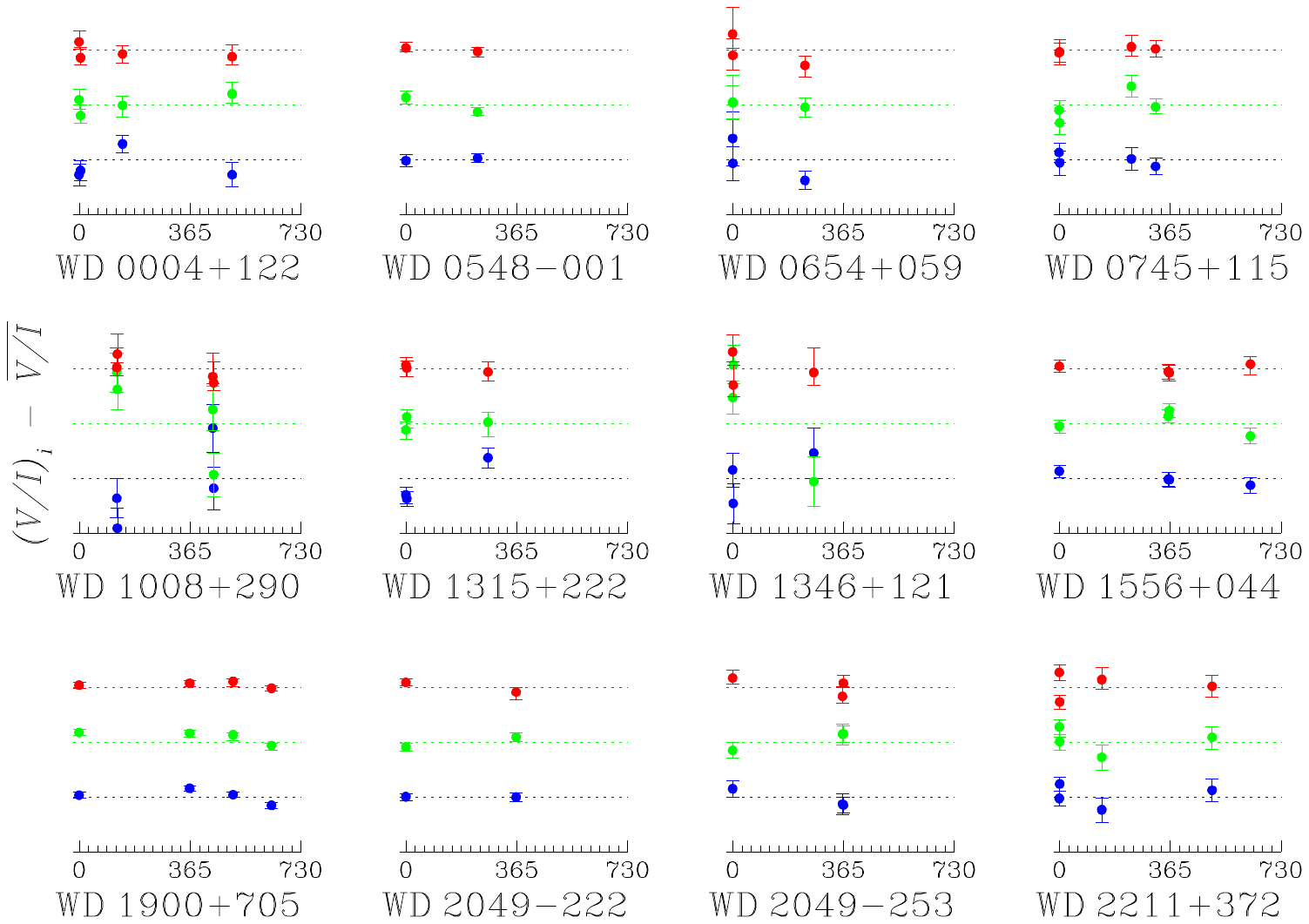}
\end{center}
\caption{\label{Fig_BBCP} Variation in the broadband circular polarisation measurements observed in three filters B$'$V$'$R$'$ of the DiPol-UF instrument at the NOT \citep{Beretal22,Beretal23,Beretal24}. In each panel, the three dotted
lines represent the average of all data obtained in a particular filter: from top to bottom, the R$'$ filter (red dotted line), the V$'$ filter (green dotted line) and the B$'$ filter (blue dotted line). These lines are separated by $\Delta V/I = 0.3$\,\%. The $x$-axis represents the time of the observations in days from the first measurement. The solid circles represent the difference between each measurement and the average of all measurements in a given filter (red circles for filter R$'$, green circles for filter V$'$, and blue circles for filter B$'$.}
\end{figure*}

\subsection{WD~0004+122 = LP 464-57 (non-variable)}\label{Sect_First} 
\begin{figure}
\begin{center}
\includegraphics[angle=0,width=8.8cm,trim={1.7cm 6.0cm 1.0cm 11.0cm},clip]{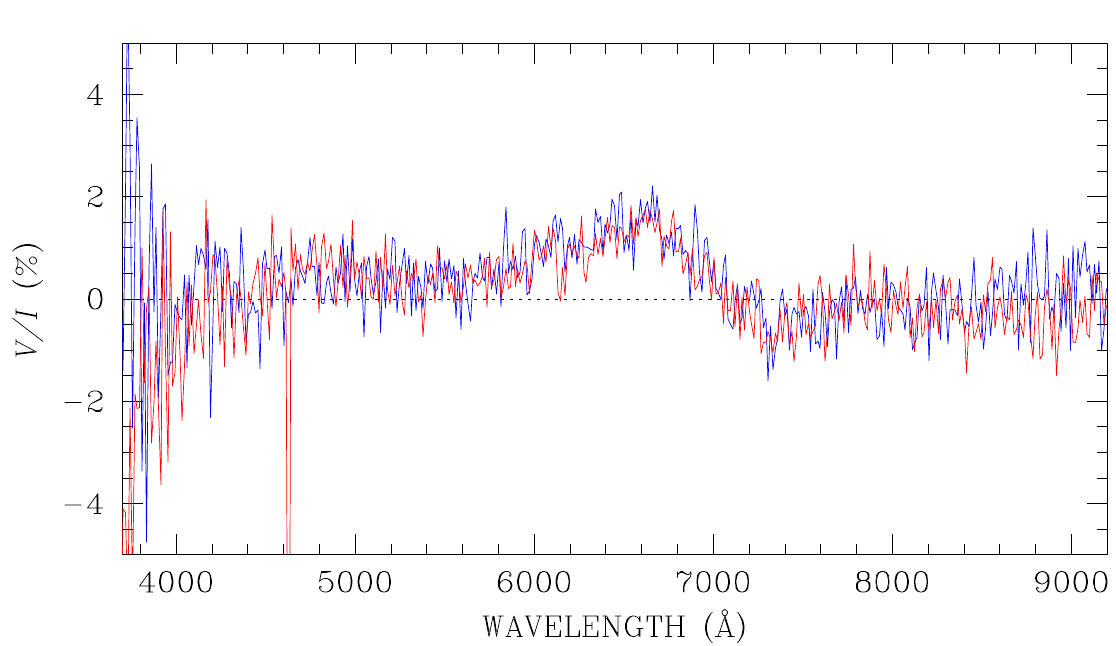}
\end{center}
\caption{\label{Fig_WD0004}  WD\,0004+122: Stokes $V/I$ from two pairs of observations obtained within 15\,m on 2019-10-07.}
\end{figure}
This star was discovered to be magnetic, using spectropolarimetry, by \citet{BagLan20}, and later observed four times by \citet{Beretal24} in broadband circular polarisation. These measurements are shown in the top left panel of Fig.~\ref{Fig_BBCP}. None of them deviated more than $\simeq 1\,\sigma$ from the mean value of the measurements obtained in the same filter, except for a $\sim 2.5\,\sigma$ difference between on measurement and the other three in the B filter, which might reflect a real small variability. We reanalysed the FORS2 spectropolarimetry published by \citet{BagLan20} to check for differences between the $V/I$ profiles obtained from the first and the second pair of exposures, sampling an interval in time of about 15\,m, without finding any difference (see Fig.~\ref{Fig_WD0004}). We conclude that the star does not show convincing sign of variability. We estimate a field strength of $\simeq 100$\,MG.

\subsection{WD~0009+501 = EGGR 381 (variable, $P \simeq 8$\,h)} 
This white dwarf was well monitored and modelled as a magnetic variable with \bs\ between 150 and 250~kG by \citet{Valetal05}, who presented a model obtained by adopting a rotational period of $0.3337 \pm 0.0031$ d.
Recent unpublished ESPaDOnS data obtained by us confirm the rotational period. 

\subsection{WD~0011--134 = G 158-45 (variable, $\pphot \simeq 0.73$\,h)}
Discovered magnetic (with a field of about 9\,MG) by \citet{Beretal92}, and as a magnetic variable by \citet{Putney97}.
From photometry, \citet{Lawetal13phot} found a rotational period of $44 \pm 0.43$~min. While there is no guarantee that the photometric period reflects the rotational period of the star, the star is definitely a magnetic variable, and in fact, this very short photometric period is completely consistent with several of the many acceptable periods found from the six widely spaced \bz\ data values published by \citet{Putney97} from spectropolarimetric observations made between 1994 Sep 29 and 1994 Dec 31. This star is one of the  magnetic white dwarfs that clearly shows longitudinal field reversal. 

\subsection{WD~0041--102 = Feige 7 (variable, $P \simeq 2$\,h)}
\citet{Lieetal77} showed that the spectrum is rich in faint lines of H and He, and identified a magnetic field of about 18--20 MG from the spectrum. They found, on the basis of broadband circular polarisation measurements, that the field varies with a period of 131.6 min. (Note that the broadband circular polarisation changes sign as the star rotates.)  A decentred dipole model of the field structure based on a series of flux spectra was derived by \citet{Achetal92}; both H and He apparently vary substantially in abundance over the surface. \citet{Achetal92} also discovered that the star varies photometrically (in a double wave) with the same period as the spectrum and broad-band polarisation. 

\subsection{WD~0051+117 = PHL 886 (variable, $P \simeq 4.7$\,d)} 
From unpublished spectropolarimetric data (Landstreet \& Bagnulo, in prep., hereafter LB25) it is found that the star is periodically variable ($P=4.703$~d) with $\bs \simeq 270$~kG. The field \bz\ reverses sign as the star rotates. 
Modelling will be presented in a forthcoming paper. 

\subsection{WD~0058--044 = GD 9 (variable, $P \simeq 2$~d?)}
From our unpublished spectropolarimetric data we found that the star is periodically variable (the period is not determined uniquely from available data, but the most probable period is $P=1.96$~d), with $\bs \simeq 300$~kG (LB25). The field \bz\ varies sinusoidally but does not (quite) reverse sign. Modelling will be presented in a forthcoming paper. 

\subsection{WD~0232+525 = EGGR 314 (variable)}
This is a well-known star that was repeatedly included in studies of bright white dwarfs and been searched for a magnetic field \citep{Bycetal91,SchSmi95,Fabetal97} without any detection being reported. Recent particularly sensitive observations \citep{BagLan22} succeeded in detecting a weak field. The three published measurements (two with ISIS, one with ESPaDOnS) reveal a very weak, and probably variable longitudinal field (the measurement obtained with ESPaDOnS has the opposite sign as those obtained with ISIS).

\subsection{WD~0233--242A = LP 830-14 = NLTT 8435A (variable, $\pphot \simeq 1.5$\,h)} 
The star is discussed by \citet{BagLan21}, who suggested the presence of a variable field of the order of 3.8\,MG. In fact, \bs\ does not vary between two FORS observations, while \bz\ clearly changes sign. According to \citet{Venetal18} it has a photometric variation period of 95\,min. 

\subsection{WD~0236--269 = PHL 4227 (n.v.: -- only 2 meas.)}
\citet{Schetal01} report two spectropolarimetric detections of circular polarisation at about the 1.2\% level. The two observations were obtained three days apart, with no changes detected. Furthermore no significant change in the spectrum-average circular polarisation was observed. We tentatively assume that is not a magnetic variable, but the evidence for non-variability is not compelling (the original paper just reports: "consistent results were obtained on the two occasions").

\subsection{WD~0253+508 =  KPD~0253+5052 (variable, $P \simeq 4$\,h)}
The field was discovered by \citet{DowMar83}, who estimated $\bs \approx 13$\,MG, and modelled by \citet{AchWic89} using only flux spectra.  
The star shows continuum polarisation up to 0.4\% that is periodically variable according to \citet{SchNor91} with $P=0.170$~d. The sign of the broadband continuum circular polarisation briefly reverses during each cycle. 

\subsection{WD~0301+059 = SDSS J030350.63+060748.9 (n.v.: -- only 2 meas.)}
The star was discovered to be magnetic by \citet{LanBag20}, who did not find any obvious sign of variability between two observations obtained 1 day apart, nor between individual exposures of the same observing series (exposure times of single frames were 225\,s). We tentatively assume that it is not magnetically variable. The star therefore appears as a strongly magnetic, massive, non-variable star. 

\subsection{WD~0313--084 = GALEX J031613.8-081637 (non-variable)}
\begin{figure}
\begin{center}
\includegraphics[angle=0,width=8.8cm,trim={1.6cm 6.0cm 1.0cm  3.0cm},clip]{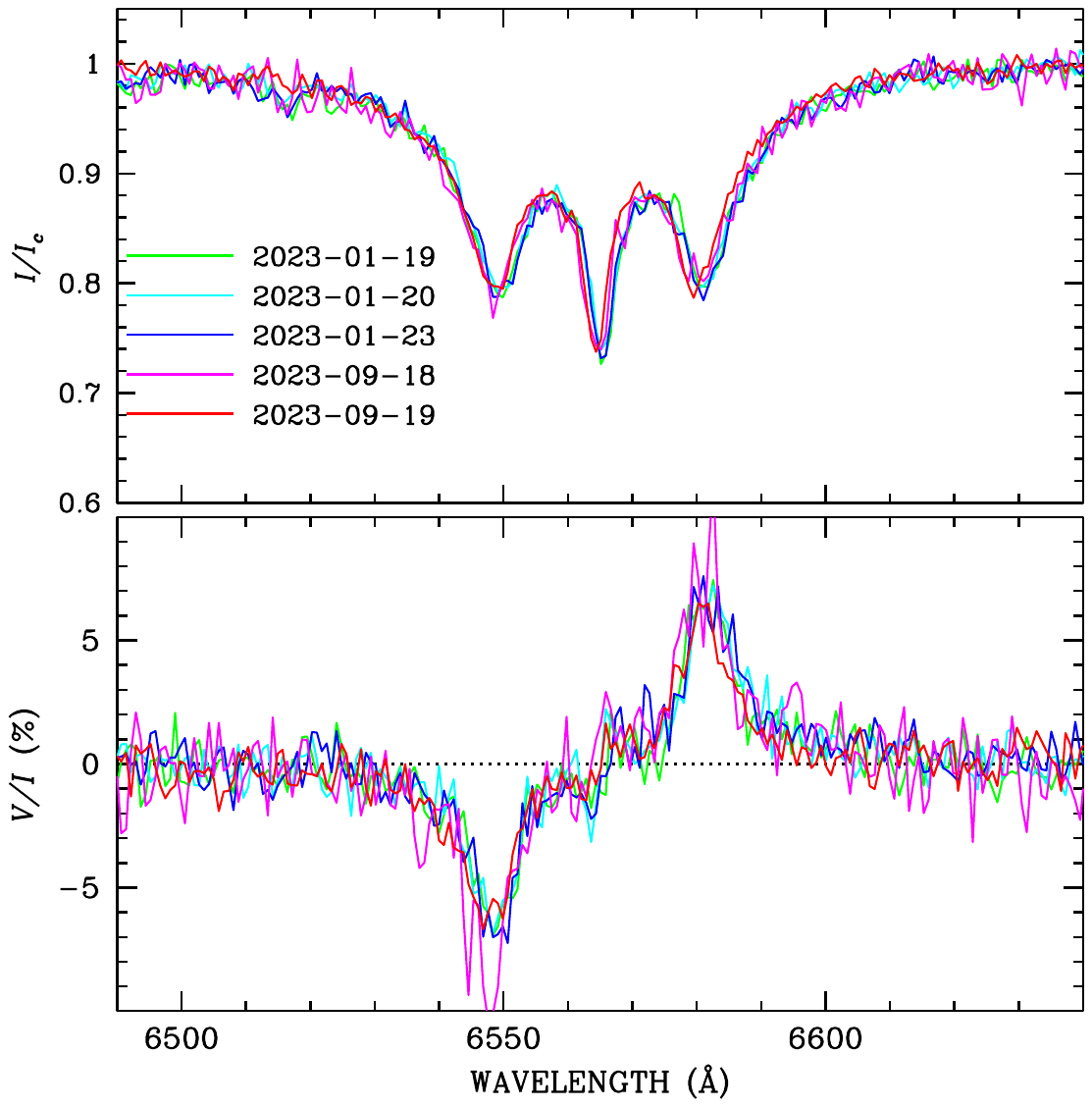}\\
\includegraphics[angle=0,width=8.8cm,trim={1.6cm 6.0cm 1.0cm 11.0cm},clip]{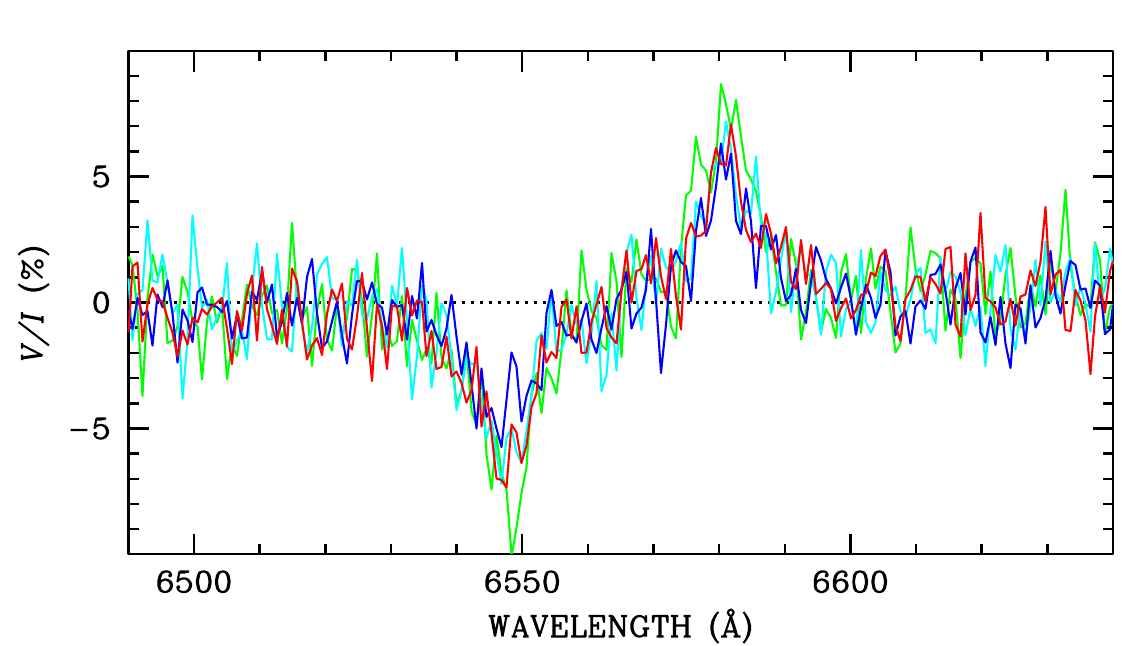}
\end{center}
\caption{\label{Fig_WD0313} {Top and middle panels:} WD~0313$-$084 observed with FORS2 in five different epochs. Bottom panels: Stokes $V/I$ from four pairs of observations obtained within 1\,h on 2023-09-19.}
\end{figure}
The star was discovered to be magnetic by \citet{OBretal23}. It was observed five times with FORS2 (see Sect.~\ref{Sect_New_Observations}), and no variability was found within timescales of 1\,h, 1\,d, and 8 months (see Fig.~\ref{Fig_WD0313}). The spectrum indicates that $\bs \simeq 800$\,kG.

\subsection{WD~0316--849 = EUVE J0317--855 (variable, $P \simeq 0.2$\,h)}
Studied by \citet{Feretal97} \citet{Buretal99}, and \citet{Venetal03}, the star has a rotation period $P = 725$~s. A detailed model based on UV spectra was obtained by \citet{Buretal99}, who found that the field varies in strength from about 180 to 800\,MG over the surface. \citet{Venetal03} shown that the star has a light curve with a minimum that corresponds to the maximum of the polarisation. This is one of the very few white dwarfs that show magnetic and photometric variability, for which we have a firmly established phase relationship between photometry and circular polarisation. This star is erroneously missing in the list of such stars compiled by \citet{Bagetal24b}.

\subsection{WD~0322--019 = EGGR 566 (variable)} 
The star was discussed by \citet{BagLan21}. Two FORS2 measurements by \citet{Faretal18} obtained during two consecutive nights ($\bz=-5.4 \pm 3.0$~kG and $-16.5 \pm 2.3$~kG) suggest that the longitudinal field may change on a timescale of a few days.

\subsection{WD~0330--000 = HE 0330-0002 (only 2 meas., and contradicting data)}
\citet{Schetal01} report two spectropolarimetric observations obtained one day apart, and concluded that no significant differences had been detected in the overall spectrum. However, the mean circular polarisation averaged over the observed wavelength range reported in their Table~1 differs by 0.5\%, contradicting the assessment of non-variability. Because of this, this star will not be considered in this paper. 

\subsection{WD~0410--114 = G 160-51 = NLTT~12758 (variable, $P \simeq 0.3\,$\,h)} %
A magnetic field of about 1.7\,MG was discovered by \citet{KawVen12}. \citet{Kawetal17} found that circular polarisation in the H$\alpha$ sigma components varies with $P = 22.6$~min. This is another example of a star for which the phase relationship between light curve and magnetic field are known \citep{Kawetal17} and it is also missing from the list compiled by \citet{Bagetal24b}.

\subsection{WD~0446--789 = WG 47 (var.:)}
\citet{BagLan18} show that this is a weak-field star with $\bz \simeq -5$\,kG; it probably has a dipolar field with axis nearly parallel to the rotation axis, because only very mild variations in \bz\ occur on a timescale of days.

\subsection{WD~0548--001 = EGGR 248 = G 99-37 (non-variable)}
\begin{figure}
\begin{center}
\includegraphics[angle=0,width=8.8cm,trim={1.3cm 6.0cm 1.0cm 3.0cm},clip]{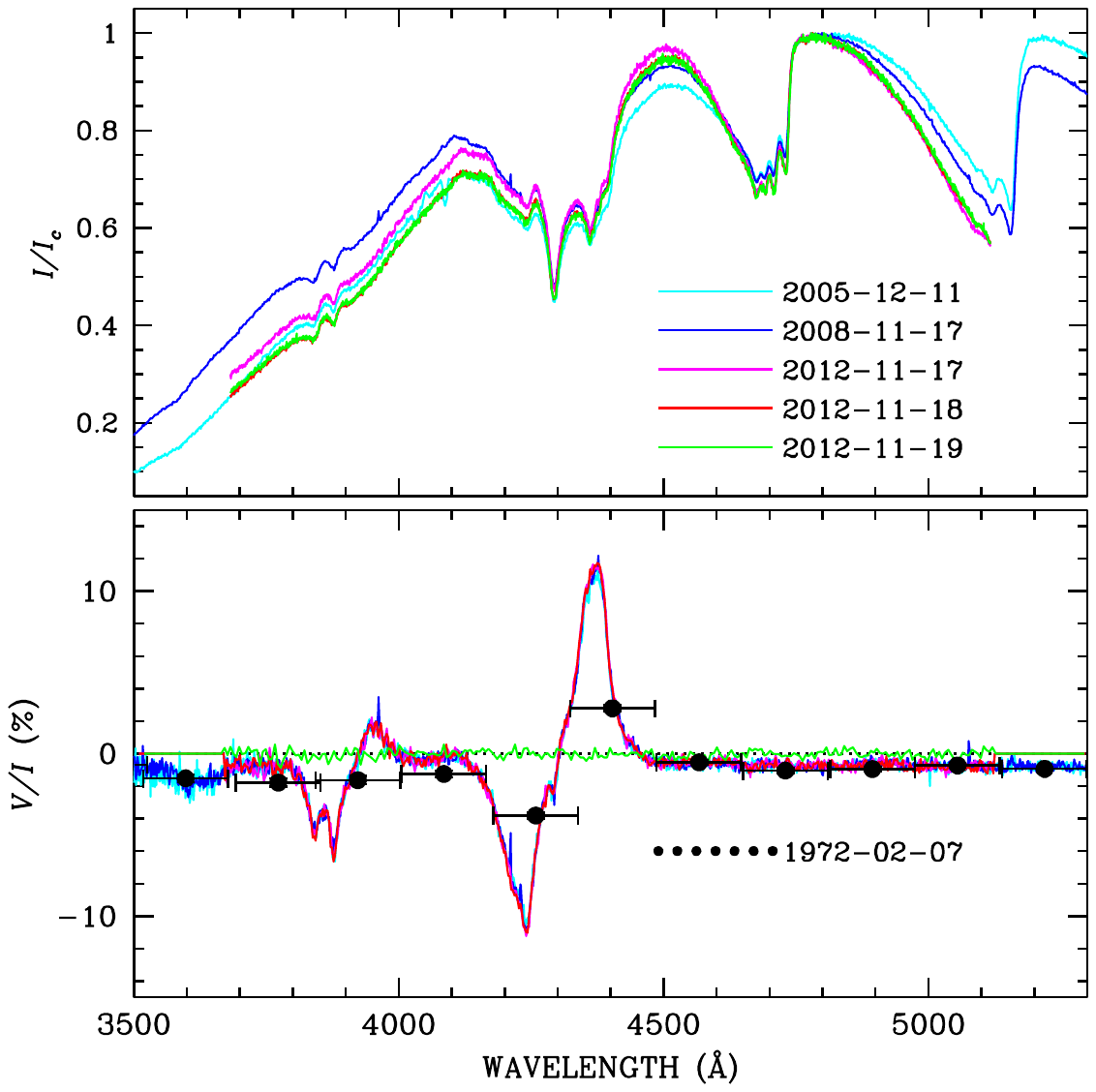}
\end{center}
\caption{\label{Fig_WD0548} Upper panel: WD~0548$-$001 = G\,99-37 flux spectra  obtained with FORS1 and FORS2 compared. The Stokes~$I$ FORS spectra have not been calibrated, and the differences in slopes are almost certainly explained by instrument and atmospheric effects. Lower pane: FORS $V/I$ spectra compared to the low-resolution MCSP spectrum of \citet{AngLan74}.}
\end{figure}
The magnetic field was discovered from continuum circular polarisation by \citet{LanAng71}. Observations of broad-band CP on six nights showed no variations. During one hour, broad-band data were printed every 6\,s. These data were Fourier analysed without finding any evidence of short-period variability. 

\citet{AngLan74} obtained an MCSP circular polarisation spectrum in 1972, and the star was re-observed by \citet{Voretal10}, who remarked that similarity with the previous spectra, reporting also "no unambiguous variations" between several polarisation spectra taken in 2003, 2005, 2008. Using FORS1 data obtained in 2005, they also reported no detection of linear polarisation.  

We retrieved from the ESO FORS1 and FORS2 archives all circular polarisation spectra obtained in 2005, 2008 and 2012. Figure~\ref{Fig_WD0548} shows all these spectra compared to the polarisation spectrum of \citet{AngLan74} . We note that there are three spectra obtained in three consecutive nights in November 2012. The spectrum obtained during the last night (2012-11-19) shows zero circular polarisation (see green solid line), but a Stokes~$I$ flux consistent with that of previous observations. We ascribe this to a non-detected instrument failure (perhaps the retarder waveplate did not move) rather than to stellar variability. 

Observations obtained by \citet{Beretal23} and \citet{Beretal24} show no variability in broadband polarimetry between two measurements obtained 8 months apart (see the top panel of Fig.~\ref{Fig_BBCP}).

This is a nearly unique DQp white dwarf which shows not only the Swan bands in the flux spectrum, but also the CH G band. The magnetic field was modelled by \citet{AngLan74} who deduced $\bz \approx 3.6$~MG from the strong polarisation signature of the CH G-band. \citet{Berdetal07} and \citet{Voretal10} modelled FORS1 spectra and found $\bz \simeq2.5$~MG.

\subsection{WD~0553+053 = EGGR 290 = G~99-47 (non-variable)}
The magnetic field of this star was discovered by \citet{AngLan72}, who also made 14 separate BBCP measurements during 10 months, without detecting any significant variation. Low resolution MCSP spectropolarimetry was obtained by \citet{Lieetal75}, and 20 years later, a higher resolution polarised spectrum was obtained also by \citet{PutJor95}. No significant variation has been detected on any timescale up to decades \citep[for details, see][]{BagLan21}. \citet{Brietal13} observed a light curve with 26.8\,m period and semi-amplitude = 0.3\,\%, which suggest that polarimetric observations could have missed short-term variability. However, \citet{AngLan72} had carried out Fourier-analysis of a polarimetric run, and found no significant variability for any period between 11 seconds and 2\,h. They claimed that periodic variation with an amplitude of 0.17\,\% would certainly have been detected. Our conclusion is therefore that the star is non-variable.

\subsection{WD~0637+478 = GD 77 (variable, $P \simeq 1.4$~d)}
Magnetic variability was reported by \citet{SchSmi95}. Eight unpublished ISIS polarisation spectra and one ESPaDOnS spectrum show that \bz\ varies with $P \simeq 1.362$~d. The value of $\bs \approx 1.0$\,MG hardly changes with rotation, but \bz\ varies sinusoidally between about $+400$ and $-250$\,kG (LB25).

\subsection{WD~0654+059 = 2MASS J06572938+0550479 (non-variable)}
Three BBCP observations obtained 8 months apart by \citet{Beretal24} show no variability of the polarisation in the three B$'$V$'$R$'$ filters (see Fig.~\ref{Fig_BBCP}). In all three filters, all three observations have $V/I \simeq -0.3$\,\%. 
We assume that the star is not variable.

\subsection{WD~0708--670 = SCR~J0708--6706 (n.v.: -- only 2 meas.)}%
Two spectra published by \citet{BagLan20} show no variability over a 2 month interval. The strength and wavelength dependence of the circular polarisation spectrum suggests that the underlying field could be of the order of 60 to 200\,MG. We tentatively assume that the star is magnetically non-variable. 

\subsection{WD\,0745$+$115 = GALEX J074842.4+112502 (non-variable)}
Strong broadband circular polarisation was detected by \citet{Beretal24}. The polarisation changes sign between the B$'$ and the other bands, and its absolute value is about 1\% in all bands. Observed four times, the star does not show variability (see the top right panel of Fig.~\ref{Fig_BBCP}).

\subsection{WD~0756+437 = EGGR 428 (variable, $\pphot \simeq 6.5$\,h )}  
The star was discovered to be magnetic and discussed in detail by \citet{Putney95}. She estimated its magnetic 
field to be $\simeq 200$~MG. Recent BBCP observations at NOT \citep{Beretal24} showed that it is rapidly variable. The first two observations, about 3\,h apart, report the largest and smallest polarisation seen in the full data set of five observations; this suggests a rotation period of the order of 6\,h. 
In fact, the photometric study of \citet{Brietal13} found a unique, very large amplitude ($\pm 4$\,\%) light variation with a period of $P=6.68$\,h.  
The similarity of this period with the period range deduced from polarisation measurements strongly confirms that 6.68\,h is the rotation period of this white dwarf. A further remarkable fact is the combination of high field (200\,MG), high mass ($1.04 M_\odot$) with advanced age (4.45\,Gyr): this star seems to be a unique example of a very old, still strongly magnetic and rapidly rotating WD-WD merger.

\subsection{WD~0810--353 = UPM J0812--3529 (non-variable)}
No variability detected among six polarised spectra obtained over a four year period, as described in a detailed study by \citet{Lanetal23}.
According to their modelling, the star shows two regions of different field strength: one with magnetic field of predominantly 30\,MG strength, outward, and one showing a field strength of 45\,MG, inward.

\subsection{WD~0816--310 = SCR J0818-3110 (variable, $P \simeq 10$\,d)}
 This is a DZ white dwarf with strong flux, spectrum and \bz\ variations observed in five FORS polarised spectra in 2023 obtained with grism 1200B. It is clear that the surface abundances vary over the surface, and appear to be locked to the surface magnetic field. The rotational period is of the order of 10~d, and $\bs \simeq 100$~kG \citep{Bagetal24a}. 

\subsection{WD~0850+192 = LB 8915 (variable, $P \sim$\,hours?)}
This DBA white dwarf shows very weak, variable H lines, slightly variable He, and variable \bz\ \citep{Wesetal01}. 
It is found that $\bs \simeq 850$~kG, and the rotation period is relatively short, probably some hours.  
We note that the correct identification of this star is LB 8915, and not LB 8827 as given in title of the paper by \citet{Wesetal01}.

\subsection{WD\,0907+213 = GALEX J091016.5+210555 (variable, $P \simeq 10$\,h)}
\citet{Mosetal24} discovered that this is a spectroscopically and magnetically variable DBA star with a rotation period of either 7.7 or 11.3\,h (the ambiguity is due to aliasing). They have modelled the field and abundance geometry with a simple model like that used for Feige 7 = WD\,0041--102. The line splitting in the flux spectra suggests a field of $\bs \simeq 0.5$\,MG.

\subsection{WD~0912+536 = EGGR 250 = G~195-19 (variable, $P \simeq 1.3$\,d)} 

This is the second magnetic white dwarf discovered \citep{AngLan71-Second}, and the first magnetic white dwarf to be discovered to be rotationally variable \citep{AngLan71-Periodic}. An improved ephemeris was provided by \citet{Angetal72-eph}.
The star shows a large variation of its circular polarisation with a period of 1.33~d \citep{Angetal72G}. 
\citet{Heretal24} measured the period of the photometric variability from TESS data as $1.3304 \pm 0.0054$\,d (note that high accuracy of period relies on two widely separated TESS observational data sets).
Six MCSP $V/I$ spectra roughly uniformly distributed in phase were obtained by Landstreet \& Angel, (unpublished). Between 4000 and 5000\,\AA, $V/I$ reverses sign during rotation; redwards of this, the strong variations retain one sign. 

\subsection{WD~1008--242 =  UCAC4~328-061594 (n.v.: -- only 2 meas.)}%
\begin{figure}
\begin{center}
\includegraphics[angle=0,width=8.8cm,trim={1.7cm 6.0cm 1.0cm 11.8cm},clip]{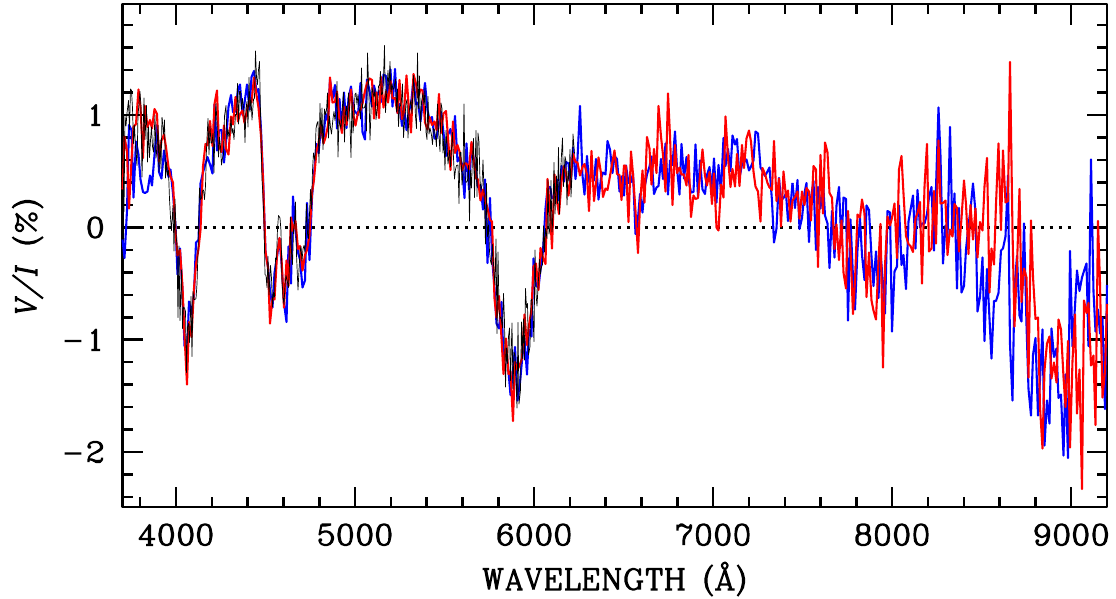}
\includegraphics[angle=0,width=8.8cm,trim={1.7cm 6.0cm 1.0cm 11.0cm},clip]{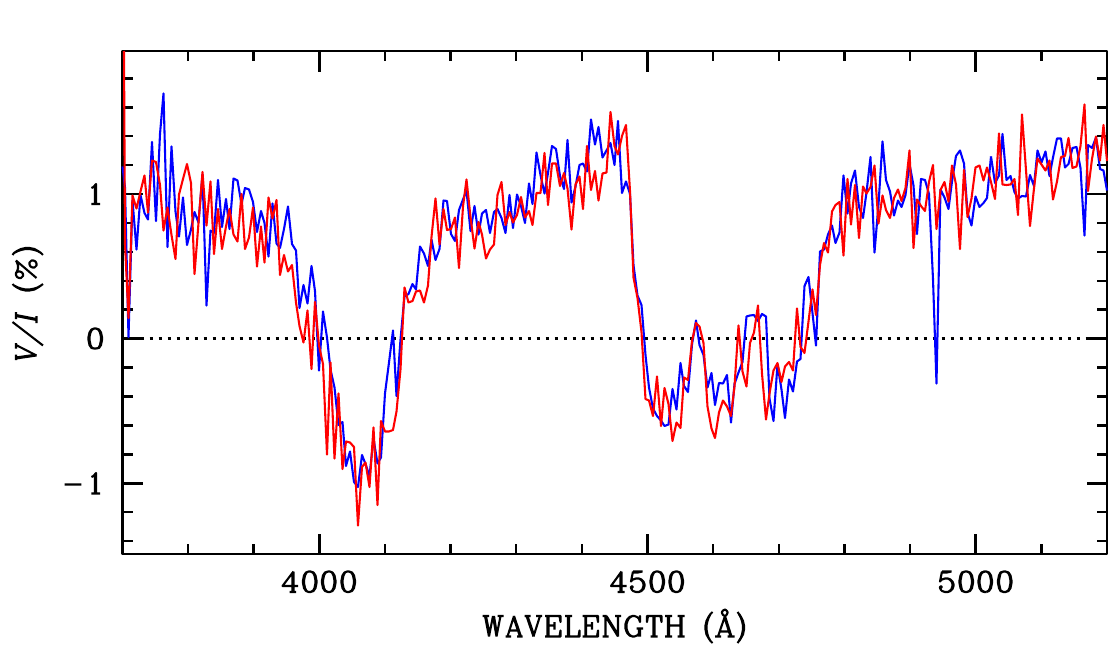}
\end{center}
\caption{\label{Fig_WD1008} Circular polarisation spectra of WD~1008$-$242 obtained with FORS2 and the 300V grism on 2022-02-10 (top panel) and with the 600B grism on 2022-01-01.}
\end{figure}
Observed twice with FORS2 in spectropolarimetric mode by \citet{BagLan22}, this white dwarf did not show any variation between two spectra obtained 40\,d apart, nor within the same observing series (see Fig.~\ref{Fig_WD1008}). Probably the field is of order 100~MG or more. We consider that it is probably not variable.
The star therefore appears a young, strongly magnetic, ultra-massive, non-variable star.

\subsection{WD~1008+290 =  LP 315-42 = LHS~2229 (non-variable)}%
This white dwarf is a cool peculiar DQ star discovered to be magnetic by \citet{Schetal99}, who tentatively suggested that the field strength is of the order of 100\,MG or more. Four observations obtained by \citet{Beretal24} show little to no variability of broadband circular polarisation (see Fig.~\ref{Fig_BBCP}).  We assume that it is not magnetically variable.

\subsection{WD~1009--184 = WT 1759 (var.:)} %
The star was discovered to be a  magnetic white dwarf by \citet{BagLan19b}, who published one measurement obtained with FORS2 and one obtained with ISIS, showing that the star has a weak and variable field ($\bz \simeq 50$\,kG). Additional unpublished data suggest also weak variability.

\subsection{WD~1015+014 = PG 1015+014 (variable, $P = 98.75$\,m)}
This is a very strongly polarised white dwarf with $V/I \simeq 1.5$~\%. The flux and polarisation spectra are strongly variable with $P = 98.75$\,m and a polar field strength of about 120\,MG \citep{Ang78,WicCro88}. \citet{SchNor91} report that BBCP also varies with P = 98.75~min, approximately sinusoidally, with extrema of $+1$ and $-1$\,\% polarisation. \citet{Brietal13} found that the star's light curve has a period consistent, within uncertainties, with that obtained from polarimetry.  
\citet{Eucetal06} obtained a series of polarised spectra with FORS using the 300V grism. They describe strong $I$ and $V$ spectrum variations, and model the field structure using a multipole field expansion. This white dwarf displays spectral features originating from regions with typical field strengths between about 50 and 90\,MG.

\subsection{WD~1031+234 =  Ton 527  = PG~1031+234 (variable, $P \simeq 3.5$\,h)}
\citet{Schetal86} report strongly variable intensity, circular and linear polarisation spectra with a 3\,h 24\,min period, and proposed a simple magnetic model with field strength in the range of 200 -- 500\,MG. Further broadband polarisation observations through the rotation period were obtained by \citet{PiiRei92}, who measured also a light variation in anti-phase with circular polarisation (that is, the star appears darker when the absolute value of the polarisation is maximum). \citet{Brietal13} found $\pphot \simeq 3.5$\,h. 

\subsection{WD~1036--204 = LP~790-29 (non-variable)}%
\begin{figure}
\begin{center}
\includegraphics[angle=0,width=8.8cm,trim={1.3cm 6.0cm 1.0cm  3.0cm},clip]{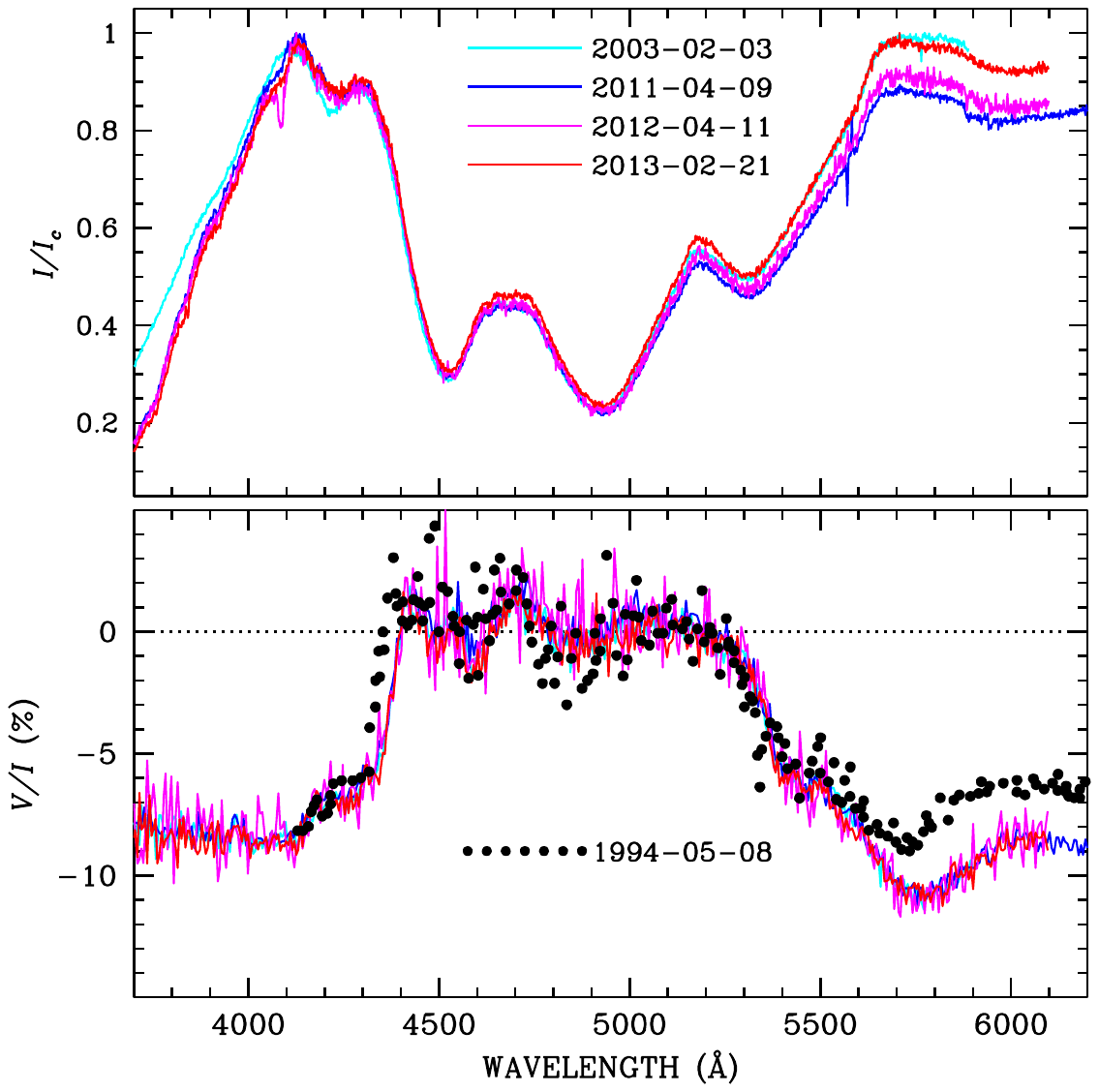}\\
\includegraphics[angle=0,width=8.8cm,trim={1.3cm 6.0cm 1.0cm 11.0cm},clip]{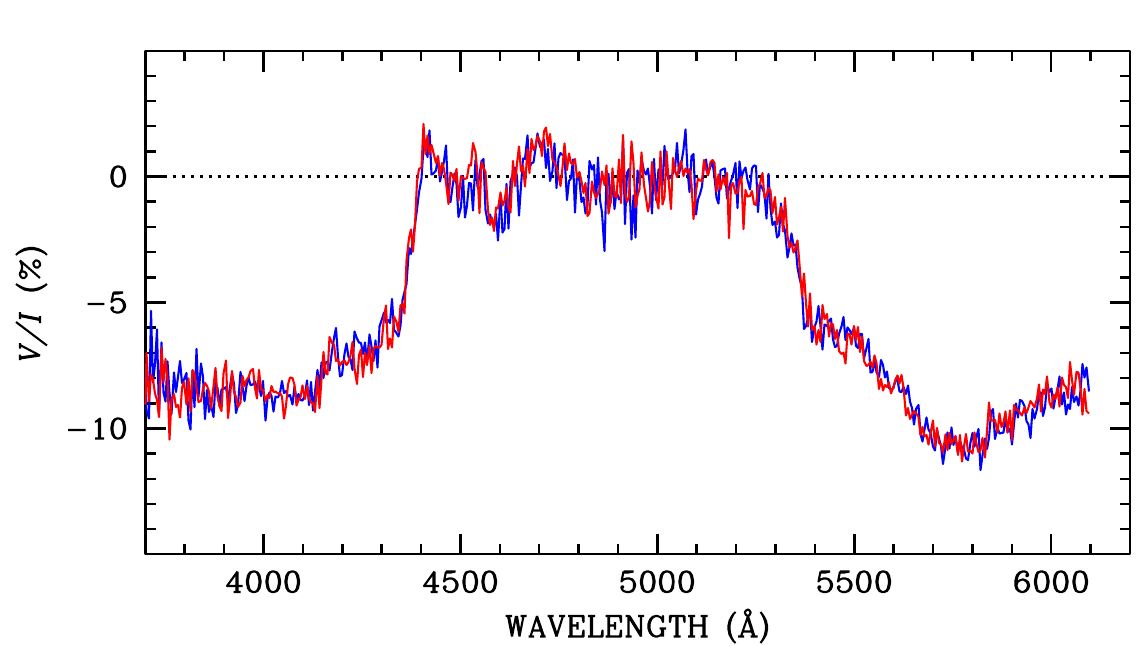}
\end{center}
\caption{\label{Fig_WD1036} {Top and middle panels:} Intensity and circular polarisation spectra of WD~1036$-$204 obtained with FORS1 and FORS2 at four different epochs; circular polarisation from \citet{Schetal95} is also shown with small circles. Bottom panel: Stokes $V/I$ from four pairs of observations obtained within 10\,m from each other on 2013-02-21.}
\end{figure}
The star was discovered to be magnetic via spectropolarimetry by \citet{Lieetal78}, and repeatedly studied \citep{Wes89,Schetal95,Schetal99,BeuRei02,JorFri02}.  \citet{BeuRei02} have monitored the star with EFOSC in spectropolarimetric mode to search for short-term variability, without finding any significant variations. \citet{BeuRei02}, however, pointed out that \citet{Schetal95} measured a polarisation signal of $\simeq -6$\,\% around 6500\,\AA, a measurement at odds with broadband polarimetric measurements obtained in 1977, 1986, 1994 and 2000, which were all about $-9$\,\% (see their Table~1; we recall here that we use the opposite definition for the sign of circular polarisation). \citet{Schetal95} therefore suggested the possibility of a very long rotation period of the star. \citet{JorFri02} carried out a similar study, with similar results regarding short term variation. They also proposed a rotation period in the range of about 24 -- 29 yr.
We have reduced FORS1 and FORS2 archival polarisation spectra from 2003, 2011, 2012, and 2013, {finding absolutely no hint of variability among them}. Figure~\ref{Fig_WD1036} shows a comparison of archive observations obtained with grism 600B. The flux is not corrected for atmospheric and instrument transmission, and the discrepancies in the slope of the flux measured in 2003 can be explained by the use of a different CCD. A comparison between FORS spectra with those obtained by \citet{Schetal95} on May 7 and 8, 1994, (wavelength range 4160--7460\,\AA) does not show obvious long-term variability, except in the range $5700-6200$\,\AA. In that range, our data are instead consistent with most of the literature and point to a value of $\simeq -9$\,\%. In addition to the polarisation spectra, FORS1 and FORS2 archive contains another two broadband circular polarisation (BBCP) observations in the R filter: a FORS1 BBCP measurement obtained in April 2006, and one obtained with FORS2 in March 2024, using a similar (but not perfectly identical) R filter (program ID 112.25C9.001). We found both measurements consistent with a polarisation signal of about $-9.4$\,\%. These measurements rule out any significant change in the region around 6500\,\AA\ over an interval of 18 years. Finally, \citet{Beretal24} have published the series of BBCP observations obtained in November 2022, November 2023 and February 2024, all consistent among themselves. In the R$'$ filter they report a polarisation signal of $\simeq -9.1$\,\%. For the reasons explained at the beginning of this section, it is not possible to accurately compare BBCP measurements obtained with different instruments, but it is clear that the only deviant point of a series of polarimetric observations obtained in the course of almost half a century is a small portion of a spectrum obtained in 1994. Remarkably, data obtained with the same instrument (FORS) over nearly two decades are fully consistent among themselves, and point strongly to a constant circular polarisation spectrum. Our conclusion is that the star is actually not variable. From the measured signal of circular polarisation of $\simeq 10$\,\% we estimate a longitudinal field of the order of 100\,MG. 

\subsection{WD~1043--050 = HE 1043-0502 (n.v.: -- only 2 meas.)} 
This DBA star was discovered magnetic by \citet{Schetal01}, who proposed that the field is $\simeq 800$~MG (although its continuum is not highly polarised). Two observations were taken a few days apart, and \citet{Schetal01} state that they did not show variability; the reported wavelength integrated circular polarisation of about 1.5\% is also essentially unchanged between the two spectra. So we consider it as candidate non-variable magnetic white dwarf.

\subsection{WD~1045--091 = HE~1045--0908 (variable, $P \simeq 3$\,h)}
This white dwarf was discovered to be magnetic by \citet{Reietal96} from a flux spectrum. Circular polarisation was confirmed, and shown to be variable by \citet{Schetal01}. 
\citet{Eucetal05} obtained a series of $I$ and $V$ spectra and, assuming a rotational period of about 2.7\,h, derived a detailed surface field model with a dominant field strength of 16\,MG, but local field strength ranging between about 10 and 75\,MG.

\subsection{WD~1105--340 = SCR J1107-3420A (non-variable)}
\begin{figure}
\begin{center}
\includegraphics[angle=0,width=8.8cm,trim={0.7cm 6.0cm 1.1cm 3.0cm},clip]{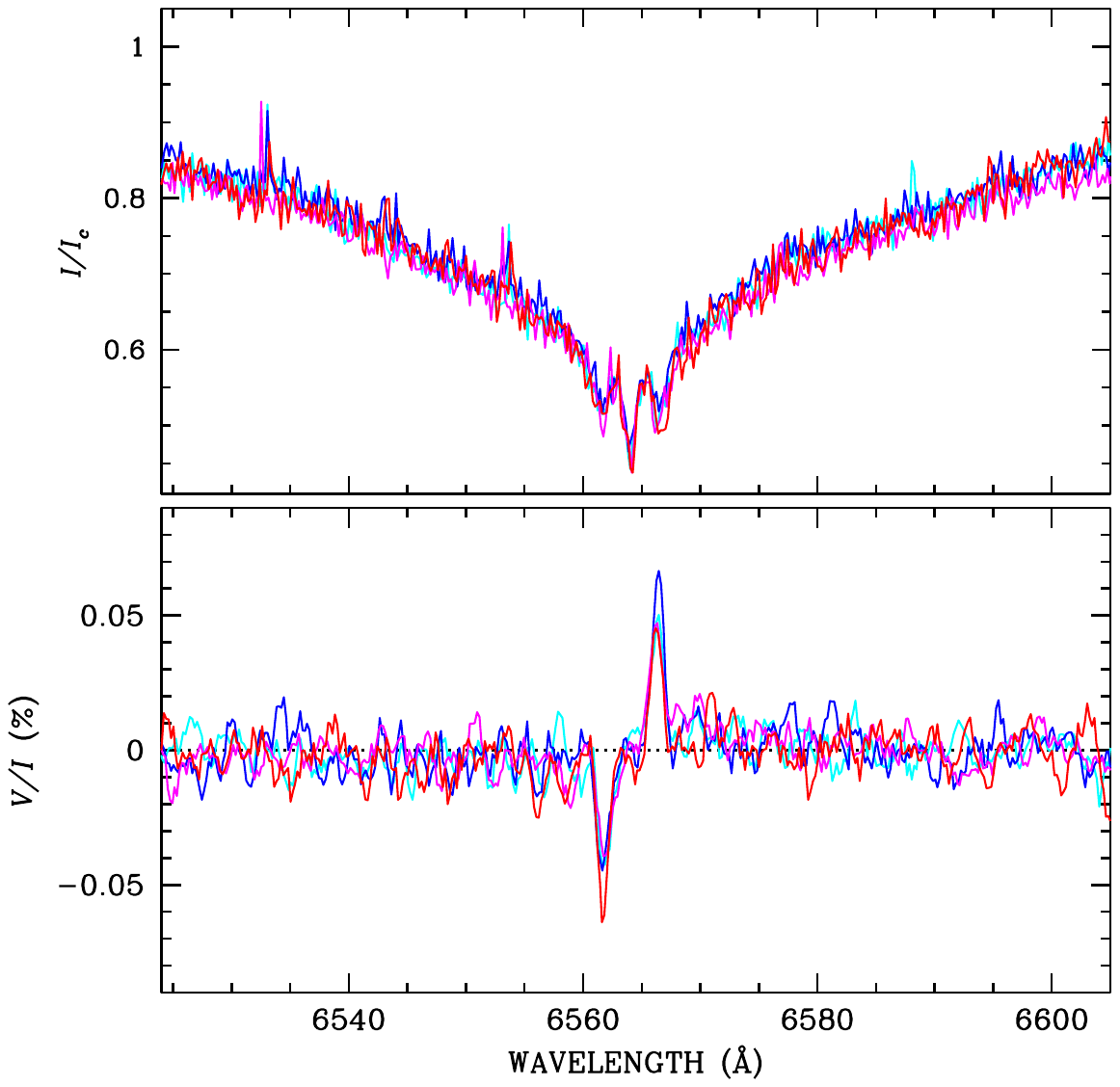}\\
\end{center}
\caption{\label{Fig_WD1105-340}  WD~1105$-$340: Stokes $I$ and $V/I$ profiles of H$\alpha$ observed with ESPaDONs at four different epochs: 2019-03-21 (two observations), 2019-05-31, and 2022-02-22. Only negligible changes are seen between spectra.}
\end{figure}

Eleven ESPaDOnS spectra taken between 2018 and 2022, with ten of them during 1 week in 2019 (see Table~\ref{Tab_Log}), show that the star has a weak, non-variable field with $\bz \approx -22 \pm 4.5$\.kG and $\bs \approx 125 \pm 5$\,kG (see Fig.~\ref{Fig_WD1105-340}). We also have two FORS spectra, one taken with 1200B and one with 1200R, with lower resolving power but higher signal-to-noise ratio (S/N). Owing to their difference resolution, these spectra were not used in this work. This star is perhaps the best studied non-variable weak-field magnetic white dwarf.

\subsection{WD~1105--048 = EGGR 76 (ultra-weak field, var.:) }
The star was repeatedly observed, and a longitudinal field of the order of 1\,kG was detected only in two measurements \citep{BagLan18}. The star seems to have a very weak and variable field, but additional measurements should be obtained to confirm the existence of a magnetic field. 

\subsection{WD~1116--470 = SCR~J1118--4721 (non-variable)}
\begin{figure}
\begin{center}
\includegraphics[angle=0,width=8.8cm,trim={1.3cm 6.0cm 1.0cm 11.2cm},clip]{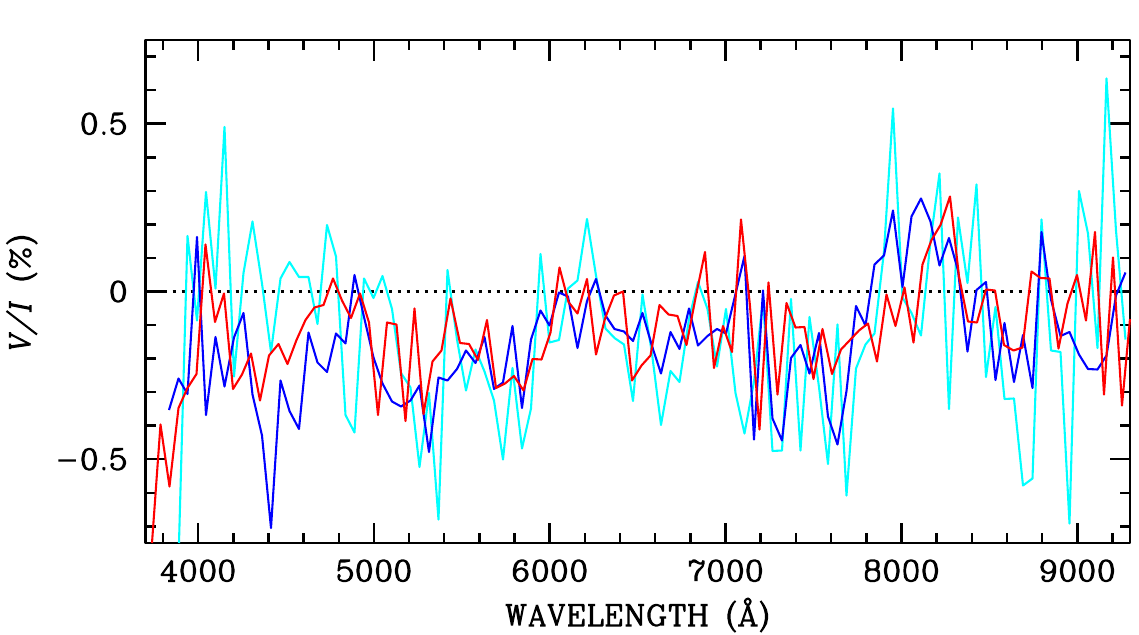}
\includegraphics[angle=0,width=8.8cm,trim={1.3cm 6.0cm 1.0cm 11.0cm},clip]{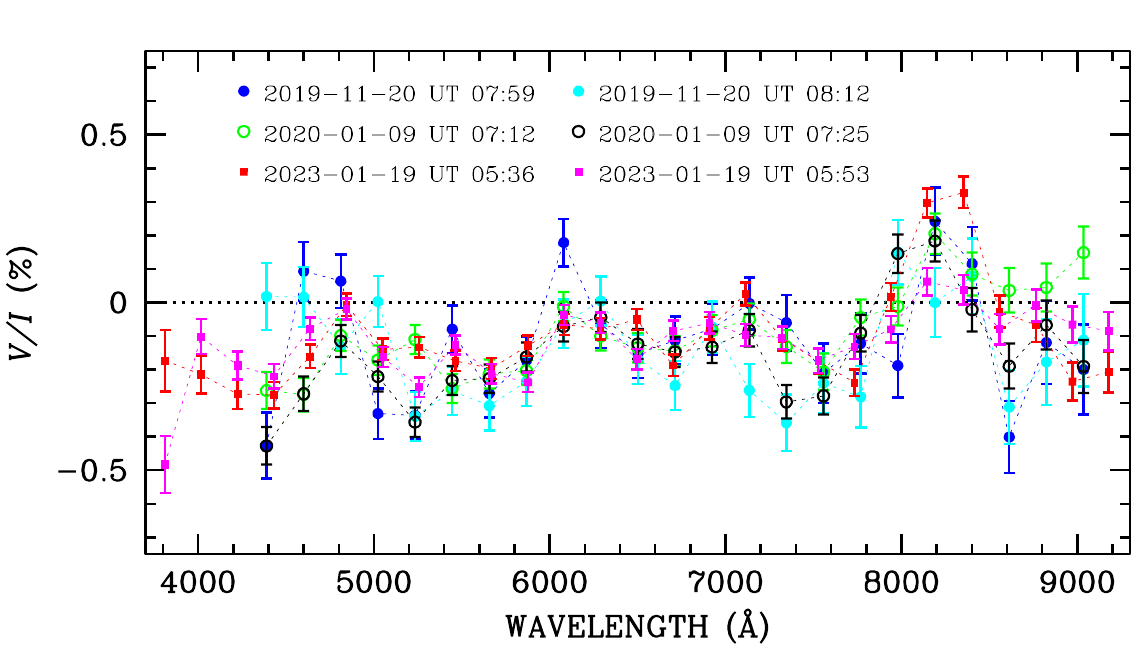}
\end{center}
\caption{\label{Fig_WD1116} {Top panel:} WD\,1116$-$470 observed with FORS2 in three different epochs. {Bottom panel:} Stokes $V/I$ (re-binned at 410\,\AA) from individual pairs of exposures as shown in the legend.}
\end{figure}
This white dwarf was observed twice with FORS2 by \citet{BagLan21}, who flagged it as suspected magnetic star. Both observations show a similar signal of circular polarisation at $-0.2$~\%, close to the FORS2 instrumental detection limit. A third observations was obtained in January 2023, and the $V/I$ spectrum is again consistent with that measured previously. The star is definitely magnetic, and most likely not variable. This confirmation (see Table~\ref{Tab_Log}) brings the number of magnetic white dwarfs in the 20\,pc volume to 34.

\subsection{WD~1211--171 = HE~1211--1707 (variable, $P \simeq 2$\,h)}
A magnetic field was suspected in this white dwarf by \citet{Reietal96}, which was confirmed with polarimetry by \citet{Schetal01}. Both papers show varying flux spectra. \citet{Schetal01} estimates $P \simeq 100 - 120$ min and $\bs \simeq 50$~MG. \citet{Brietal13} measured a photometric period of 1.79\,h, consistent with the previous estimates from polarimetric data. The star is polarised at a level that varies between 0 and 3\% during the rotation cycle. Modelling by \citet{Schetal01} strongly suggests a He dominated atmosphere with $\te \simeq 12000$~K, but \citet{Reietal96}, using IUE data, estimates 23000~K. \citet{Genetal21} gives 30000~K and $1.20 M_\odot$. We tentatively assume $\teff = 23000$\,K , $M = 1.2 M_\odot$, and an He-rich atmosphere. 

\subsection{WD~1217+475 = SDSS J121929.45+471522.8 (DAHe variable, $\pphot=15.26$\,h)} 
DAHe with 18.5\,MG field strength and a photometric period of about 15.25\,h \citep{Ganetal20}. Spectroscopy reveals Zeeman components of the Balmer lines varying in strength but with constant splitting. Published data do not demonstrate that the star is magnetically variable but cannot rule out this possibility either. Therefore we have decided not to include this star in our sample.

\subsection{WD~1249--022 = GALEX J125230.9-023417 (DAHe variable, $\pphot=0.09$\,h)}
This is a DAHe white dwarf that shows variable H$\beta$ and H$\alpha$ lines that appear sometimes in emission and sometimes in absorption. Field strength has been estimated 5\,MG and rotational period = 0.09\,h \citep{Redetal20}.

\subsection{WD~1312+098 = PG~1312+099 -- (variable, $P \simeq 5.4$\,h)} %
Variable continuum circular polarisation was detected in this hot DAH by \citet{SchNor91}, who present over 100 measures, and find $P = 5.43$\,h. 
Circular polarisation varies approximately between +1\% and --1\%. 

\subsection{WD~1315--781 = LAWD 45 (n.v.:  -- only 2 meas.)}
No change between two $\bs\ \simeq 5.5$\,MG and $\bz\ \simeq 0$\,MG  measurements from FORS 300V spectra taken five nights apart by \citet{BagLan20}. We tentatively assume that it is magnetically non-variable. 

\subsection{WD~1315+222 = LP 378-956 (non-variable)}
Two observations by \citet{Beretal23} and one by \citet{Beretal24} show no obvious sign of variability (see Fig~\ref{Fig_BBCP}). %

\subsection{WD~1328+307 = G~165--7 (variable)} 
This star is found to host a magnetic field of $\bs \simeq 650$\,kG based on line splitting observed in a good S/N SDSS spectrum \citep{Dufetal06}. These authors have also obtained low-resolution polarised spectra. They 
state that three 600 sec polarisation spectra were taken on 2005-12-30 at Steward Obs, all yielding essentially the same $\bz \approx 150$\,kG, but  that the polarisation amplitude in similar Steward polarised spectra from 2006-05-03 (apparently not measured)  is at least two  times  weaker than in 2005 spectrum. We conclude that the star is magnetically variable. 


\subsection{WD~1346+121 =  LP 498-66 (non-variable)}
Observed three times by \citet{Beretal23} and \citet{Beretal24}, we assume the star, which shows circular polarisation ranging between --1 and 0\,\%\ in the three DIPol-UF bands, is magnetically non-variable (see Fig.~\ref{Fig_BBCP}). The field strength of this white dwarf is probably in the tens of MG.

\subsection{WD~1350--090 = PG 1350-090 = GJ 3814 (non-variable)}%
Discovered by \citet{SchSmi94}, who measured $\bz = 85 \pm 9$~kG. We have obtained four polarised spectra with ESPaDOnS that show \bs\ $\simeq 450 - 465$~kG. There is no strong evidence of field variability, see also  \citet{SchSmi94}. Data will be published in a forthcoming paper (LB25). 

\subsection{WD~1532+129 = G~137-24 (variable)}%
Originally classified as DZ white dwarf by \citet{Kawetal04}, the star was discovered to be a magnetic white dwarf by \citet{BagLan19b}, who published two FORS2 measurements and one ISIS measurement. The star is variable, with \bz\ values from  FORS2 measurements of $ -21 \pm 1$~kG and $-4 \pm 1$~kG, while \bs\ is not strong enough to split spectral lines, leading to  $\bs \la 300$~kG.  
\begin{figure}
\begin{center}
\includegraphics[angle=0,width=8.8cm,trim={0.8cm 1.5cm 1.0cm 1.0cm},clip]{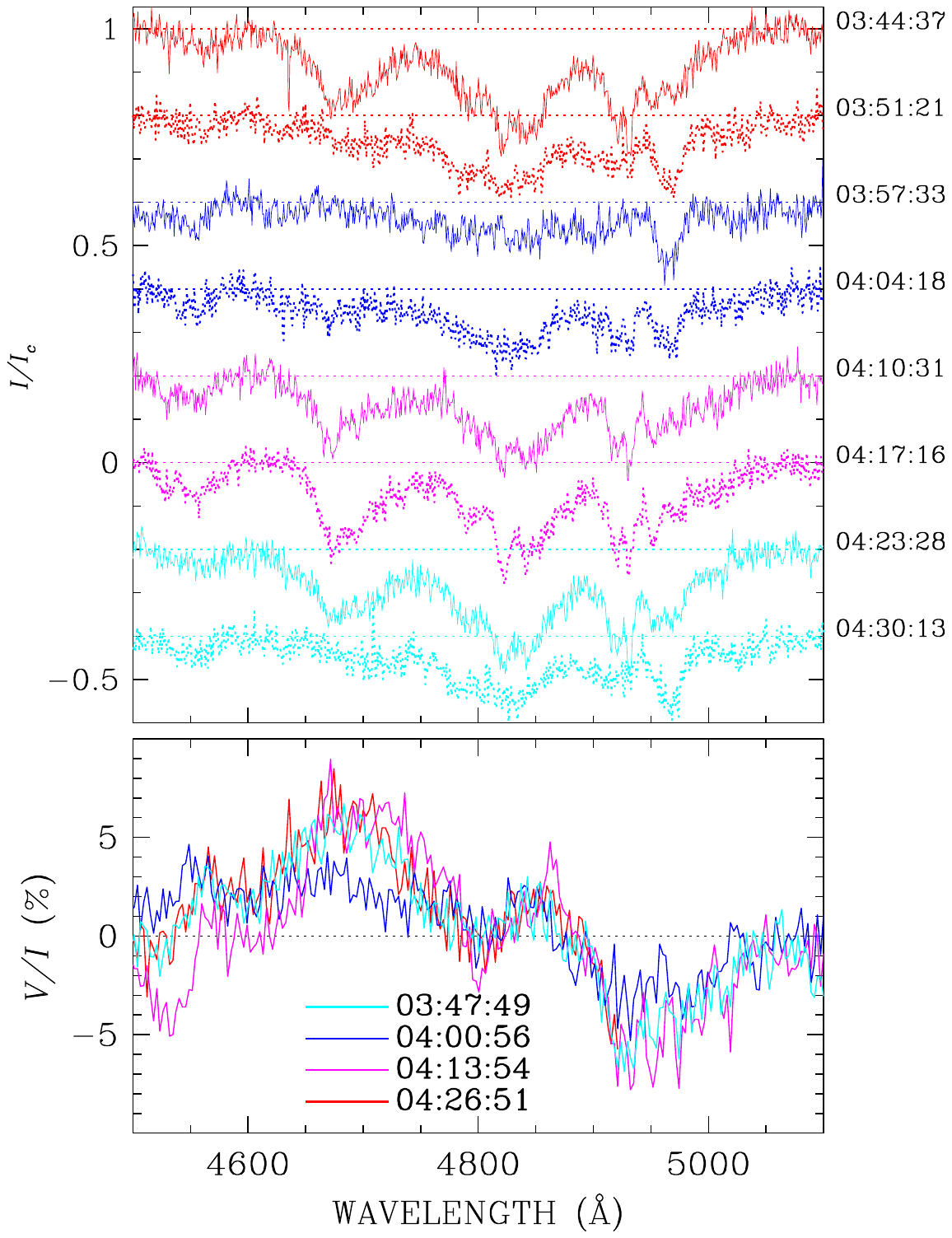}\
\end{center}
\caption{\label{Fig_WD1619} Spectra of WD\,1619$+$046 around H$\beta$ obtained with FORS2 on 2023-06-16.
}
\end{figure}
\begin{figure}
\begin{center}
\includegraphics[angle=0,width=8.8cm,trim={1.3cm 6.0cm 1.0cm 3.0cm},clip]{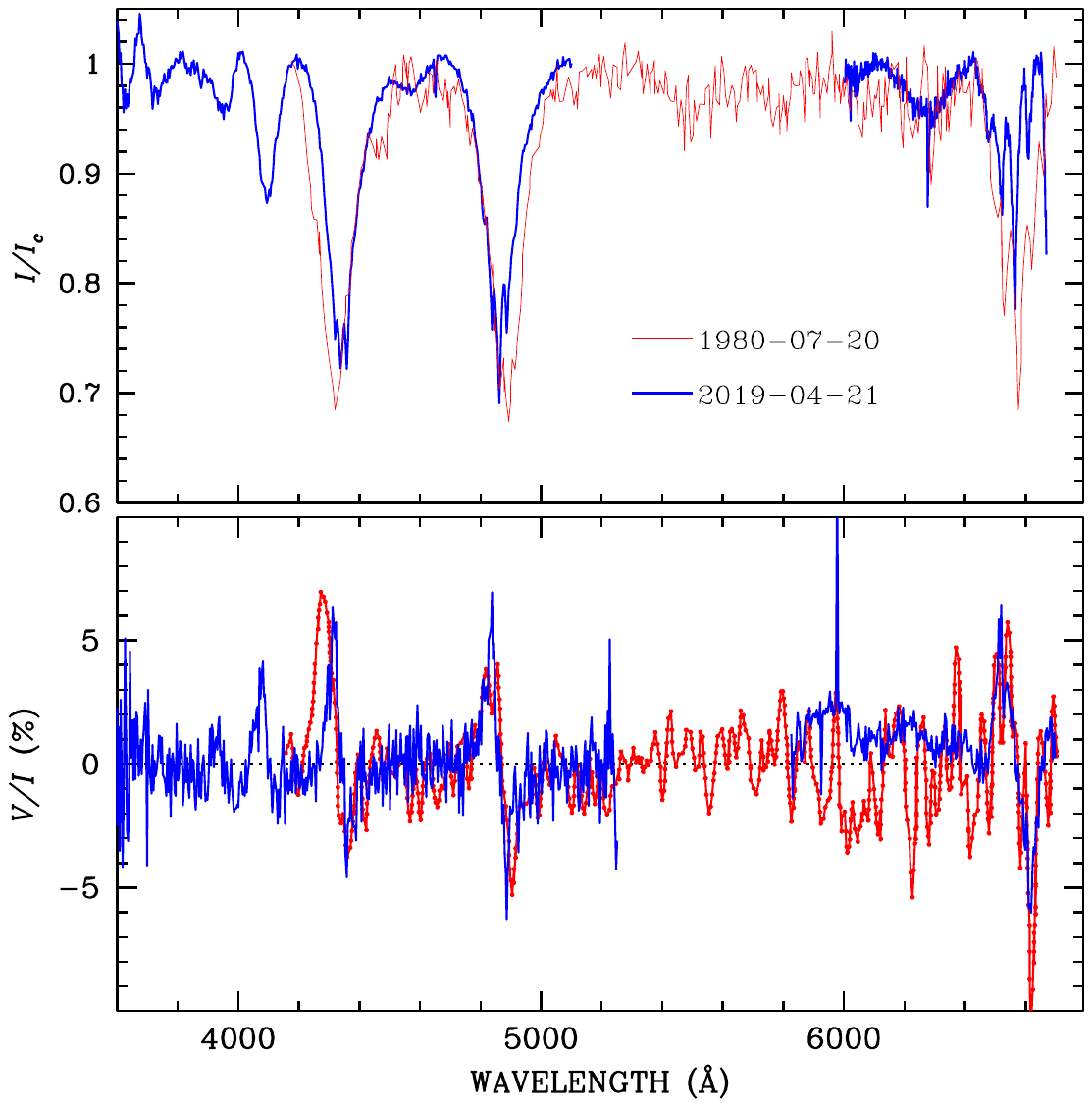}\\
\includegraphics[angle=0,width=8.8cm,trim={1.3cm 6.0cm 1.0cm 1.0cm},clip]{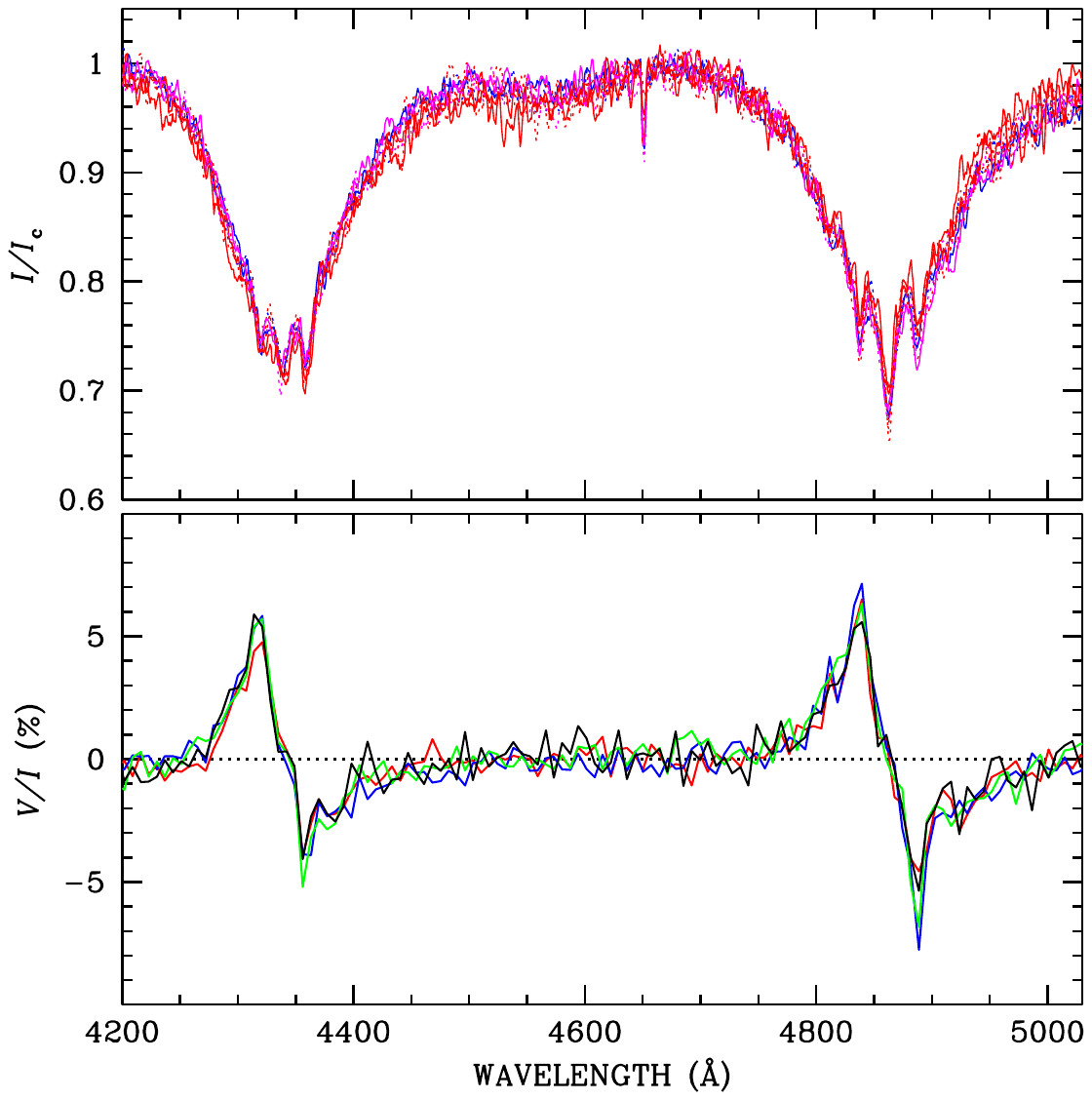}
\end{center}
\caption{\label{Fig_WD1658} {Top panels:} Comparison of spectra of WD~1658+440 obtained in 1980 
\citep{Lieetal83} and in 2019 (this work). {Bottom panels:} Overplot of eight intensity spectra obtained in sequence
with 450s exposure time, and of four circular polarisation spectra obtained from pairs of frames 
obtained in sequence.}
\end{figure}

\subsection{WD~1556+044 =  PM J15589+0417 (non-variable)}
Discovered to be magnetic by \citet{Beretal22}, the star was re-observed three more times by \citet{Beretal24}. The observed circular polarisation, which is detected at the $10 \sigma$ level, ranges between $-0.3$ and $+0.4$\%\ in the three filter bands of DIPol-UF, so the order of magnitude of the field strength is probably some tens of MG. The polarisation does not show any variability (see Fig.~\ref{Fig_BBCP}).

\subsection{WD~1615+542 = GALEX J161634.4+541011 (DAHe magnetically variable, $\pphot=1.59$\,h)} 
DAHe with a variable magnetic field (from 3.5 to 6.5\,MG) and a rotation period $P = 95.29$\,m estimate via photometry \citep{Manetal23}.

\subsection{WD~1619+046 = GALEX J162157.7+043219 (variable, $P \simeq 40$\,m?)}
This white dwarf was discovered to be magnetic and rapidly variable in this work (see Sect.~\ref{Sect_New_Observations} and Fig.~\ref{Fig_WD1619}). 
From the H$\beta$ regions, it is clear that the star is a DAH with a rather non-uniform field of $\bs \simeq 15$\,MG.
We observe clear changes in intensity between spectra obtained a few minutes apart, and also polarisation (which is measured
by the combination of two spectra) shows some variation, but without sign reversal. The close similarity between the first pair and the last pair of intensity spectra suggest that the star's rotation period is approximately 40\,min.

\subsection{WD~1639+537 =  GD\,356 (non-variable, but with a light curve and \pphot = 1.93\,h)}
This star is the prototype of DAHe white dwarfs.  It shows a light curve with low amplitude and $\pphot \simeq 1.93$\,h \citep{Waletal21}. A time series of circular polarisation spectra obtained during an entire rotational cycle shows now variability while subtle sinusoidal variability is seen in the position of the $\sigma$ components of H\,$\beta$ and H$\alpha$ lines \citep{Waletal21}. \citet{Waletal21} compared spectropolarimetry obtained in 2019, with that published by \citet{Feretal97}, finding no changes. Its field strength is of the order of 10\,MG \citep{Waletal21}. We classify the star as non-variable, but showing very small changes of its apparent magnetic field with a $\sim 2$\,h rotation period.

\subsection{WD~1658+440 = PG~1658+441  (n.v.: -- only 2 meas., one very old)} 
The star was discovered to be magnetic by \citet{Lieetal83}, who measured $\bz \simeq 0.7$\,MG, and $\bs \simeq 2.3$\,MG
from spectropolarimetry of $H\alpha$, H$\beta$ and H$\gamma$ (see their Fig.~5). Their observations were obtained on July 20 1980, with a spectral resolution of 10\,\AA. The same authors measured a signal of broadband circular polarisation = $-0.016 \pm 0.033$\% and $-0.044\pm 0.019$\,\% in the range 3300-8600\,\AA\ and concluded that a $3\,\sigma$ upper limit of 0.10\% semi-amplitude could be set for any presumed sinusoidal variation with a period between 0.5 and 5\,h. For possible periods between 4 minutes and 0.5 hours, a less stringent limit of 0.30\% polarisation semi-amplitude may be deduced. A comparison with our ISIS spectra obtained on April 21, 2019 (Sect.~\ref{Sect_New_Observations}) is shown in the top panels of Fig.~\ref{Fig_WD1658}. We observe a wavelength shift possibly due to bad calibration; most remarkable is the different strength of H$\alpha$, but perhaps this is instrumental, because the ISIS spectrum was obtained with a setting that puts H$\alpha$ at the edge of the CCD, and no good flat-fielding correction could be applied. Although our comparison between 1980 and 2019 data is not conclusive, we certainly do not see any convincing change of the spectral features that demonstrate long-term variability. A comparison between the four pairs of Stokes $V/I$ profiles obtained with ISIS in 2019 show no differences within the error bars (see the bottom panels of Fig.~\ref{Fig_WD1658}); therefore we rule out variability on a timescale of 10-15 minutes. This star seems to be a young, strongly magnetic, massive, tentatively classified as non-variable star. Photometric studies lead to inconsistent results: \citet{Brietal13} found that the star is photometrically variable with a period between 6\,h and 4\,d, while, using TESS photometry, \citet{Olietal24} derived a period shorter than 1\,h. \citet{Heretal24}, instead, did not detect periodicity in TESS data.

\subsection{WD~1703--266 =  UCAC4 317-104829 (variable) }
This DA white dwarf was discovered to have a magnetic field by \citet{BagLan20}, with $\bs \approx  8$\,MG.  They also found that two FORS2 polarised spectra four days apart are significantly different and yield different values of \bz, so the star is variable on a timescale of some days or less. 

\subsection{WD~1712--590 =  Gaia DR3 5915797694789556096 (variable)}\label{Sect_WD1712}
\begin{figure}
\begin{center}
\includegraphics[angle=0,width=8.8cm,trim={1.7cm 6.0cm 1.0cm 3.0cm},clip]{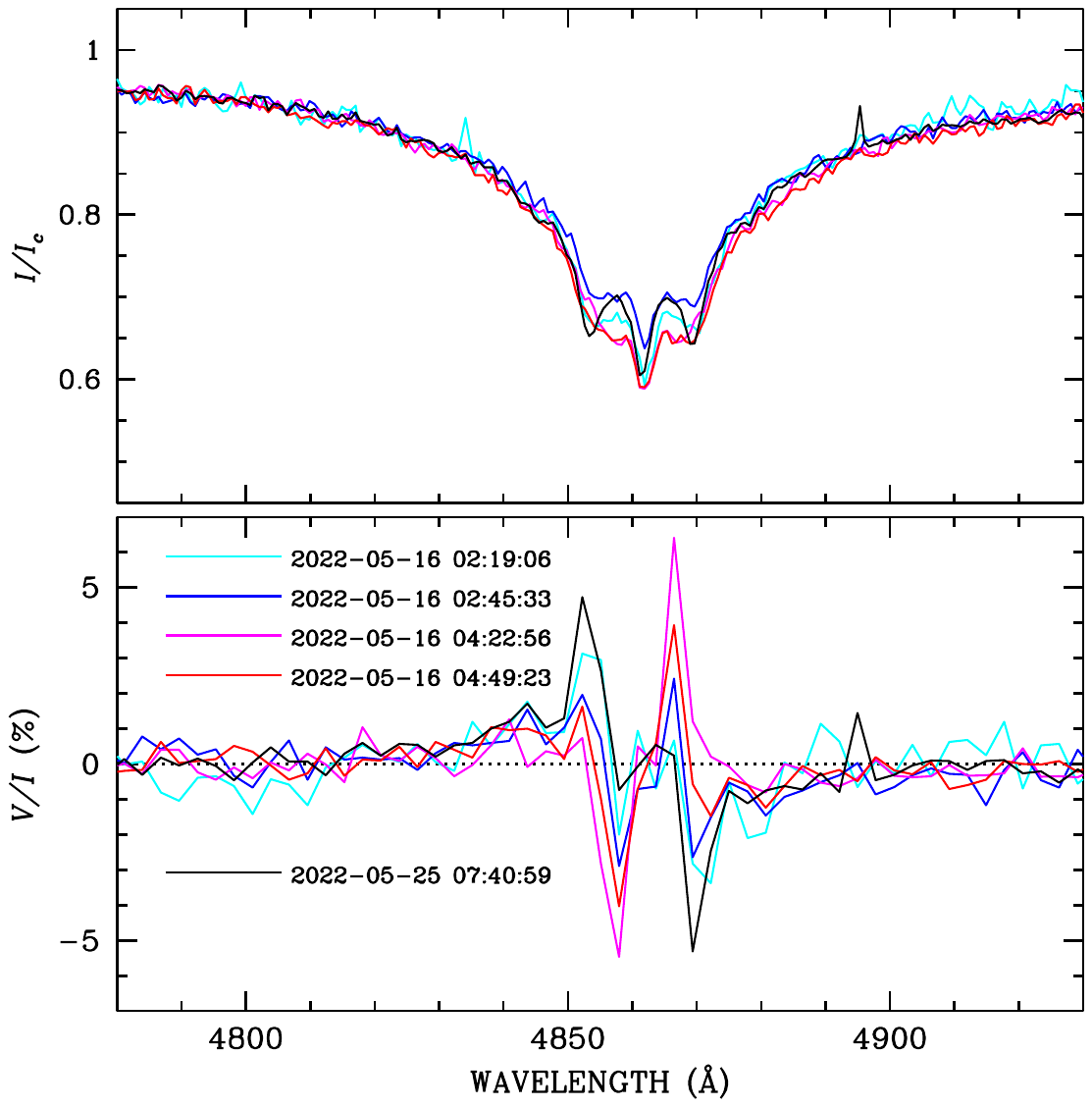}\\
\end{center}
\caption{\label{Fig_WD1712} WD~1712$-$590 observed with FORS2 in the dates shown in the bottom panel. H$\beta$ detail.}
\end{figure}
Discovered to be magnetic by \citet{OBretal23}.
We observed this white dwarf in polarimetric mode with grism 1200B three times, twice during the same night. Intensity spectra show very similar splitting of Balmer lines, but the flux distribution within the line cores changes quite significantly, rather similarly to WD\,2359--434 \citep{Lanetal17}. The deduced field strength is $\bs \la 0.8$\,MG. The Stokes $V$ spectra also show that the star is strongly variable within 1-2\,h (see Fig.~\ref{Fig_WD1712}), and that the longitudinal field reverses its polarity during rotation. 

\subsection{WD~1743--521 = L 270-31  = BPM\,25114 (variable, $P=2.84$\,d)}
\citet{WicBes76} reported a magnetic field of about 35\,MG in this southern DA star from close examination of the peculiar flux spectrum. \citet{Weg77} found variability of light and of the flux and polarisation spectrum with $P \approx 2.84$\,d, and modelled the spectrum variations, finding results consistent with a magnetic dipole field. Field modelling was carried out in more detail by \citet{MarWic78}, who deduced a dipole field strength of 36\,MG, corresponding to $\bs \simeq 18$\,MG. 

\subsection{WD~1748+708 = EGGR 372 = G~240-72 (s.v.:) }%
\begin{figure}
\begin{center}
\includegraphics[angle=0,width=9.0cm,trim={1.2cm 2.3cm 1.0cm 1.0cm},clip]{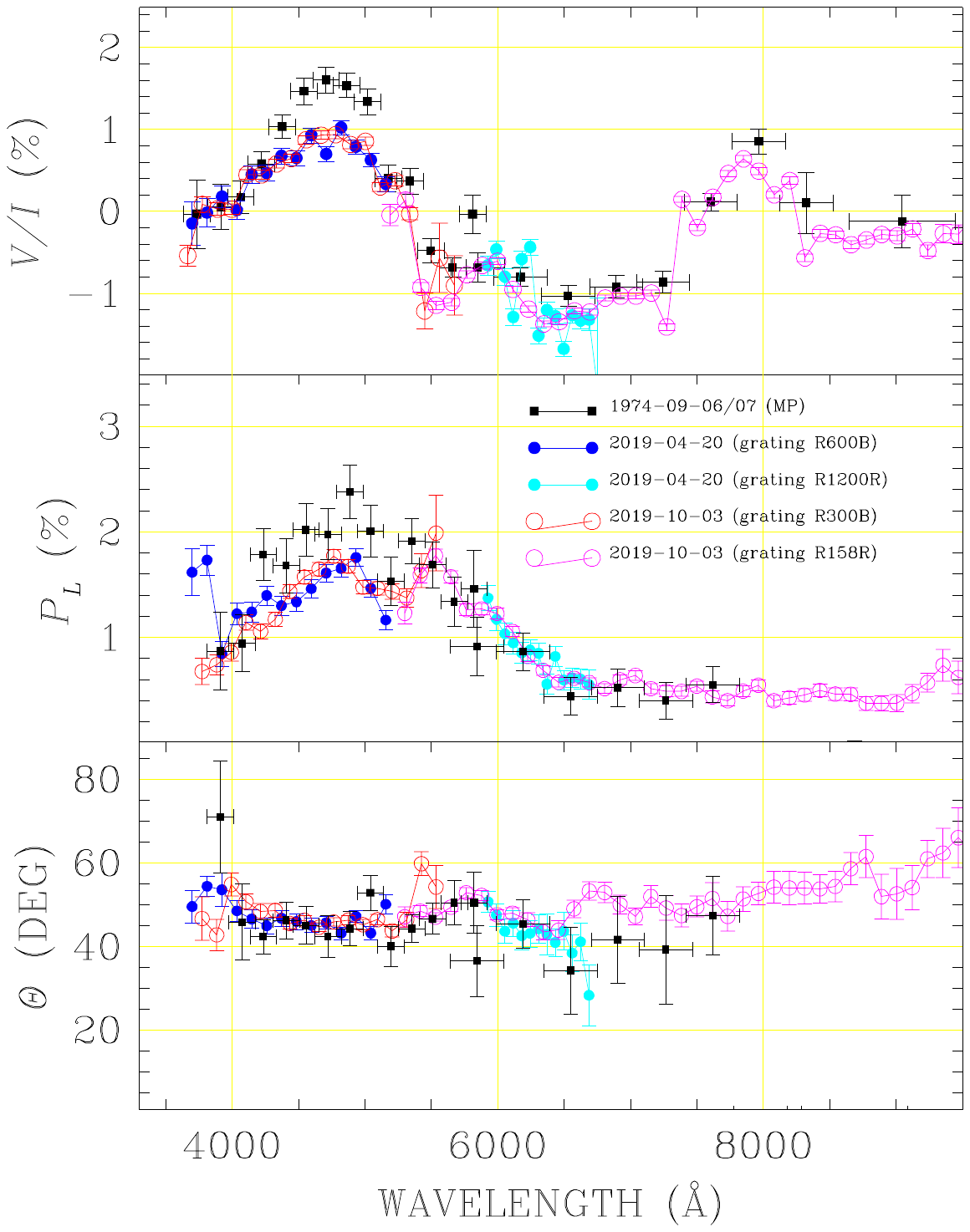}
\end{center}
\caption{\label{Fig_WD1748} Polarisation spectra of WD~1748+708 = G240-272.}
\end{figure}
Intrinsic broadband circular and linear polarisation of this magnetic white dwarf were discovered by \citet{Angetal74b}, who reported no variations in repeated observations of broad-band circular polarisation over a month, and linear polarisation over two nights.  The star was later observed in broadband circular and linear polarimetry by \citet{West89}, who generally found similar polarisation levels and position angles to earlier work, and in broadband linear polarimetry by \citet{BerPii99}. \citet{BerPii99} found evidence of clear change of the position angle of linear polarisation (mainly rotation of the polarisation angle) on a timescale of 20 years. We have observed the star with ISIS both in circular (twice) and in linear polarisation, and compared these observations with MCSP low-resolution spectra of $I$, $V$ and $P$ obtained by Angel \& Landstreet on September 6 and 7, 1974, that were never published until now. Our ISIS spectra are consistent among themselves. When compared with spectropolarimetry obtained in 1974 and the 1990s, our new ISIS linear polarisation data find small changes in the $V$ and $P$ polarisation amplitude in the blue and visual, whereas the overall position angle behaviour is almost identical to that observed observed in 1974, thus we cannot confirm the changes in the polarisation position angle seen in 1997 by \citet{BerPii99}. On the other side, \citet{Antetal16} found that the star is photometrically variable with a period between 5 hours and two days. \citet{Brietal13} detected no short period variability, but claim that photometry varied over a ten month interval. We conclude that the star show some signs of subtle long term variability that should be further investigated.

\subsection{WD\,1750$-$311 = UCAC4 295-140552 = [MTR2015] OW J175358.85-310728.9 (n.v.: -- but $\pphot \simeq 0.5$\,h)}\label{Sect_WD1750}
\begin{figure}
\begin{center}
\includegraphics[angle=0,width=8.8cm,trim={1.7cm 6.0cm 1.0cm 3.0cm},clip]{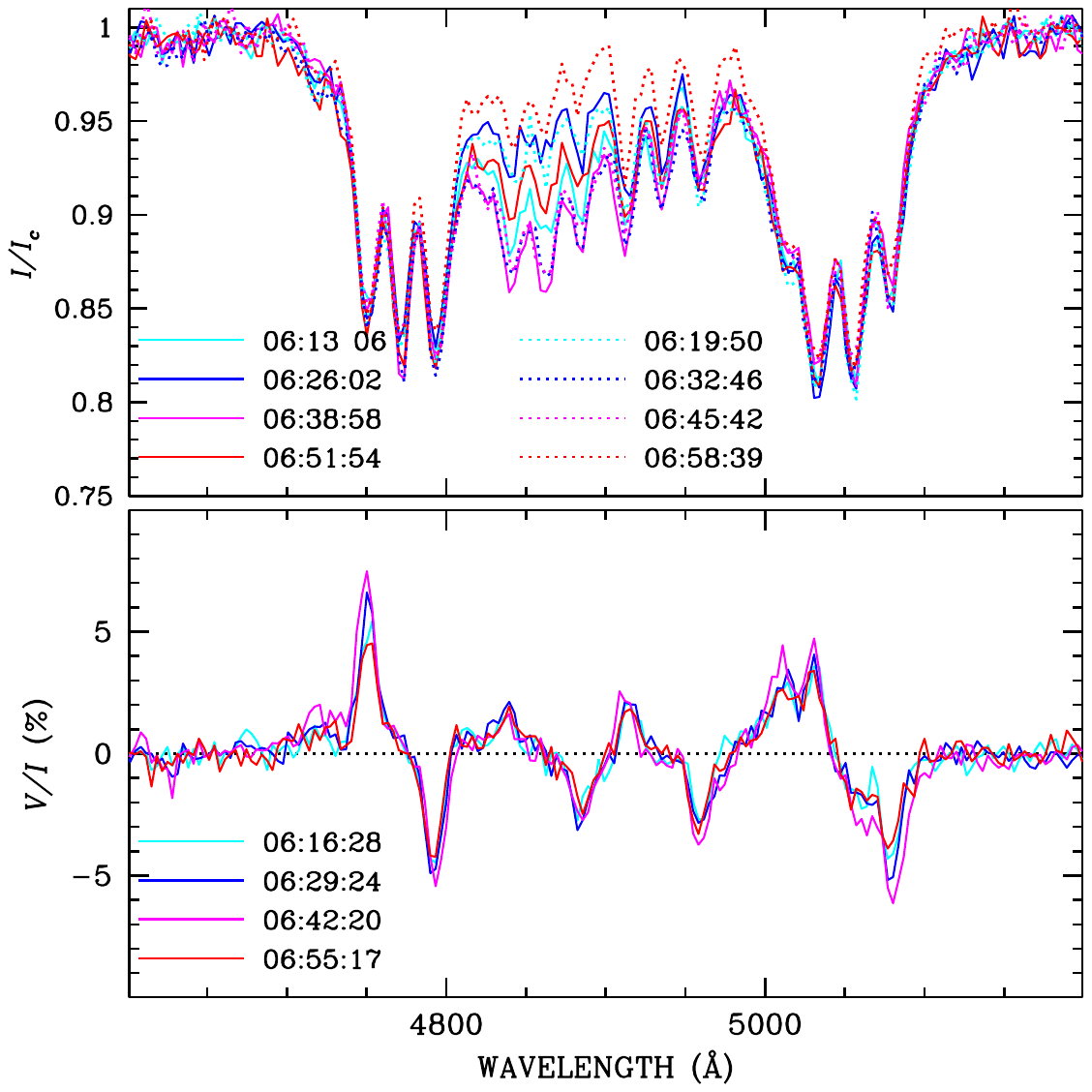}\\
\end{center}
\caption{\label{Fig_WD1750} WD~1750$-$311 observed with FORS2 on 2023-06-12 with grism 300V.}

\end{figure}
A magnetic field of $\bs \approx 2.1$\,MG was identified by \citet{Macetal17} in this hot DQ white dwarf, which has strong lines of neutral C as well as fairly strong Balmer lines. They observed clear light variability of the star with an amplitude of about $\pm 2$\,\% and a period of 35 min.  These authors discuss the origin of the light variations: they argue that the variability is not due to pulsation, as no subsidiary frequencies are observed, and conclude that the variation could be due to rotation. They do not seem to have considered testing this with the spectra that they have collected of the object by looking for spectral variations, although they do test the spectra for radial velocity variations, and find none.  

Our four FORS2 300V spectra, taken with 13\,m spacing through the light variation cycle, show virtually no changes in either flux or polarisation {except} for H$\beta$, which has a strongly variable $I$ profile depth but almost constant $V/I(\lambda)$ (see Fig.~\ref{Fig_WD1750}). Possibly this is due to a magnetic field distribution that is nearly axisymmetric about the rotation axis but a distribution of H that varies strongly around that axis. This might be related to the light variability that has been detected. We have decided to classify the star as candidate magnetically non-variable.

\subsection{WD1754$-$550 = GALEX J175845.9-550117 (variable)}
\begin{figure}
\begin{center}
\includegraphics[angle=0,width=8.8cm,trim={2.0cm 6.0cm 1.0cm 11.5cm},clip]{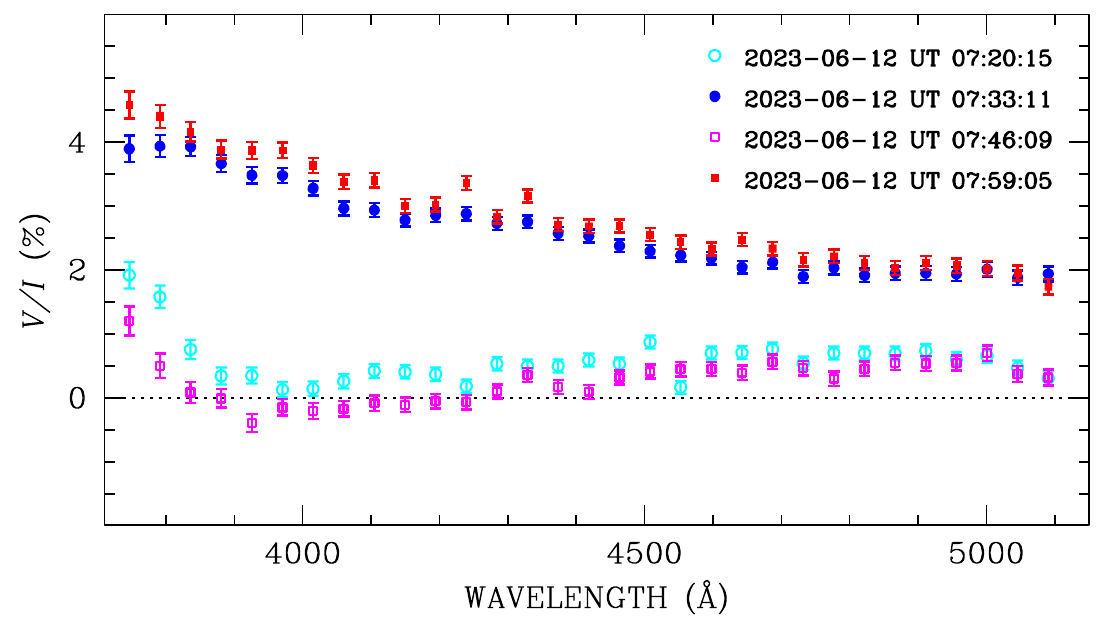}\\
\end{center}
\caption{\label{Fig_WD1754} FORS2 $V/I$ spectra of WD~1754--550, rebinned at 45\,\AA\ steps, from four
exposure pairs  obtained on 2023-06-12 at the times shown in the legend. The intensity spectrum is 
featureless.}
\end{figure}
This star was discovered to be magnetic in this work (Sect.~\ref{Sect_New_Observations}). There is a strong signal of circular polarisation
that seems variable on a short timescale between 0 and 4\%, suggesting a rotation period of the order of 15\,min: see Fig~\ref{Fig_WD1754}.
The inferred field strength is presumably of the order of 100\,MG or more. The intensity spectrum appears featureless, so the $\teff \simeq 35\,000$\,K
star could be defined a hot DC.

\subsection{WD~1814+248 = G~183--35 (variable)}
\citet{Putney97} showed that this white dwarf, formerly classified as a DC, is in fact a cool DAH that shows weak H$\alpha$ and H$\beta$, and that both lines are split by a field of about 6.8\,MG, roughly the same in two observations.  \citet{Kiletal19} observed that the white dwarf shows spectral variations. A series of spectra taken over several hours show that the separation of the $\sigma$ components of H$\alpha$ from the central $\pi$ component changes rather abruptly between about 90\,\AA\ and about 120\,\AA, equivalent to fairly sudden jumps between \bs\ values of about 4.6 and 6 MG, repeating with a period of about 4\,h. The authors suggest that this unusual form of variability may be due to a patchy distribution of H over an He-rich envelope.  A plausible model that might explain the observations could be a dipole oblique to the rotation axis, decentred in the direction of one pole so that the polar strengths at the two magnetic poles are unequal, with H-rich  patches around both poles, but little or no H in a belt around the magnetic equator. 

\subsection{WD~1829+547 = G~227-35 (non-variable)}%
\begin{figure}
\begin{center}
\includegraphics[angle=0,width=9.0cm,trim={1.3cm 1.9cm 1.0cm 1.0cm},clip]{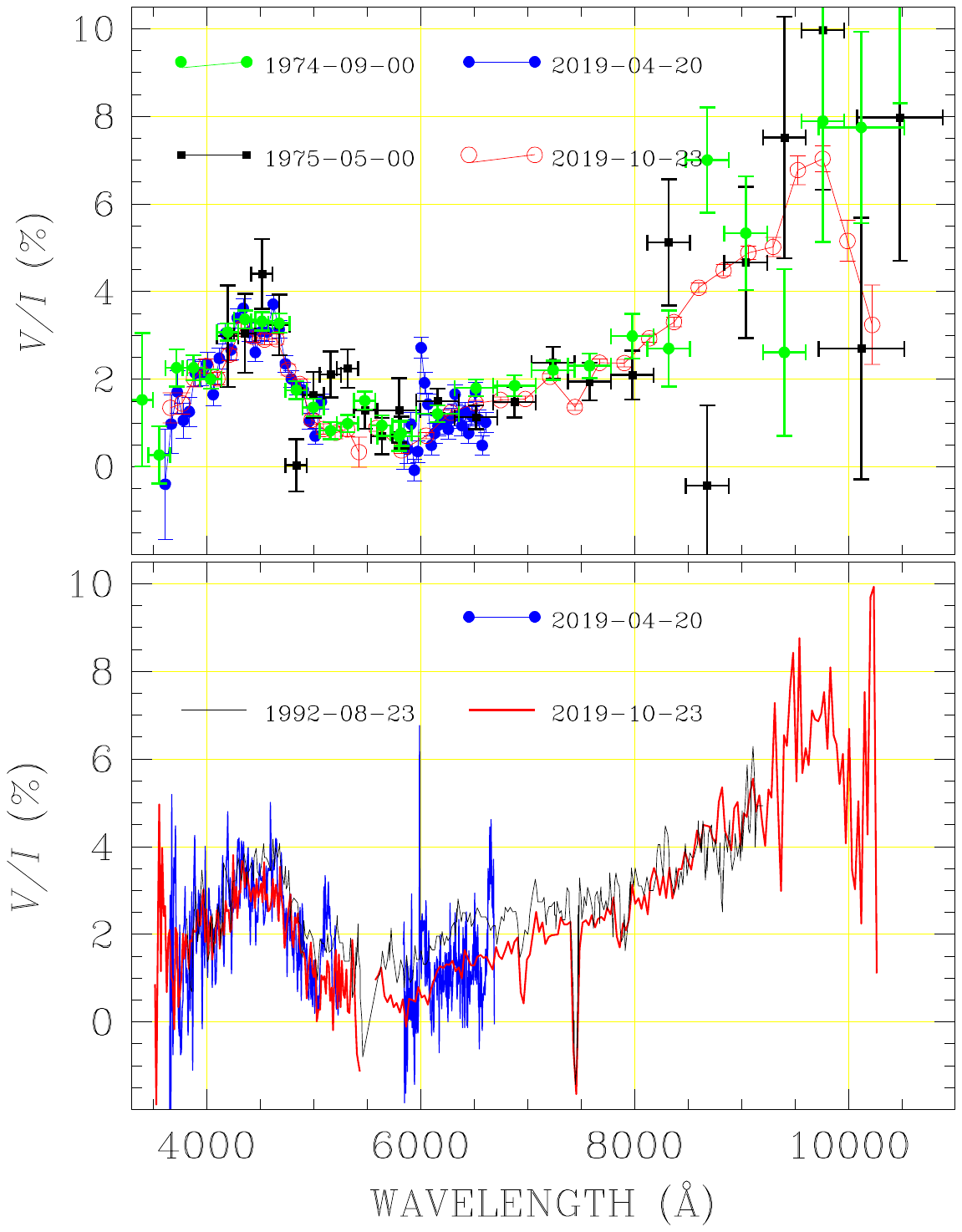}
\end{center}
\caption{\label{Fig_G227var} Polarisation spectra of WD\,1829$+$547 = G227-35. {Top panel:} Comparison between MP spectra obtained in 1974 and our ISIS spectra obtained in 2019, after degrading their resolution. {Bottom panel:} Comparison between ISIS spectra obtained in 2019 and a spectrum obtained by \citet{PutJor95} in August 1992.}
\end{figure}
This white dwarf was discovered to be strongly magnetic by \citet{Angetal75} using both broadband polarimetry and low-resolution spectropolarimetry (on September 9, 1974). Limited tests of broadband variability were negative \citep{Angetal75,Angetal81}. It was observed again in spectropolarimetric mode by \citet{Cohetal93} and by \citet{PutJor95} who estimated a dipolar field strength of 170-180~MG. We observed the same star with ISIS twice in circular polarisation and once in linear polarisation. There are strong circular polarisation features at 6960, 7450 and 7930\,\AA. Figure~\ref{Fig_G227var} shows a comparison between all these spectra. We do not see any obvious sign of variation. 

\subsection{WD~1900+705 = LAWD 73  = Grw~$+70\degr$~8247 (s.v.:)} %
The star has a magnetic field of the order of 180\,MG \citep{Jor03}, and shows hints of small variability. We refer to \citet{BagLan19a}, who present a review and analysis of its polarimetric characteristics, and in particular, a comparison with observations obtained 50 years apart show some discrepancies in both linear and circular polarisation. More recent broadband circular polarisation obtained at NOT \citep{Beretal22,Beretal23} show no variability on a timescale of 2 years (see Fig.~\ref{Fig_BBCP}).

\subsection{WD~1953--011 = GJ 772 (variable, $P \simeq 1.5$\,d)}%
\citet{Maxetal00} modelled a series of Stokes~$I$ spectra in terms of a global dipolar field with polar field strength of order 100~kG, together with a spot with a field of order 500~kG. This basic modelling was confirmed by the analysis of a series of polarised spectra obtained with FORS1 and on the Russian 6 m telescope, which also revealed a rotation period of 1.448\,d \citep{Valetal08}.

\subsection{WD~2010+310 = GD 229 (s.v.:) }

\begin{figure}
\begin{center}
\includegraphics[angle=0,width=9.0cm,trim={0.7cm 1.2cm 1.0cm 1.0cm},clip]{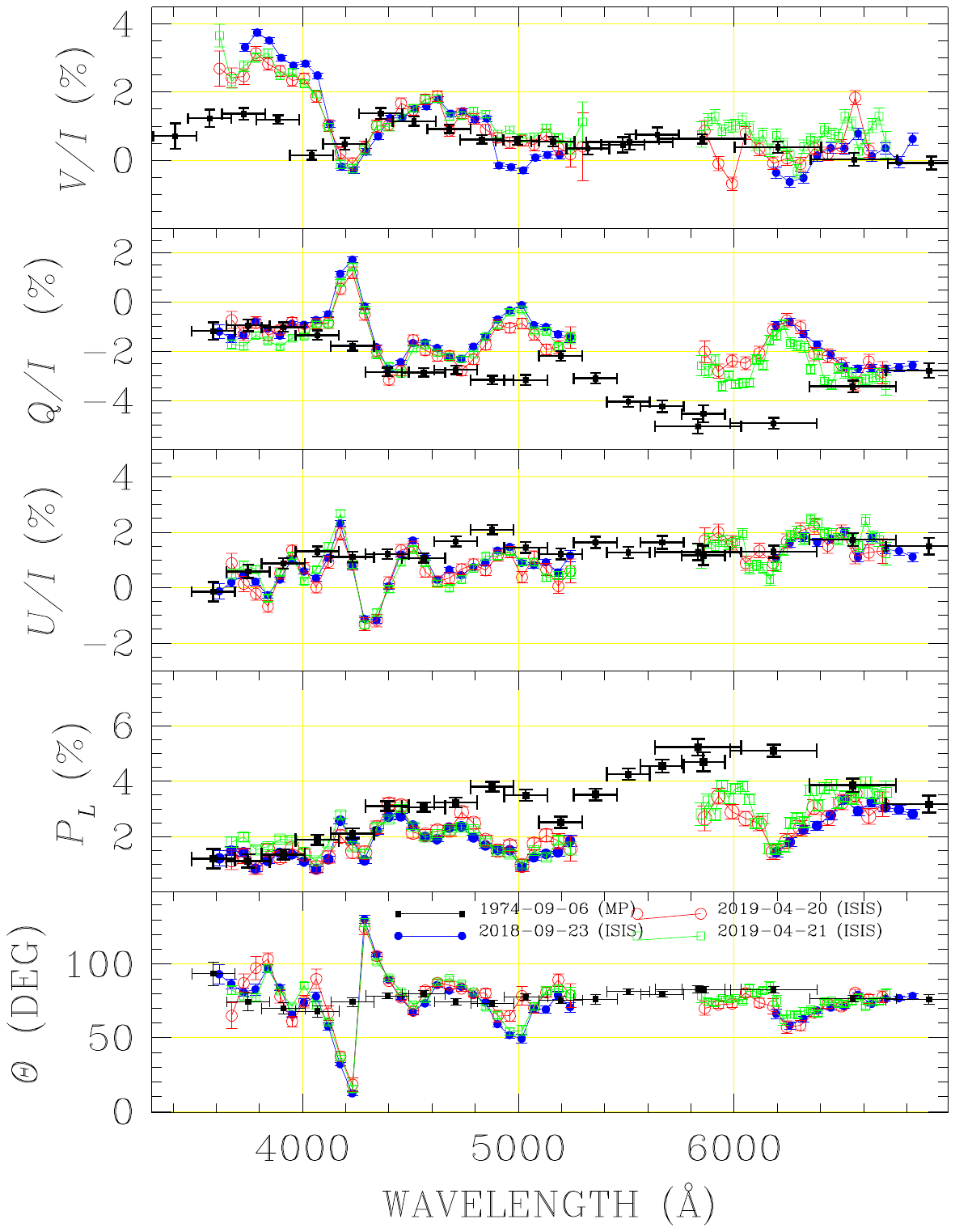}
\end{center}
\caption{\label{Fig_WD2010} MCSP linear and circular polarisation spectra of WD\,2010+310 = GD\,229.}
\end{figure}

This star was the fifth circularly polarised white dwarf to be discovered. \citet{Sweetal74} reported a series of measurements indicating the presence of elevated levels of both circular polarisation and linear polarisation, and strongly suggesting polarisation variability. The claim of variability was then questioned by \citet{Kemetal74}, who argued that the initial measurements had been badly contaminated by polarised foreground zodiacal light contamination. 

Linear and/or circular polarisation of GD\,229 has been measured by \citet{LanAng74}, \citet{Efimov81}, \citet{Angetal81}, \citet{West89}, \citet{Berdyugin95}, and \citet{BerPii99}.

\citet{AngLan74} obtained three low-resolution $I$ and $V$ spectra of the star on the nights of 1973 Nov 6--8 using the MCSP. Synthetic broad-band polarimetry derived from these low-resolution spectra revealed no variations over three nights. 

\citet{Wes89} observed linear polarisation in GD\,229 again in 1986 using broadband filters, but found no strong evidence of variation, not even of the position angle of linear polarisation. 

Observations of GD\,229 using broadband polarimetry was also carried out for linear polarisation by \citet{Berdyugin95} and for both circular and linear polarisation, with higher precision,  by \citet{BerPii99}, who compared their results to earlier work. Their results appear to show some quite significant changes in both circular and linear polarimetry compared to earlier observations by other groups. A major difficulty of comparing the various measurements is that each group has been made with different instrumentation. The consequences of using different filter passbands, different detector sensitivities as functions of wavelength, and even different (and possibly incorrect) calibration of instrumental polarimetric efficiency, make it very hard to compare these data. However, \citet{BerPii99} do offer strong evidence for significant rotation of the  angle of linear polarisation, by about $30^\circ$ over 20 years, which is difficult to explain by instrumental effects. This is probably the most robust effect that emerges from comparison of the many kinds of polarimetric observations 

Linear and circular polarisation spectra of GD\,229 were obtained in 2018 and 2019 using the ISIS spectropolarimeter on the \textit{William Herschel} Telescope (Sect.~\ref{Sect_New_Observations}). These new data are shown in Fig.~\ref{Fig_WD2010}, and compared to the circular polarised spectra by \citet{AngLan74}, and to previously unpublished linear spectropolarimetry of Angel and Landstreet (see Table~\ref{Tab_Log}). Remarkably, the angle of linear polarisation appears to have returned to its value during the 1970s. Other differences compared to the spectra of Angel and Landstreet are observed, but some of these are undoubtedly due to greatly different resolving power, and some may be due to uncertainties in the calibration of the 1970s data, particularly in the UV. 

It is thus difficult to establish clearly how much variability has occurred in GD\,229. There is evidence for variations, but these appear to have relatively small amplitude compared to the overall scale of polarisation, except possibly for position angle rotation of linear polarisation (which however has not been confirmed by our most recent linear spectropolarimetry). Furthermore, because of very limited sampling with a variety of instruments, it is not possible to establish clearly any particular timescale for variations.

We note that the spectrum of GD\,229 has been modelled as due to He in a field of hundreds of MG by \citet{Joretal98,Joretal01} and \citet{Jor03}.

\subsection{WD~2047+372 = EGGR 261 (variable, $P \simeq 6$\,h)}%
Originally, the star was observed polarimetrically by \citet{SchSmi95}, who did not detect its weak and sign reversing field (their measurements were $\bz = -42 \pm 59$\,kG and $-2.5 \pm 4.6$\,kG). 

This star was discovered to be magnetic by \citet{Lanetal16}, and monitored and modelled by \citet{Lanetal17}. It is currently the weakest white dwarf field ($\bs \approx 60$\,kG) that has been modelled in detail,  mainly on the basis of a series of 18 ESPaDOnS spectra. The rotation period, determined from the variation of \bz, is 0.243~d. The observed variations of \bz\ and \bs\ are well modelled using a simple dipole model. 

No rotational variation is detected in TESS photometry \citep{Heretal24}.  

\subsection{WD~2049--222 =  LP 872-48 (var.:)}\label{Sect_WD2049}
\begin{figure}
\begin{center}
\includegraphics[angle=0,width=8.8cm,trim={1.0cm 6.0cm 1.0cm 11.5cm},clip]{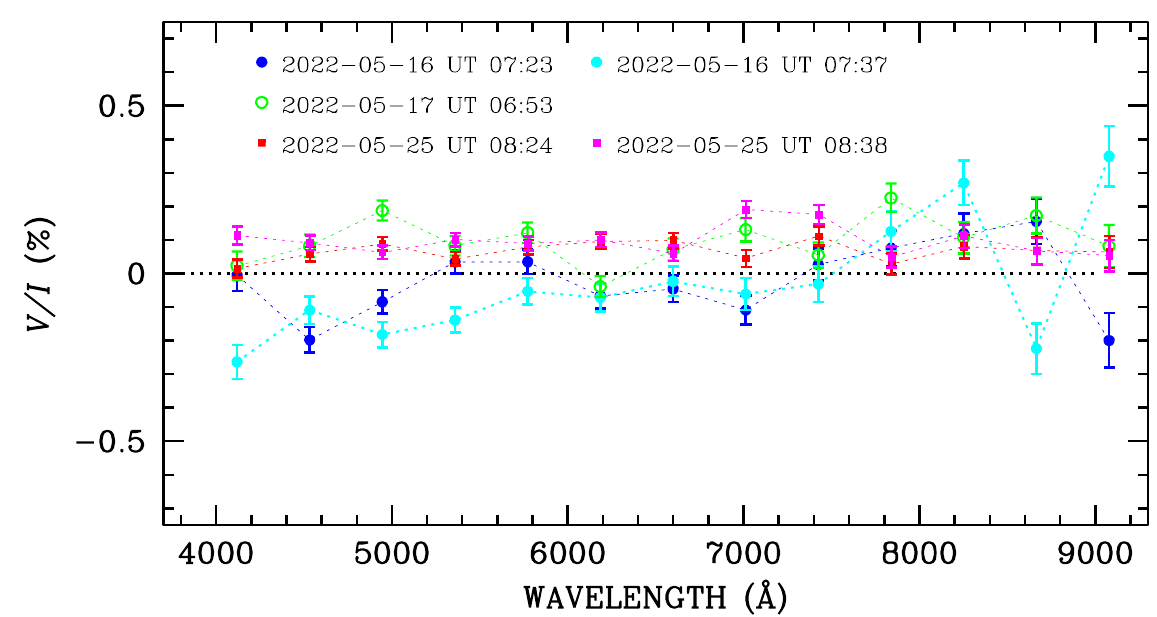}
\end{center}
\caption{\label{Fig_WD2049} Circular polarisation spectra of WD\,2049$-$222, rebinned at $\simeq 410$\,\AA, and obtained
at the time specified in the legend.}
\end{figure}
Discovered to be magnetic by \citet{Beretal22} with BBCP measurements. The star has $V/I \simeq +0.1$\,\%, and is one of the weakest polarisation levels securely detected. The inferred \bz\ field strength is only of the order of a few MG. Broad-band measurements were repeated in July 2022 \citep{Beretal24} to check for variability, which was not detected (see Fig.~\ref{Fig_BBCP}). We also observed the white dwarf three times with FORS2 with grism 300V (Sect.~\ref{Sect_New_Observations}). The data are barely above the
threshold of instrumental polarisation \citep[$\simeq 0.1$\%; see][]{Sieetal14}, and therefore it is hard to establish whether the hints of variability seen in the spectra are real or not (see Fig.~\ref{Fig_WD2049}). However, star WD\,1116--440 shows a similar level of polarisation, and constant over about 4 year (see Fig.~\ref{Fig_WD1116}), suggesting that the tiny variability observed in WD\,2049--222 may be real and not an instrumental artefact. 

\subsection{WD~2049--253 = UCAC4 325-215293 (non-variable)}%
Discovered to be magnetic by \citet{BagLan20}, who observed continuum circular polarisation of order 0.4\%\ and deduced a field strength of order 20\,MG. This white dwarf was re-observed in broadband circular polarisation once by \citet{Beretal22} and two more times by \citet{Beretal24}, but shows no sign of variability (see Fig.~\ref{Fig_BBCP}).

\subsection{WD~2051-208 = BPS CS 22880-0134 (variable, $P \simeq 1.5$\,h)}
The magnetic field of this white dwarf was discovered from the shape of the H$\alpha$ line by \citet{Koeetal09}. 

We have two series of five polarised spectra each, one using FORS grism 1200B and one with grism 1200R (LB25). These unpublished data clearly reveal very rapid rotation of this massive magnetic white dwarf. The stellar rotation period is 0.0594~d = 1.425\,h. The value of \bz\ ranges approximately sinusoidally between about +50~kG and --30~kG, while the corresponding values of \bs\ increase from about 200~kG to nearly 300~kG, in good agreement with the two values of <|B|> (220 and 290~kG) obtained by \citet{Koeetal09} from UVES SPY spectra. 

\subsection{WD~2105--820 = LAWD 83 (variable)}
This star was suspected to have a weak magnetic field by \citet{Koeetal98}, who observed that the core of H$\alpha$ was abnormally broad, but they could not decide whether this was due to rapid rotation of $v \sin i \approx 65$~km/s or to a magnetic field of about 43~kG. Five FORS1 polarised spectra of the star by \citet{Lanetal12} detected an apparently nearly constant magnetic field of $\bz \approx 10$\,kG. Later, FORS2 polarised spectra by \citet{BagLan18} and \citet{Faretal18} reveal that \bz\ sometimes decreases to $\bz \approx 4$~kG, so the star is apparently variable \citep[see][]{BagLan21}.

\subsection{WD~2138--332 = L 570-26 (variable, $P \simeq 6.19$\,h)}
The DZA star was discovered to be magnetic, with a variable \bz\ of the order of 10\,kG \citep{BagLan19b} and a rotational period of $P=6.19$\,h \citep{Heretal24, Faretal24}. The same period was found from the analysis of the equivalent width and \bz\ curves by \citet{Bagetal24b}, who proposed a magnetic model with a dipolar field with $\bs \simeq 50$\,kG. The star shows photometric and \bz\ curves with light minimum corresponding to \bz\ maximum.

\subsection{WD~2150+591 = UCAC4 747-070768 (variable, $P \simeq 2.4$\,d)}%
The star was discovered to be magnetic by \citet{LanBag19a}, who reported two ISIS measurements that showed clearly that the field is variable with a period of hours or days.  We subsequently monitored the star with one ESPaDOnS observation and several more ISIS spectra, and confirmed variability. These observations and a model of the star's magnetic field will be presented in a forthcoming paper (LB25). 

\subsection{WD~2211+372 =  LP 287-35 (non-variable)} 
This DC white dwarf was discovered to be magnetic by \citet{Beretal23}, who found circular polarisation in excess of 1\%\ in the blue. It was re-observed by \citet{Beretal24}. It does not seem to be variable (see Fig.~\ref{Fig_BBCP}). The field strength \bs\ is probably of the order of 60\,MG or more.

\subsection{WD~2316+123 = KUV~23162+1220 (variable, $P \simeq 18$\,d)}
Discovered magnetic by \citet{Sioetal84}. \citet{SchNor91} reported BBCP of amplitude up to nearly 1\% that varies sinusoidally with P = 17.86 d. Both the flux spectrum and the linear and circular polarisation spectra have been modelled repeatedly \citep{Lieetal85,AchWic89,Frietal93,PutJor95}; all agree that the global field strength is of the order of 30\,MG. This star has the longest rotational period firmly established for a white dwarf.

\subsection{WD\,2359$-$434 = LAWD 96 (variable, $P \simeq 2.7$\,h) }\label{Sect_Last}
WD\,2359$-$434 was suggested to be a magnetic star by \citet{Koeetal98} on the basis of the peculiar profile of the H$\alpha$ core, and \bz\ was found to be non-zero by \citet{Aznetal04}. A series of polarised ESPaDoNS spectra revealed a rotation period of 0.1123\,d \citep{Lanetal17}. The star is also a photometric variable with the same period. The field structure has been modelled approximately, and found to be distinctly more complex than a simple dipole, with \bs\ varying approximately between 50 and 100\,kG. 
This white dwarf offers one of the clearest examples known of a field structure that is substantially more complex than a simple co-linear multipole expansion \citep{Lanetal17}.

\end{document}